# *E. coli* chemotaxis is information-limited

H.H. Mattingly[1,2,†], K. Kamino[1,2,†], B.B. Machta[3,4,*], T. Emonet[1,2,3,*]

Affiliations:

[1] Department of Molecular, Cellular, and Developmental Biology, Yale University.
[2] Quantitative Biology Institute, Yale University.
[3] Department of Physics, Yale University.
[4] Systems Biology Institute, West Campus, Yale University.

† Equal contribution.
* Correspondence to: thierry.emonet@yale.edu and benjamin.machta@yale.edu

## Abstract

Organisms must acquire and use environmental information to guide their behaviors. However, it is unclear whether and how information quantitatively limits behavioral performance. Here, we relate information to behavioral performance in *Escherichia coli* chemotaxis. First, we derive a theoretical limit for the maximum achievable gradient-climbing speed given a cell's information acquisition rate. Next, we measure cells' gradient-climbing speeds and the rate of information acquisition by the chemotaxis pathway. We find that *E. coli* make behavioral decisions with much less than the 1 bit required to determine whether they are swimming up-gradient. However, they use this information efficiently, performing near the theoretical limit. Thus, information can limit organisms' performance, and sensory-motor pathways may have evolved to efficiently use information from the environment.

Organisms' survival depends on their ability to perform behavioral tasks. These tasks can be challenging because they require that the organism measure signals in its environment and respond appropriately, in real time. Information theory is a natural language for quantifying the fidelity of measurements and responses, but it is unclear how an abstract quantity like information might limit an organism's performance at real-world tasks. Past studies have used information theory to understand the maximum amount of information biological systems can acquire and transmit about environmental signals [1–6] and have shown that they can approach certain theoretical limits [6–9]. But high information transfer is not sufficient for high performance because not all of the information contained in the signal is behaviorally relevant, and not all of it is appropriately acted on [10]. What limits does information place on performance, and how efficiently do organisms use the information they acquire relative to these limits?

We address these questions using one of the best-understood behaviors in biology: bacterial chemotaxis. The bacterium *Escherichia coli* alternates between "runs," which propel the cell forward, and "tumbles," which randomly reorient its swimming direction [11] (Fig. 1). Along its path, *E. coli* continuously senses the concentration $c(t)$ of chemoattractant it encounters using transmembrane receptors. Relative changes in concentration [12] $s(t) = \frac{d}{dt}\log(c)$ along the cell's swimming trajectory, which we define to be the "signal", induce changes in activity $a(t)$ of receptor-associated CheA kinases. CheA activity goes on to modulate transitions in the cell's motor behavior $m(t)$ between run and tumble states via a signal transduction pathway [13,14]. If the relative concentration of attractant is increasing ($s(t) > 0$), the cell tends to run for longer on average, thereby biasing its random motion up the gradient [11]. However, noise in sensing and signal transduction corrupt the signal [15–18]. Since the goal of chemotaxis is to climb chemical gradients, performance can be quantified by the cell's drift velocity $v_d$ up a static gradient. *E. coli* chemotaxis has been studied extensively, but the amount of information a cell acquires about chemical signals and its relationship to chemotactic performance are unclear.

Quantifying information transfer during behavioral tasks presents new challenges. First, bacteria continuously make decisions about whether to tumble. Thus, unlike most other studies in biological systems [1–3,5,6], we cannot use the one-shot or instantaneous mutual information [19] between signal $s(t)$ and behavior $m(t)$ as a measure of information transfer. Instead, we need an information rate [20]. Second, the signals are generated by the cell's own motion in the gradient [21]. Thus, a natural extension, the mutual information rate $\dot{MI}$ between signal $s(t)$ and behavior $m(t)$, is unsuitable because signal-motor correlations make it non-zero even for non-responsive cells (Fig. 1AB). We address these challenges by isolating the information that flows from signal to behavior. The mutual information rate can be decomposed into the sum of two directed information terms [22] (SI), the transfer entropy [23] rates from behavior to signal $\dot{I}_{m \to s}$ and from signal to behavior $\dot{I}_{s \to m}$, or $\dot{MI} = \dot{I}_{m \to s} + \dot{I}_{s \to m}$. The first term $\dot{I}_{m \to s}$ quantifies the feedback of behavior onto signal. The second term, $\dot{I}_{s \to m}$, which we refer to here as the information rate, in bits/s, measures the directed influence of the signal on behavior and must be nonzero for the cell to climb the gradient.

We first used rate-distortion theory [6,24,25] to derive the theoretical bound on performance imposed by information transfer. To do this, we constructed a model of run-and-tumble navigation that is abstracted from *E. coli*'s molecular implementation (Fig. 1A; SI). During runs, cells swim with an effective, constant speed $v_0$ (Fig. 1A) and lose direction with rotational diffusion coefficient $D_r$. During tumbles, they reorient randomly with directional persistence $\alpha$. In the absence of a gradient, cells switch from run to tumble and vice versa with constant rates $\lambda_{R0}$ and $\lambda_T$. In a gradient, tumbles are instead initiated at a rate $\lambda_R(\{s\})$, which depends on the past history of signals $\{s\}$. Throughout, we consider navigation in shallow gradients, where information is most likely to be a limitation on chemotactic performance.

Using this model, we derived how performance $v_d$ and information transfer $\dot{I}_{s \to m}$ depend on the tumble rate response $\lambda_R(\{s\})$. While any response to signals implies information transfer, it does not necessarily imply high drift speed. We derived the maximum drift speed $v_d$ possible given an information rate $\dot{I}_{s \to m}$ by optimizing over responses $\lambda_R(\{s\})$. In the SI, we show that in this model the optimal response only depends on the current rate of change of log-concentration, $s(t)$. Because bacteria necessarily make comparisons of concentrations over a finite time to infer $s(t)$ from stochastic arrivals of ligand molecules at the cell surface [15], they can't implement the optimal behavioral response. Nevertheless, the performance achieved by any response is bounded by (Fig. 1C):

$$v_d/v_0 \leq F(\boldsymbol{\theta}) \left( \frac{\log(2) \, \dot{I}_{s \to m}}{12 \, D_r} \right)^{\frac{1}{2}}, \text{ where } 0 \leq F(\boldsymbol{\theta}) \leq 1, \boldsymbol{\theta} = \{\lambda_{R0}, \lambda_T, \alpha\}. \quad (1)$$

$F(\boldsymbol{\theta}) = \frac{(1-\alpha) \lambda_{R0}}{(1-\alpha) \lambda_{R0} + 2 D_r} \left( 8 \frac{D_r}{\lambda_{R0}} P_{run} \right)^{\frac{1}{2}}$ is a dimensionless function of the behavioral parameters (see Fig. 1C for a definition). This expression makes rigorous the intuition that information transfer sets a limit on how fast a cell can climb a gradient [15,26,27].

To quantify the bound above, we measured the rotational diffusion coefficient $D_r$ and behavioral parameters $\boldsymbol{\theta}$ from trajectories of swimming *E. coli* cells (Figs. S1, S2). Individual cells in a clonal population exhibit nongenetic differences in behavioral parameters $\boldsymbol{\theta}$ [28–30], which in *E. coli* are highly correlated with $P_{run}$, the fraction of time the cell is running $P_{run}$ [28,30] (Fig. S1,S2). From this data, we find $F(\boldsymbol{\theta}) = 0.531 \pm 0.005$ for the median phenotype ($\pm$ one standard error; parameters are in Table S1). Perhaps surprisingly, the bound predicts that run-and-tumble navigation is theoretically possible with very small information rates: a hundredth of a bit per second is sufficient to climb gradients at ~6% of the run speed. This is far less than the 1 bit per run (~1 bit/s) required to distinguish whether concentration is currently increasing or decreasing before every tumble decision [20].

Our central questions are: how much sensory information do *E. coli* acquire, and how efficiently do they use that information? Eqn. 1 was derived by considering a model of *E. coli* in which signaling details have been coarse-grained away. But information about signals is acquired upstream in the *E. coli* chemotaxis pathway and must be communicated to the motors. Since this signal transduction process can only lose information [31], the information available at any intermediate step within the signaling pathway also places a limit on the cell's performance, as in Eqn. 1. Intuitively, in an information-efficient cell, most sensory information acquired upstream would be preserved at the motors and contribute to gradient-climbing. If so, the cell should climb gradients at speeds approaching the theoretical bound (Fig 1C). Alternatively, cells could acquire abundant information about the signal but either lose much of it during transfer to behavior or fail to act on it appropriately. For these cells, performance would be further away from the bound. To quantify where *E. coli* lie on this spectrum, we define information efficiency, $\eta$, as the ratio of the cell's chemotactic performance $v_d$ to the maximum performance possible with the amount of sensory information it acquires, dictated by Eqn 1.

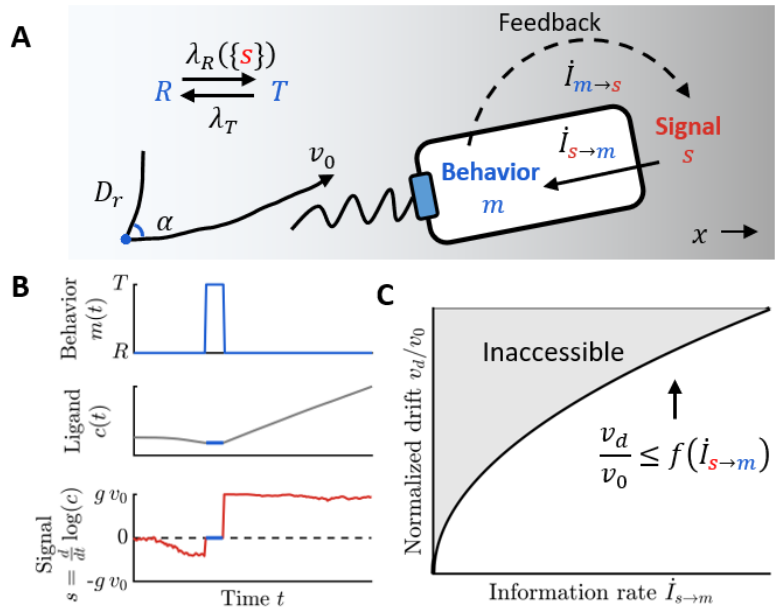

**Figure 1. Information flow from signal to swimming behavior sets a limit on chemotaxis performance.**
**A)** A cell navigates chemical gradients by sensing relative changes $s(t) = \frac{d \log(c)}{dt}$ in attractant concentration $c$ over time. The cell stochastically transitions between behavioral states $m$: a run state $R$, in which the cell swims with constant speed $v_0$ and rotational diffusion $D_r$; and a tumble state $T$, in which it reorients randomly with directional persistence $\alpha$ (see Fig. S1). The cell responds to past signals $\{s\}$ by changing its transition rate $\lambda_R(\{s\})$ from run to tumble. The average tumble rate $\lambda_{R0}$ and run rate $\lambda_T$ define the fraction of time the cell spends in the run state, $P_{run} = \frac{\lambda_T}{\lambda_{R0}+\lambda_T}$. The cell's motion creates the signal that it experiences (dashed arrow). The transfer entropy rate from behavior to signal $\dot{I}_{m \to s}$ quantifies this information flow, which is non-zero even for a non-chemotactic cell. Signal affects the cell's behavior through $\lambda_R(\{s\})$ (solid arrow), and the transfer entropy rate from signal to behavior $\dot{I}_{s \to m}$ quantifies this information flow needed to perform chemotaxis. **B)** Time courses of behavior $m(t)$ (top), concentration $c(t)$ (middle), and signal $s(t)$ (bottom) for the trajectory shown in (A). In static gradients, signals arise entirely from the cell's motion, $s(t) = g\, v_x(t)$, where $g = d \log(c)/dx$ is the steepness of the log-concentration gradient and $v_x$ is the cell's up-gradient velocity. **C)** The information sent from signal to behavior $\dot{I}_{s \to m}$ sets an upper limit on chemotaxis performance, defined as the up-gradient drift speed $v_d$ relative to $v_0$. With information rate $\dot{I}_{s \to m}$, the cell's drift speed is bounded by Eqn. 1.

Quantifying *E. coli*'s information efficiency requires measuring the rate at which cells acquire information during chemotaxis. The activity $a(t)$ of receptor-associated CheA kinases is among the first steps of *E. coli*'s signaling pathway that computes concentration changes [14]. Therefore, we consider the rate of information transfer from signals to kinase activity $\dot{I}_{s \to a} \geq \dot{I}_{s \to m}$ to be the cell's information acquisition rate. To quantify $\dot{I}_{s \to a}$ we measured the signal statistics cells experience during navigation, as well as the response and noise properties of the kinases in immobilized cells [20,32] (Fig. 2). This approach eliminates the feedback from kinase activity onto the signals cells experience, allowing us to combine these measurements to estimate the information acquisition rate $\dot{I}_{s \to a}$ during navigation (Fig. 2A; SI).

The input signal statistics are characterized by their power spectrum $S(\omega)$. It is often difficult to know the natural signal statistics an organism experiences [7,33]. But during bacterial chemotaxis in static gradients, the signal is generated from the cell's own motion in the gradient. Thus, the signal power spectrum is $S(\omega) = g^2 V(\omega)$, where $V(\omega)$ is the power spectrum of the cell's velocity along the gradient direction and $g = d \log(c)/dx$ is the steepness of the log-concentration gradient. Furthermore, in shallow gradients, the statistics of the cell's motion are nearly identical to those in the absence of a gradient (SI). To quantify the $x$-velocity autocorrelation function $V(t)$, we tracked individual swimming trajectories in a constant 100 µM background of the attractant α-methyl-aspartate (MeAsp) (Fig. 2CD; Fig. S2; Methods; SI; ~$10^4$ s of total trajectory time in the bin of median $P_{run}$, 7 s average trajectory duration). For different values of $P_{run}$, we fit each of the measured $V(t)$ with decaying exponentials, $V(t) = a_v \, e^{-\lambda_{tot}|t|}$. For the median phenotype, the best-fit parameters were $a_v = 157.1 \pm 0.5$ (µm/s)$^2$ and $\lambda_{tot} = 0.862 \pm 0.005$ s$^{-1}$. $V(\omega)$ was then computed by taking the Fourier transform of $V(t)$ (Fig. S7).

Having quantified the input statistics, we turned to measuring the response and noise properties of CheA kinase activity (SI) using Förster resonance energy transfer (FRET) between the kinase's substrate CheY and the phosphatase CheZ inside single cells [17,18,34] (Fig. 2B). In shallow gradients, cells experience weak signals, and therefore the average kinase response depends linearly on recent signals. In this regime, the response is fully characterized by how it amplifies different frequencies in the signal, $K(\omega)$, or equivalently, the response to an impulse (delta function) of signal, $K(t)$. To measure the impulse response $K(t)$ to MeAsp, we used a microfluidic device that allows us to rapidly switch (in ~100 ms) the concentration of attractant delivered to hundreds of immobilized cells [34] (Figs. S3). To ensure cells were in the log-sensing regime [12,35], we first adapted them to a background of 100 µM MeAsp. We then delivered 10 positive and 10 negative 10% step changes of MeAsp concentration (corresponding to delta functions of signal $s$) (Fig. 2E; Methods), small enough that the cells' responses were in the linear regime [36,37] (Fig. S5). Individual cell responses exhibited a stereotypical shape (Fig. 2F) that was well-described by a phenomenological model $K(t) = G \, e^{-t/\tau_2}(1 - e^{-t/\tau_1}) \, H(t)$, where $G$ is the gain, $\tau_1$ is the rise time, $\tau_2$ is the adaptation time, and $H(t)$ is the Heaviside step function. We fit this model to each cell's average responses to the positive and negative stimuli simultaneously and then determined the population median values of the parameters (± one standard error; $n = 442$ cells) (Table S1; Fig. S5): $G = 1.73 \pm 0.03$, $\tau_1 = 0.22 \pm 0.01$ s, and $\tau_2 = 9.9 \pm 0.3$ s. The value of $\tau_1$ that we inferred includes the CheA kinase response time, the stimulus concentration switching time, and the kinetics of CheY/CheZ binding, making it longer than the actual kinase response time. The kinase response time alone has been measured before to be $\tau_1 \sim 0.05$ s [38], but is not directly accessible using this FRET system. After verifying that our results are not sensitive to the value of $\tau_1$ (SI; Fig. S8), we used the literature value in our estimate of the information rate.

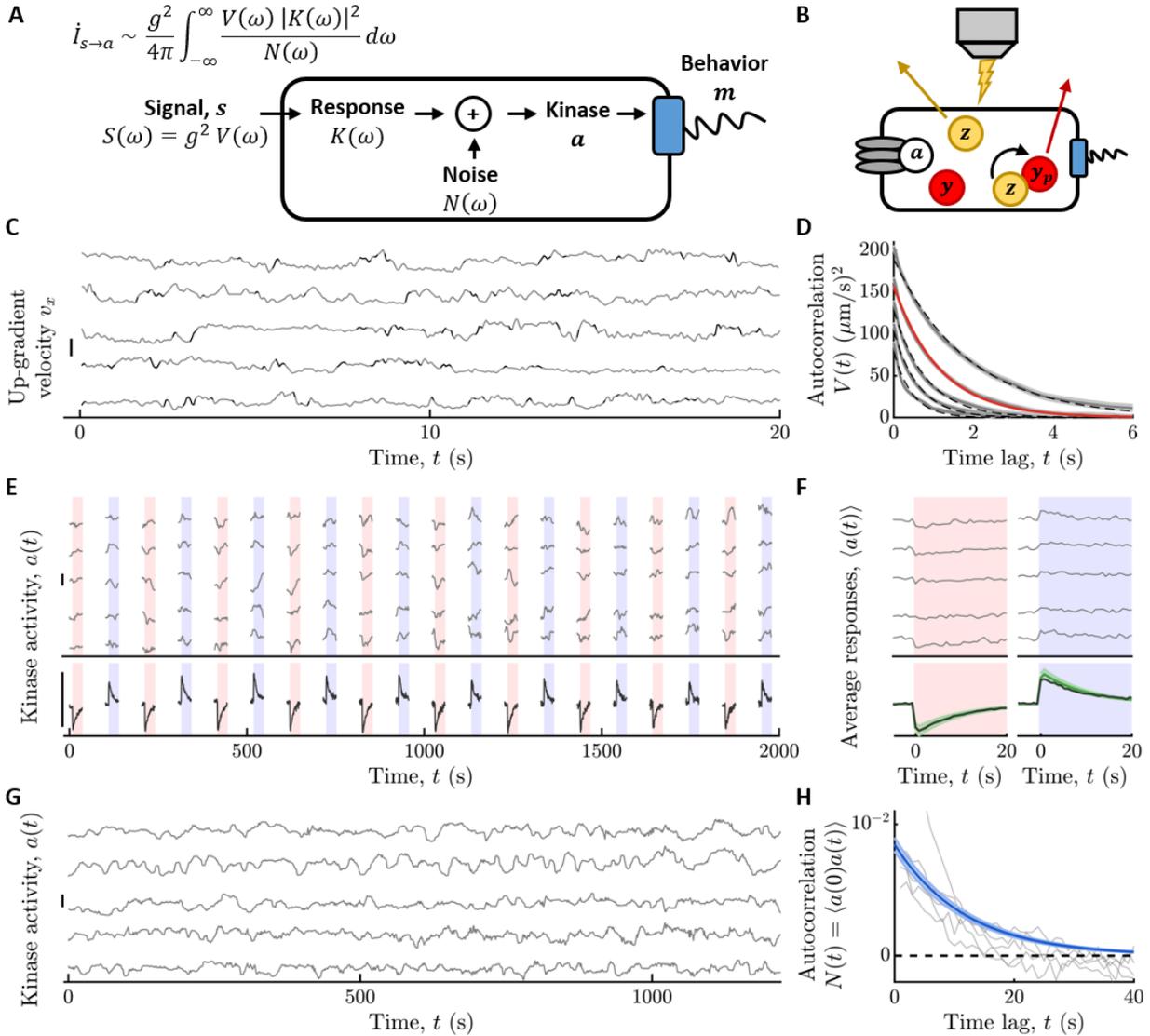

**Figure 2. Measuring the rate of information transfer from signal to intracellular kinase. A)** Information rate $\dot{I}_{s \to a}$ from signal $s$ to kinase activity $a$ depends on the signal power spectrum $S(\omega)$, the kinase frequency response $K(\omega)$, and the kinase noise power spectrum $N(\omega)$. The signal is $s(t) = g\, v_x(t)$, where $g$ is the gradient steepness and $v_x$ is the cell's up-gradient velocity. $S(\omega) = g^2 V(\omega)$, where $V(\omega)$ is the power spectrum of $v_x$. **B)** Kinase activity $a$ was quantified from the FRET between the substrate of the kinase, CheY-mRFP, and its phosphatase, CheZ-mYFP (Methods; SI; Figs. S3-S6). **C)** (Gray) Individual cells' $v_x(t)$ in a uniform concentration of 100 μM MeAsp. (Black: tumbles; scale bar 40 μm/s.) **D)** (Gray) Average autocorrelation of $v_x$, $V(t)$, for $P_{run} = 0.93, 0.89, 0.84, 0.79, 0.74$ (top to bottom; throughout, shading is ± one standard error; black dashed lines are exponential fits to $V(t) =$

$a_v \exp(-\lambda_{tot} |t|)$). (Red) Best fit to the population median bin, $P_{run} \sim 0.89$ (Fig. S2; Table S1; SI). From this we computed $V(\omega)$. **E)** Immobilized cells were delivered 10 μM steps, up (red shading) and down (blue shading) (background 100 μM MeAsp). (Top, gray) Kinase activity $a(t)$ for five cells (here and in (G), smoothed with 10th order median filter, and scale bars represent $\Delta a = 0.3$; Methods; SI; Fig. S5). (Bottom, black) Population average $a(t)$ ($n = 442$ cells). **F)** Single-cell average (top, gray) and population-average (bottom, black) response functions $K(t)$ to positive and negative stimuli. (Green) $K(t) = G \exp(-t/\tau_2)(1 - \exp(-t/\tau_1)) H(t)$, where $H(t)$ is the Heaviside step function and $G$, $\tau_1$, and $\tau_2$ are the population median parameters extracted from fits to single-cell responses. From this we computed $K(\omega)$. **G,H)** (Gray) Kinase activity (G) and corresponding autocorrelations (H) in single cells adapted to a constant background of 100 μM MeAsp (Methods; SI; Fig. S6). (Blue, H) $N(t) = \langle a(t)a(0) \rangle = \sigma_n^2 \exp(-|t|/\tau_n)$, where $\sigma_n$ and $\tau_n$ are the population median parameters extracted from fits to single-cell traces ($n = 262$ cells). From this we computed $N(\omega)$. $V(\omega)$, $K(\omega)$, and $N(\omega)$ shown in (Fig. S7).

We quantified the statistics of noise in kinase activity by measuring FRET in single cells in a constant background of 100 μM MeAsp (Fig. 2G; Fig. S6). These fluctuations were well-approximated by an Ornstein-Uhlenbeck process, consistent with previous measurements [16,18]. Using Bayesian filtering (SI), we inferred the single-cell parameters of the noise model directly from the time series. These parameters determined the noise autocorrelation function $N(t) = \sigma_n^2 e^{-|t|/\tau_n}$ (Fig. 2H; SI), from which we computed the power spectrum $N(\omega)$ (Fig. S7). The median parameter values in the population (± one standard error; $n = 262$ cells) were $\sigma_n = 0.092 \pm 0.002$ AU, the long-time standard deviation of the noise, and $\tau_n = 11.75 \pm 0.04$ s, the correlation time (Table S1). Note that these measurements include the effects of all noise sources upstream of the kinases, including Poisson single-molecule arrivals at the cells' transmembrane receptors [15].

With the input statistics, response function, and noise, we then computed the information rate from the signal to kinase activity $\dot{I}_{s \to a}$ (Fig. 3A). Since the signal power is proportional to $g^2$, information rate is, as well: $\dot{I}_{s \to a} = \beta g^2$. Using our measurements above, we estimate that the *E. coli* chemotaxis system transfers information to the kinases at a rate $\beta = 0.22 \pm 0.03$ bits/s per mm$^{-2}$ of squared gradient steepness (SI). Thus, in shallow gradients, where concentration varies on millimeter to centimeter length scales, cells only get on the order of $10^{-2}$ bits/s. The bound in Eqn. 1 predicts that this should be sufficient for a run-and-tumble navigator to climb gradients at a few percent of its swimming speed. However, it is unclear how much of this information is communicated to the motors and used to navigate.

To determine how efficiently *E. coli* use this information to navigate, we measured their drift speeds by tracking individual cells' motion in gradients of varying steepness. Static, linear MeAsp gradients were constructed (Methods) in a background concentration of 100 μM, with length scales ranging from 10 mm ($g = 0.1$ mm$^{-1}$) to 2.5 mm ($g = 0.4$ mm$^{-1}$). From >$10^5$ seconds of trajectories in each gradient condition, we estimated the average drift speed $v_d$ as the time-averaged up-gradient velocity over all cells in each experiment (SI). As expected from theory in shallow gradients, the drift speeds increased linearly with gradient steepness $v_d = \chi g$, with a proportionality constant of $\chi \sim 4.30 \pm 0.15$ μm/s per $\text{mm}^{-1}$ of gradient steepness (Fig. 3B; Fig. S9), consistent with previous measurements [39].

With measurements of both the information acquisition rate and the performance, we are now in a position to quantify *E. coli*'s information efficiency, $\eta$. For each gradient $g$, we plotted the drift speed $v_d(g)$ against the information rate $\dot{I}_{s \to a}(g)$ (blue curve in Fig. 3C). On the same plot, we show the maximum drift speeds, given by the bound in Eqn. 1 (green curve in Fig. 3C). The ratio of these two curves is the information efficiency, $\eta = v_d(g) / \left[ F(\boldsymbol{\theta}) \left( \frac{\log(2) \, \dot{I}_{s \to a}(g)}{12 \, D_r} \right)^{\frac{1}{2}} \right]$. We find that *E. coli* achieve an efficiency of $\eta = 0.66 \pm 0.05$—that is, they climb gradients at ~66% of the maximum possible speed given the rate at which their kinases acquire information about the gradients, despite having to infer concentration changes over a finite time. The observation that *E. coli*'s efficiency is order 1 implies that much of the information about the signal contained in kinase activity is preserved in behavior and used to navigate. Thus, *E. coli* cells efficiently use information from environmental signals to perform chemotaxis, indicating that information is likely difficult for their receptor-associated kinases to acquire, limiting their performance.

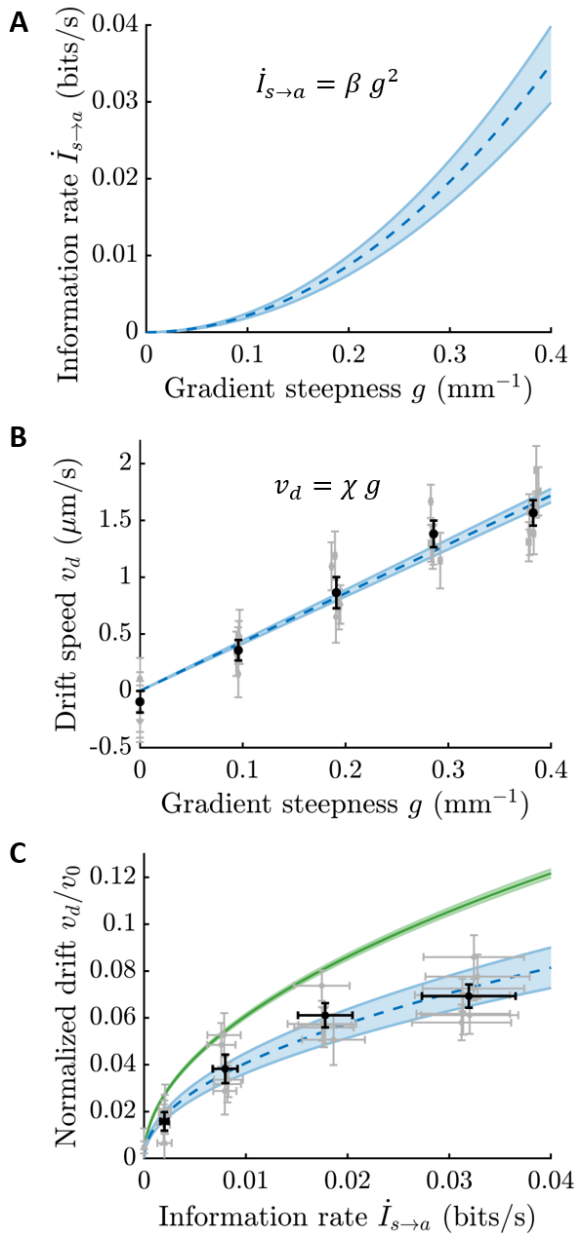

**Figure 3. *E. coli* use information efficiently to navigate.** **A)** The rate of information transfer from signal to kinase activity depends on gradient steepness $g$: $\dot{I}_{s \to a} = \beta \, g^2$ ($\beta = 0.22 \pm 0.03$ bits/s / $\text{mm}^{-2}$; Fig. 2; SI). Throughout, shading and error bars indicate ± one standard error. **B)** Chemotactic performance as a function of gradient steepness $g$, in a background of 100 μM MeAsp. Gray dots: average drift speeds in individual experiments. Black dots: averages over experiments. Error bars on $g$ are smaller than the markers. Population average drift speed increases linearly with gradient steepness, $v_d = \chi \, g$ ($\chi = 4300 \pm 150$ $\mu\text{m}^2$/s; blue dashed line and shading; SI). **C)** From measurements of *E. coli* cells' information rates and chemotactic drift speeds, we compared their performance to the theoretical bound (Eqn. 1). Green: predicted maximum performance given information acquisition rate $\dot{I}_{s \to a}$ (Eqn. 1). Blue: measured performance $v_d/v_0$ ($v_0 = 22.61 \pm 0.07$ μm/s) versus information rate $\dot{I}_{s \to a}$, obtained by eliminating $g$ from the fits of $v_d(g) = \chi \, g$ and $\dot{I}_{s \to a}(g) = \beta \, g^2$ to plot

$v_d/v_0 = \chi/v_0 \left(\dot{I}_{s \to a}/\beta\right)^{1/2}$. Black and gray dots are data points from (B). Taking the ratio of the blue and green curves, we find that *E. coli* achieve drift speeds within $66 \pm 5\%$ of the theoretical limit.

Many studies of information theory in biology have focused on the maximum amount of information signaling pathways can [1–6,40–42]. Here, we instead ask how much information an organism needs, and how much it uses, while carrying out a task, by studying information transfer in its functional context. Achieving high efficiency requires the signaling cascade to acquire, transmit, and act on information that is relevant to their behavioral task [10], but in many cases, which bits are relevant is not clear *a priori*. By using rate-distortion theory [6,24,25], we find that the bits that are relevant to bacterial chemotaxis are those that indicate how fast the attractant concentration is currently changing (see SI). This is demonstrated by our finding that maximum efficiency is achieved when the behavioral response $\lambda_R(\{s\})$ is instantaneous. Furthermore, previous works measured one-shot information transfer, in bits, by biochemical networks [1–6,40–42]. Here, we measured the rate at which *E. coli* transfer information, in bits/s, between time-varying inputs and outputs, with natural input statistics driven by the cells' own motion. Combining this with measurements of *E. coli*'s chemotactic performance and comparing to the theoretical limit (Eqn. 1), we found that most of the information about concentration changes that flows through CheA is both relevant to and used for navigation.

The information-performance bound in Eqn. 1 depends on the cell's swimming behavior and the physical properties of the environment in which it swims. We quantified this bound for a typical wild type cell swimming in liquid, but individual cells have different behavioral parameters $\boldsymbol{\theta}$ [16,28,30,29], and the distribution of swimming phenotypes in the population depends on growth conditions. This raises the question of what is the maximum performance that can be achieved with a given information rate by any phenotype. Maximizing $F(\boldsymbol{\theta})$ in Eqn. 1 with respect to $\boldsymbol{\theta}$ (SI), we find that the optimal agent changes direction by tumbling at the same rate as rotational diffusion [21,43,44] (Fig. S10). The median phenotype in our conditions tumbles more frequently than this. However, in other environments, such as semisolid agar, in which reorientations are required to escape traps [45], more frequent tumbling may be optimal.

*E. coli*'s proximity to the theoretical limit raises the question of how information is transmitted so reliably from continuous kinase activity into discrete transitions at the flagellar motors. Additionally, while stochastic arrival of ligand molecules at the cell's surface [15,46] reduce how efficiently *E. coli* can use information by forcing them to integrate and respond to signals over a finite time, they also place an upper limit on the amount of relevant information a cell can possibly acquire. Future work will clarify what this fundamental bound on sensing is and how close *E. coli* are to it.

Information transfer is not necessarily the end-goal of biological tasks, but it is needed to perform many of them. Our results suggest organisms may be under selective pressure to efficiently use information from environmental cues to perform tasks necessary for their survival.

**Methods**

Strains and plasmids

The strain used for the FRET experiments is a derivative of *E. coli* K-12 strain RP437 (HCB33), a gift of T. Shimizu, and described in detail elsewhere [18,34]. In brief, the FRET acceptor-donor pair (CheY-mRFP and CheZ-mYFP) is expressed in tandem from plasmid pSJAB106 [18] under an isopropyl β-D-thiogalactopyranoside (IPTG)-inducible promoter. The glass-adhesive mutant of FliC (FliC*) was expressed from a sodium salicylate (NaSal)-inducible pZR1 plasmid [18]. The plasmids are transformed in VS115, a *cheY cheZ fliC* mutant of RP437 [18] (gift of V. Sourjik). The crosstalk coefficient for spectral bleedthrough was measured using a strain expressing CheZ-YFP from a plasmid, and that for cross-excitation was measured using a strain expressing CheY-mRFP from a plasmid, both of which are gifts from T. Shimizu. RP437, the direct parent of the FRET strain and also a gift from T. Shimizu, was used to measure behavioral parameters and chemotactic drift speeds. A mutant that can't tumble due to an in-frame deletion of the *cheY* gene, VS100 (gift of V. Sourjik), was used to measure the rotational diffusion coefficient $D_r$. All strains are available from the authors upon request.

Cell preparation

Single-cell FRET microscopy and cell culture was carried out essentially as described previously [18,34]. In brief, cells were picked from a frozen stock at -80°C and inoculated in 2 mL of Tryptone Broth (TB; 1% bacto tryptone, 0.5 % NaCl) and grown overnight to saturation at 30°C and shaken at 250 RPM. Cells from a saturated overnight culture were diluted 100X in 10 mL TB and grown to OD600 0.45-0.47 in the presence of 100 μg/ml ampicillin, 34 μg/ml chloramphenicol, 50 μM IPTG and 3 μM NaSal, at 33.5°C and 250 RPM shaking. Cells were collected by centrifugation (5 min at 5000 rpm, or 4080 RCF) and washed twice with motility buffer (10 mM KPO4, 0.1 mM EDTA, 1 μM methionine, 10 mM lactic acid, pH 7), and then were resuspended in 2 mL motility buffer. Cells were left at 22°C for 90 minutes before loading into the microfluidic device. All experiments, FRET and swimming, were performed at 22-23°C.

For swimming and chemotaxis experiments, cells were prepared identically. Saturated overnight cultures were diluted 100X in 5 mL of TB. After growing to OD600 0.45-0.47, 1 mL of cell suspension was washed twice in motility buffer with 0.05% w/v of polyvinylpyrrolidone (MW 40 kDa) (PVP-40) added. Washes were done by centrifuging the suspension in an Eppendorf tube at 1700 RCF (4000 RPM in this centrifuge) for 3 minutes. After the last wash, cells were resuspended with varying concentrations of MeAsp (see below).

Microfluidic device fabrication and loading for FRET measurements

Microfluidic devices for the FRET experiments [34] were constructed from polydimethylsiloxane (PDMS) on 24 x 60 mm cover glasses (#1.5) following standard soft lithography protocols [47]. Briefly, the master molds for the device were created with a negative SU-8 photoresist on 100-mm silicon wafers. Approximately 16-μm-high master molds were created. To fabricate the device, the master molds were coated with a 5-mm-thick layer of degassed 10:1 PDMS:curing agent mixture (Sylgard 184, Dow Corning). The PDMS layer was cured at 80 °C for 1 hour, and then cut and separated from the wafer, and holes were punched for the inlets and outlet. The punched PDMS layer was further cured at 80 °C for > 2 hours. Then, the PDMS was cleaned with transparent adhesive tape (Magic Tape; Scotch) followed by rinsing with (in order) isopropanol, methanol, and Millipore-filtered water. The glass was rinsed with (in order) acetone, isopropanol, methanol, and Millipore-filtered water. The PDMS device was tape-cleaned an additional time before the surfaces of the device and coverslip were treated in a plasma bonding oven (Harrick Plasma). After 1 min of exposure to plasma under vacuum, the device was laminated to the coverslip and then baked at 80°C hotplate for > 30 min to establish a covalent bond.

Sample preparation in the microfluidic device was conducted as follows. Five inlets of the device (Fig. S3) were connected to reservoirs (Liquid chromatography columns, C3669; Sigma Aldrich) filled with motility buffer containing various concentrations of α-methyl-aspartate (MeAsp) through polyethylene tubing (Polythene Tubing, 0.58 mm id, 0.96 mm od; BD Intermedic). The tubing was connected to the PMDS device through stainless steel pins that were directly plugged into the inlets or outlet of the device (New England Tubing). Cells washed and suspended in motility buffer were loaded into the device from the outlet and allowed to attached to the cover glass surface via their sticky flagella by reducing the flow speed inside the chamber. The pressure applied to the inlet solution reservoirs was controlled by computer-controlled solenoid valves (MH1; Festo), which rapidly switched between atmospheric pressure and higher pressure (1.0 kPa) using a source of pressurized air. Only one experiment was conducted per device.

Single-cell FRET imaging system
FRET imaging in the microfluidic device was performed using an inverted microscope (Eclipse Ti-E; Nikon) equipped with an oil-immersion objective lens (CFI Apo TIRF 60X Oil; Nikon). YFP was illuminated by an LED illumination system (SOLA SE, Lumencor) through an excitation bandpass filter (FF01-500/24-25; Semrock) and a dichroic mirror (F01-542/27-25F; Semrock). The fluorescence emission was led into an emission image splitter (OptoSplit II; Cairn) and further split into donor and acceptor channels by a second dichroic mirror (FF580-FDi01-25x36; Semrock). The emission was then collected through emission bandpass filters (FF520-Di02-25x36 and FF593-Di03-25x36; Semrock) by a sCMOS camera (ORCA-Flash4.0 V2; Hamamatsu). RFP was illuminated in the same way as YFP except that an excitation bandpass filter (FF01-575/05-25; Semrock) and a dichroic mirror (FF593-Di03-25x36; Semorock) were used. An additional excitation filter (59026x; Chroma) was used in front of the excitation filters. To synchronize image acquisition and the delivery of stimulus solutions, a custom-made MATLAB program controlled both the imaging system (through the API provided by Micro-Manager [48]) and the states of the solenoid valves.

Procedure for measuring the linear response functions
All experiments were performed in a background MeAsp concentration of $c_0 = 100$ μM. Measurements were made in single cells. First, the FRET level at minimum kinase activity was measured by delivering a saturating stimulus (1 mM MeAsp plus 100 μM serine [49]) for 10 seconds. Immediately afterwards, the FRET level at maximum kinase activity was measured by delivering motility buffer with no attractant (0 μM MeAsp, 0 μM serine) for 5 seconds. When cells are adapted to 100 μM MeAsp, removing all attractant is sufficient to elicit a maximal response [18,36].

Donor excitation interval (i.e., measurements of $I_{DD}$ and $I_{DA}$; see SI) was 0.5 seconds and acceptor excitations (i.e., measurements of $I_{AA}$; See SI) were done before and after the set of donor excitations. After this, the concentration of MeAsp was returned to the background $c_0$, and no serine was delivered to the cells for the rest of the experiment. Imaging was then stopped and cells were allowed to adapt to the background for 120 seconds.

After this, a series of stimuli were delivered to the cells in the microfluidic device (see Figure 2E for stimulus protocol). Importantly, the cells were only illuminated and imaged for part of the experiment in order to limit photobleaching. First, cells were imaged for 7.5 seconds in the background concentration $c_0$. Then, the concentration of MeAsp was shifted up to $c_+ = 110$ µM for 30 seconds and imaging continued. Donor excitation interval was 0.75 seconds and acceptor excitations were done before and after the set of donor excitations. After this time, imaging was stopped and the MeAsp concentration returned to $c_0$ for >60 seconds to allow cells to adapt. Then, the same process was repeated, but this time shifting MeAsp concentration down to $c_- = 90$ µM. Alternating up and down stimuli were repeated 10 times each.

FRET levels at minimum and maximum kinase activity were measured again at the end of the experiment. The whole imaging protocol lasted <2200 seconds. In total, cells spent <60 minutes in the device, from loading to the end of imaging. Analyses of these data are described in the SI.

Procedure for measuring the noise statistics
Spontaneous fluctuations in kinase activity were also measured in a background MeAsp concentration of $c_0 = 100$ µM. Measurements were made in single cells. FRET levels at minimum and maximum kinase activity were measured at the beginning and the end of each experiment, as described above. As above, after these measurements, imaging was then stopped and cells were allowed to adapt to the background for 120 seconds. After this, cells were imaged for about 1200 seconds. Throughout, donor excitations (i.e., measurements of $I_{DD}$ and $I_{DA}$; see SI) were done every 1.0 second, except when it was interrupted by acceptor excitations (i.e., measurements of $I_{AA}$; see SI), which were conducted every 100 donor excitations. The whole imaging protocol lasted <1400 seconds. In total, cells spent about < 60 minutes in the device, from loading to the end of imaging. Analyses of these data are described in the SI.

Procedure to measure swimming and behavioral parameters
After the second wash, cells were centrifuged again and resuspended in motility buffer containing 100 µM MeAsp. Then, the cell suspension was diluted to an OD600 of 0.00025. The cell suspension was then loaded into µ-Slide Chemotaxis devices (ibidi; Martinsried, Germany), the same type of device used to create static gradients, described below. However, instead of tracking cells in the gradient region, we tracked their swimming in one of the large reservoirs, which are roughly 750 µm deep. 1000-s movies of swimming cells were recorded on a Nikon Ti-E Inverted Microscope using a CFI Plan Fluor 4X objective (NA 0.13). This objective's depth of field is about ±18 µm, much shorter than the depth of the chamber. Adjusting the focal plane to the middle of the chamber made cells that were swimming near the ceiling or floor of the device, which could experience hydrodynamic interactions that affect their behavior [50], not visible in the movie. At the same time, this lower magnification objective allowed us to collect relatively longer swimming trajectories. Movies were captured around 30 minutes after loading cells into the chamber to mimic the gradient experiments below. Images here and below were captured using a sCMOS camera (ORCA-Flash4.0 V2; Hamamatsu). Analyses of these data are described in the SI. Five biological replicates were done for behavioral parameter measurements, and four biological replicates were done for measuring $D_r$.

Procedure to measure chemotactic drift speeds

Chemotaxis experiments were performed in µ-Slide Chemotaxis devices (ibidi; Martinsried, Germany). These devices generate a linear gradient between two concentration reservoirs that is stable for a long time. After the second wash, the cell suspension was split into two Eppendorf tubes, 0.5 mL each. After one more centrifugation, one tube of cells was resuspended in 1 mL of motility buffer with 100 µM of MeAsp, to be injected into the "low-concentration reservoir", and the other was resuspended in 1 mL of motility buffer with 2 µM of fluorescein and varying concentrations of attractant, to be injected into the "high-concentration reservoir". Cells in both tubes were diluted to OD 0.001 for each experiment. Loading cells in both reservoirs ensured that the concentration of cells throughout the experimental device was approximately uniform. This limited the effects of potential biases that could arise from observing a finite field of view.

Using a background concentration of at least 100 µM MeAsp ensured that the cells were in the log-sensing regime [12]. The "high" concentrations of MeAsp used were 110.5 µM, 122.1 µM, 135.0 µM, and 149.2 µM. With 1 mm separating the two reservoirs, these concentrations produced linear gradients that approximated shallow exponentials gradients with steepness of roughly: $g = \{0.1, 0.2, 0.3, 0.4\}$ mm$^{-1}$. $g$ was calculated from $g = \frac{\Delta \log c}{\Delta x}$, where $\Delta \log c$ is the difference in log concentrations between the two reservoirs, and $\Delta x$ is the distance between them. This is exactly the average steepness of log-concentration across with the width of the channel. To see this, the steady state concentration profile is linear, $c(x) = \frac{\Delta c}{\Delta x}(x - x_0) + c_0$, where $\Delta c$ is the difference in concentration between the two reservoirs, $x_0$ is the midpoint between them, and $c_0$ is the concentration at $x = x_0$. From this, the gradient of log concentration depends on position $x$ and can be computed from $g(x) = \frac{d \log(c(x))}{dx} = \frac{1}{\frac{1}{c_0}\frac{\Delta c}{\Delta x} x + 1} \frac{1}{c_0}\frac{\Delta c}{\Delta x}$, where we have defined a reference frame where $x_0 = 0$. Averaging over the channel by integrating over $x$ from $-\Delta x/2$ to $\Delta x/2$ and dividing by $\Delta x$ gives, $\langle g \rangle = \frac{\log\left(1 + \frac{1}{2}\frac{\Delta c}{c_0}\right) - \log\left(1 - \frac{1}{2}\frac{\Delta c}{c_0}\right)}{\Delta x} = \frac{\log\left(c_0 + \frac{1}{2}\Delta c\right) - \log\left(c_0 - \frac{1}{2}\Delta c\right)}{\Delta x} = \frac{\Delta \log c}{\Delta x} = g$. Close to the low-concentration reservoir, $g(x)$ is larger than $g$, and vice versa near the high-concentration reservoir, but these errors are small and approximately cancel each other out when we average drift speeds of cells across the channel.

To load the device, first the reservoirs were sealed with the manufacturer's tabs. Cell suspension with 100 µM MeAsp was injected into the channel where the gradient would form. Excess liquid in the inlets was removed. Then one tab from each reservoir was removed, and the gradient channel was sealed with tabs. The left reservoir was then fully unsealed, and the right reservoir was sealed with tabs. 60-65 µL of cell suspension with 100 µM MeAsp was injected into the left reservoir, and then both inlets of that reservoir were sealed with tape or tabs. Care was taken to make sure there were no bubbles in reservoir at the inlets. Then, the right reservoir was unsealed, and 60-65 µL of cell suspension with the higher concentration of MeAsp was injected. A timer was then immediately started. The right reservoir was then sealed.

Cells were imaged by phase contrast with a CFI Plan Fluor 10X objective (NA 0.30). The depth of the gradient region of the device is 70 µm, and the depth of field of the objective is about ±4 µm. Focusing on the middle of the chamber with this objective filtered out cells that could be interacting with the ceiling or floor surfaces. Images of fluorescein were taken every 5 minutes using a CFI Plan Fluor 4X objective (NA 0.13) through a YFP filter cube (Chroma 49003), illuminated by a LED (SOLA SE, Lumencor) with an exposure time of 100 ms. Since the diffusivity of fluorescein is similar to (slightly lower than) that of MeAsp (MW of fluorescein is 376 kDa; MW of MeAsp is 147 kDa; Sigma Aldrich), we used fluorescein as an indication of when the attractant gradient was stable and linear in the observation region between the two reservoirs, as has been done before [51,52]. Once the fluorescein profile was stable for several time points (typically around 50-60 minutes after loading), a 1000-second phase contrast movie was recorded at 20 FPS using the 10X phase contrast objective. Before the recording, the transmitted light illumination was adjusted to minimize the number of saturated pixels. After the recording, an additional image of the fluorescein profile was recorded, and the cells were observed to check that they were still swimming normally. Analyses of these data are described below. At least five biological replicates were performed for each gradient steepness.

**Acknowledgements:** We thank Jeremy Moore and Xiaowei Zhang for help setting up the experimental assays. We also thank Katja Taute and Marianne Grognot for helpful discussions about the gradient experiments. We thank Pieter Rein ten Wolde for providing detailed feedback on an earlier version of this manuscript, as well as Artur Wachtel, Isabella Graf, and Damon Clark for providing comments. We acknowledge Tom Shimizu and Victor Sourjik for bacteria strains and Rafael Gomez-Sjoberg, Microfluidics Lab, for providing information and software to control the solenoid valves in the microfluidic setup.

**Funding:** HM, KK, and TE were funded by NIH R01s GM106189 and GM138533. HM was funded by NIH F32 GM131583. TE and BM were funded by a Yale PEB Seed Grant. BM was funded by Simons Investigator Award 624156 and NIH R35 GM138341.

**Contributions:** HM and KK contributed equally to this work. HM, KK, BM and TE designed the research. HM and BM derived the theoretical bound with inputs from TE and KK. HM performed the experiments tracking bacteria. KK performed the single cell FRET experiments. HM, KK, and TE validated the data. HM, KK, BM, and TE discussed the data analysis. HM and KK performed the data analysis. HH, KK, BM and TE wrote the initial draft and all revisions.

**Competing interests:** Authors declare no competing interests.

# Supplementary Materials for *E. coli* chemotaxis is information-limited

H.H. Mattingly[1,2,†], K. Kamino[1,2,†], B.B. Machta[3,4,*], T. Emonet[1,2,3,*].

† Equal contribution.

* Correspondence to: thierry.emonet@yale.edu and benjamin.machta@yale.edu

# Table of Contents

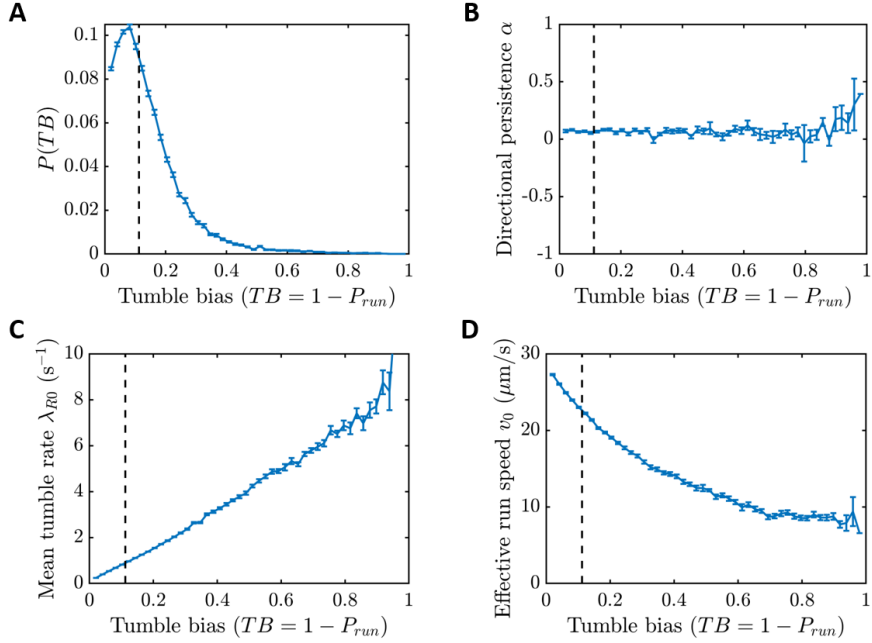

**Fig. S1. Average swimming parameters in bins of tumble bias**

**A)** Probability mass function of tumble bias $TB = 1 - P_{run}$ in populations of RP437 recorded with a 4X objective (Methods; bin size = 0.02). Each cell was weighted by its trajectory duration when generating this distribution. Error bars in each panel were determined by boot strapping (SI). In all panels, the vertical black dashed line indicates the population-median $P_{run}$. **B-D)** In bins of tumble bias: (B) Average directional persistence of tumbles, defined as $\langle \cos(\Delta\theta) \rangle$, where $\Delta\theta$ is the angle between the cell's heading vectors before and after runs; (C) Average tumble rate $\lambda_{R0}$; and (D) Effective run speed $v_0$. During transitions between run and tumble states, the swimming speed takes finite time to reach a steady value [1,2]. We lump these transitions into the run state, leading to effective run speeds that depend on how often the cell tumbles, i.e. $P_{run}$. Excluding these transitions would make cells appear to climb gradients artificially slowly compared to their run speeds. It would also lead to over-estimated information rates.

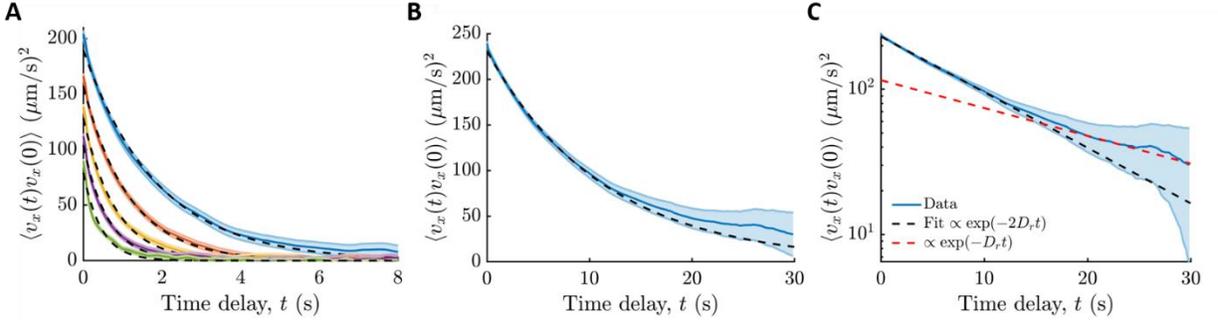

**Fig. S2. Velocity autocorrelation functions for RP437 (WT) and VS100 (non-tumbling ΔcheY mutant)**

**A)** Autocorrelations of the $x$-velocity of wild type (RP437) *E. coli* cells in the absence of a gradient, averaged over cells in bins of $P_{run}$ (bin size 0.005). From blue to green, $P_{run}$ values are: 0.93, 0.89, 0.84, 0.79, 0.74. The orange line is the correlation function the median $P_{run}$. Curves in this panel are reproduced from Fig. 2D of the main text. The black dashed lines are the best fits of a decaying exponential to each correlation function (SI). Shading is ± one standard error, which was computed from the standard deviation of the samples of $v_x(0)v_x(t)$ at each time delay $t$, divided by $\sqrt{n_i(t)}$. $n_i(t) = n(t)\frac{\Delta t}{\Delta t + 2/\lambda_{tot}}$ is the effective number of independent measurements at time delay $t$ [3], where $n(t)$ is the total number of observations at delay $t$, $\Delta t$ is the imaging interval (50 ms), and $\lambda_{tot}$ is the best fit to the decay rate of each exponential. Since samples come from multiple cells, whose velocities are uncorrelated, this underestimates the actual $n_i$. **B)** Population-averaged $x$-velocity autocorrelation function of VS100 cells, which lack the *cheY* gene and therefore cannot tumble. Their correlation function is expected to decay exponentially with rate $\lambda_{tot} = 2D_r$, from which we inferred the rotational diffusion coefficient, $D_r$. Black dashed line is an exponential fit to the first 10 seconds of time delay. **C)** Same as (B) (velocity autocorrelation function for VS100 cells), but with the y-axis on log-scale. At long time delays, there is a bias for cells that remained visible, i.e. cells that had small vertical component to their velocity and therefore remained in the depth of field of our microscope objective. These cells by chance appear to be undergoing rotational diffusion in two dimensions instead of three; therefore, they lose direction at a rate $D_r$ instead of $2D_r$. This transition did not affect our inference of $D_r$.

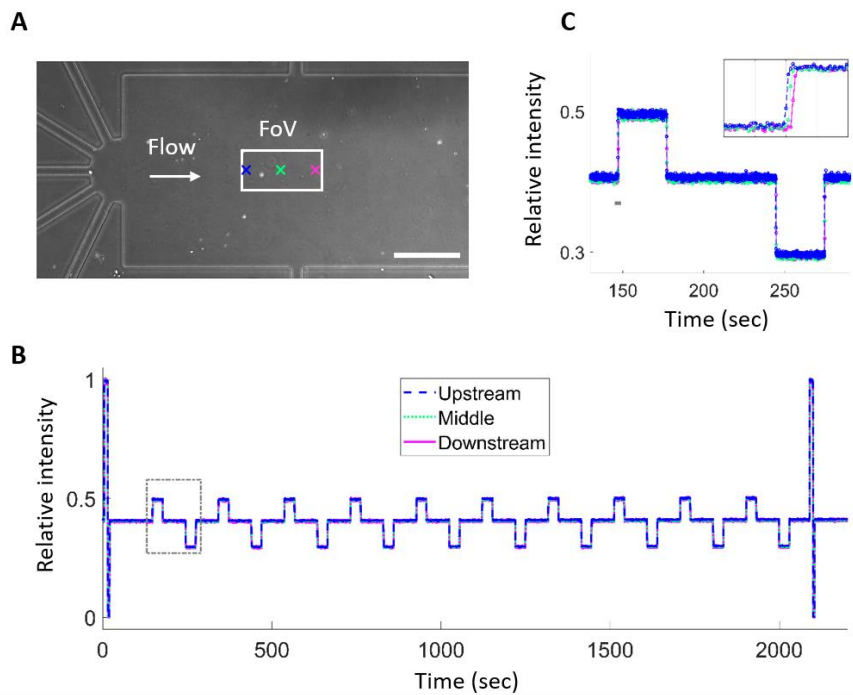

**Fig. S3. Stimulus control by microfluidics**

**A)** An image of the microfluidic device used in the FRET experiments (Kamino et al, *Sci. Adv.* 2020). The white scale bar at the bottom right is 200 µm. The direction of flow is shown by a white arrow and the position of a field of view (FoV) for FRET measurements is shown by a white rectangle. The three locations at which the temporal profiles of delivered chemical solutions were measured and plotted in panels (B) and (C) are indicated by the three colored crosses. Different solutions were delivered into the FoV from the five inlet channels shown on the left (Methods). **B)** The temporal profile of chemoattractant stimuli that the cells experience during the FRET experiment corresponding to main text Fig. 2E and Fig. S5. Each solution delivered to the FoV through each inlet channel was visualized by using different concentrations of fluorescein, and the relative fluorescence intensities of the solutions were measured at three different positions in a 60X field of view (see panel (A)) and the measurement at each location is plotted in the color of its corresponding cross in panel (A). Five solutions were delivered to the imaging region by computer-controlled solenoid valves, which control the pressure applied to each channel (Methods). **C)** A magnified view of the gray box part of panel (B). The inset shows further magnified plots of the time window indicated by the horizontal gray bar to resolve the transient rise of fluorescence intensity. The X-grid in the inset is 1 s interval. Intensities were obtained at 75 ms intervals, indicated by the circles. The transient rise and delay between positions within the FoV are on the order of 0.1 s.

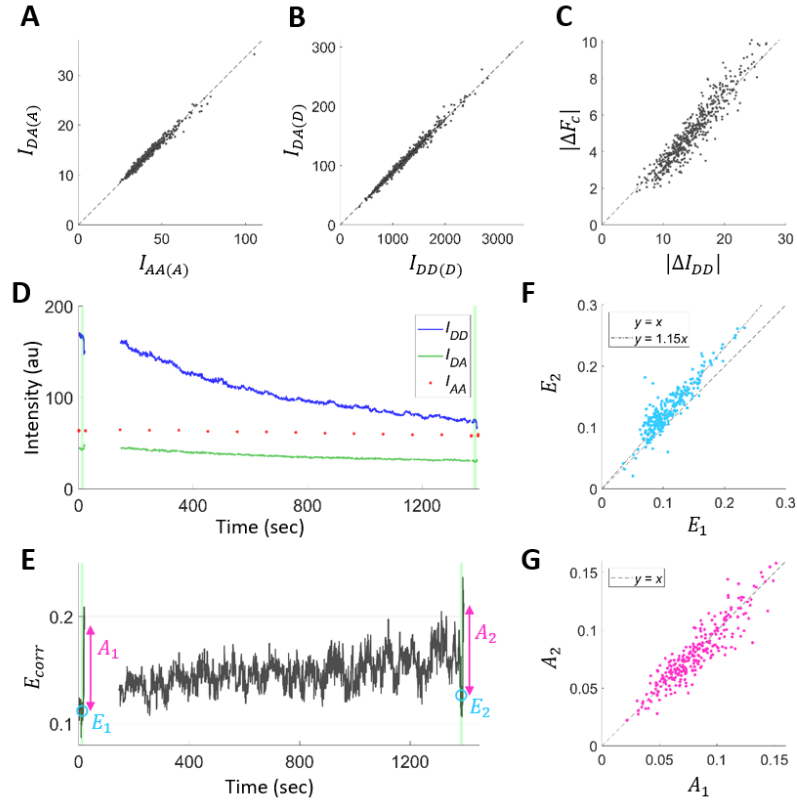

**Fig. S4. FRET analysis**

**A)** Single-cell fluorescence intensities from a strain that only expresses the acceptor (mRFP), obtained through the acceptor channel $I_{AA(A)}$ and the FRET channel $I_{DA(A)}$. The slope gives an estimate of the cross-excitation coefficient $a_E = 0.337$. **B)** Single-cell fluorescence intensities from a strain that only expresses the donor (mYFP), obtained through the donor channel $I_{DD(D)}$ and the FRET channel $I_{DA(D)}$. The slope gives the estimation of the bleedthrough coefficient $d_E = 0.089$. **C)** Absolute changes in the donor fluorescent signal $|\Delta I_{DD}|$ and the sensitized emission $|\Delta F_c|$ before and just after the removal of a saturating chemoattractant stimulus obtained from a FRET strain that expresses both the donor and acceptor. The slope gives the parameter $G_E = 0.35$. **D)** Representative time series of background-subtracted fluorescent signals $I_{DD}(t)$, $I_{DA}(t)$, and $I_{AA}(t)$ from a FRET experiment. Green shading at the start and end of the experiment indicate times when a saturating stimulus (1 mM MeAsp and 100 µM serine) was applied, which were followed by the removal of all attractants (i.e., [MeAsp] = [serine] = 0 M) for 5 seconds. After this, the rest of the experiment was performed in a background of 100 µM MeAsp. The fluorescence intensities decrease over time due to photobleaching. We show in the SI that the photobleaching and the finite precision of the parameter estimation (A-C) results in a bias in FRET estimation, but we also show that it can be corrected (E-G). **E)** The FRET index $E_{corr}(t)$ (SI) computed from the fluorescent signals $I_{DD}(t)$, $I_{DA}(t)$, and $I_{AA}(t)$ in panel (D). **F)** The median of $E_{corr}(t)$ during the saturating stimuli at the beginning ($E_1$) and at the end ($E_2$) of the measurement in panel (E). Consistent with the theoretical analysis (SI), the level of $E_2$ shows systematic deviation from the level of $E_1$ (i.e., from the line $y = x$) as a result of photobleaching of the fluorescent proteins. **G)** The change in $E_{corr}$ induced by the removal of the saturating stimulus at the beginning ($A_1$) and at the end ($A_2$) of the measurement defined in panel (E). Consistent with the theoretical analysis (SI), the signal changes show undetectable bias after photobleaching, distributing around the line $y = x$. Thus, we can correct for the slowly-increasing bias in the absolute value of $E_{corr}(t)$ over time (E and F) by subtracting the trend, while also preserving information about signaling-induced changes in $E_{corr}$ (SI).

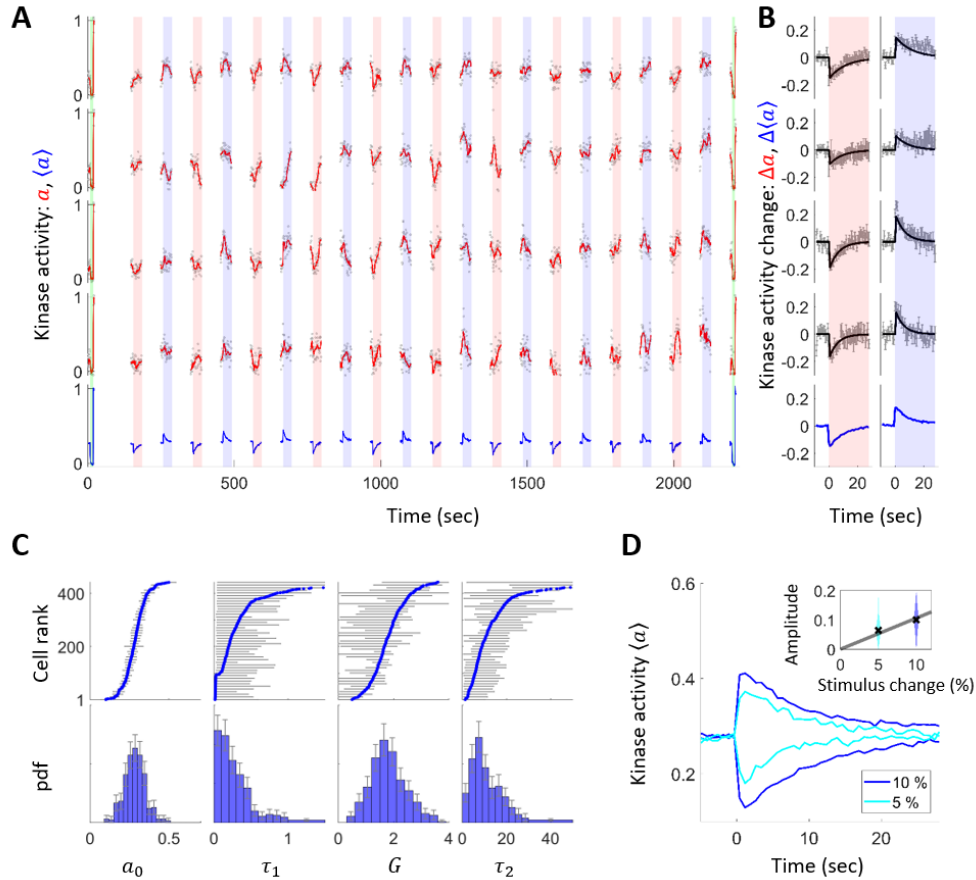

**Fig. S5. Extracting linear response function**
**A)** Representative time series of single-cell kinase activity $a$ (SI for definition; red lines) and population average $\langle a \rangle$ (blue lines). Gray dots are raw data points and red lines are smoothed curves (10th order median filter; 7.5 s time window). Red and blue shading indicates times at which [MeAsp] is elevated by 10 μM and decreased by 10 μM from 100 μM, respectively. Green shading indicates when saturating stimuli (1 mM MeAsp and 100 μM serine) were applied to measure the minimum kinase activity, which were followed by the removal of the attractants (i.e., [MeAsp] = [serine] = 0 M) for 5 seconds to measure the maximum kinase activity. **B)** Averaged kinase responses. Gray lines with error bars are the within-cell average and standard error of the mean of the change in kinase activity $\Delta a$, defined as the change in $a$ from its pre-stimulus value (left: step-up responses; right: step-down responses). Best fit model curves are shown in black (SI). Population-averaged kinase activity changes $\Delta\langle a \rangle$ are shown at the bottom in blue. **C)** (Top) Values of the extracted parameters sorted by the mean (for $a_0$) or maximum a posteriori probability (MAP) estimates (for $\tau_1$, $G$, and $\tau_2$). The errorbars are standard errors (for $a_0$) or 25 and 75 percentiles of the posterior distributions (for $\tau_1$, $G$, and $\tau_2$), and are shown for one in every ten cells. (Bottom) Marginal histograms of each parameter. 95% bootstrap confidence intervals are shown by the error bars. **D)** Time series of population-averaged kinase activity $\langle a \rangle$ upon 5% and 10% change in [MeAsp] from 100 μM background (changed at time zero; both step-up and step-down responses are shown). The inset shows the distribution of single-cell response amplitudes (defined as the average of $|\Delta a|$ over the first 3 s after the stimulus) upon step concentration changes. Both step-up and step-down responses are lumped into the same distribution. The gray line shows $y = a\,x$ fitted to the population average shown by black crosses. Consistent with previous work (Sourjik and Berg, PNAS 2002), the responses show quasi-linearity in this concentration range.

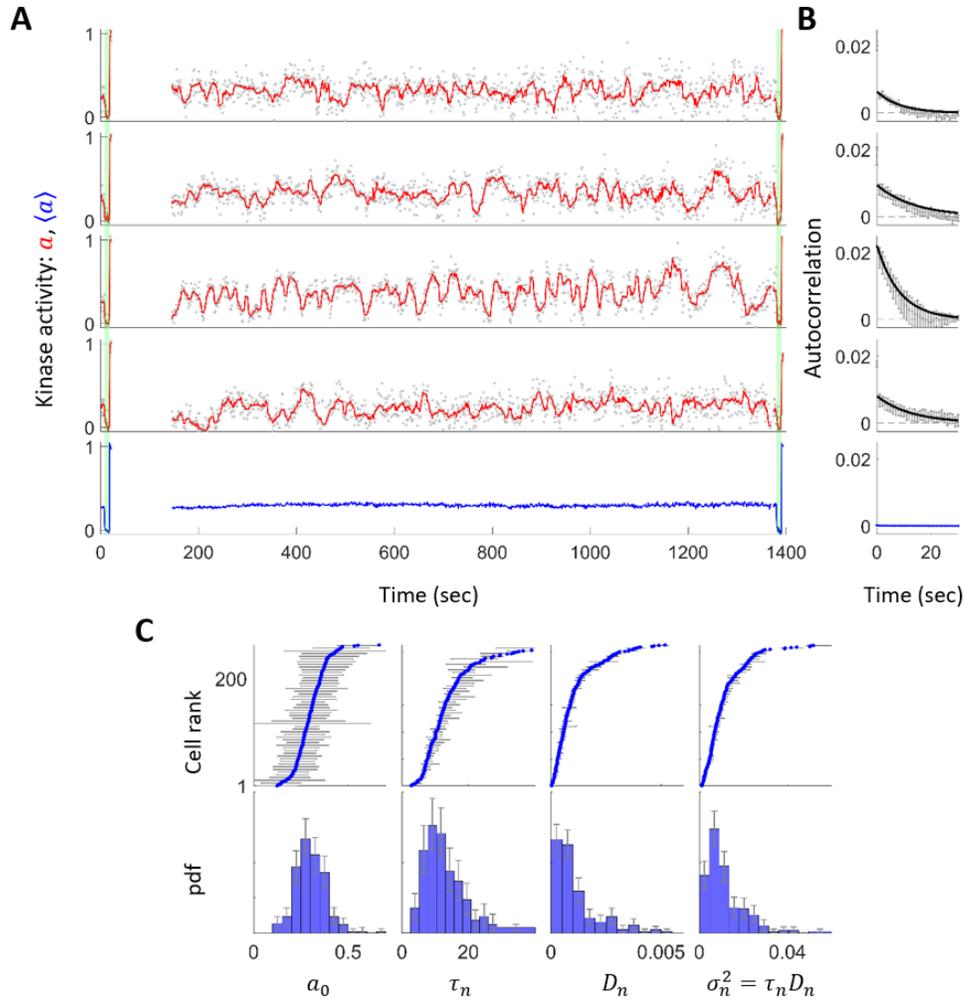

**Fig. S6. Quantifying signaling noise**
**A)** Representative time series of single-cell kinase activity $a$ (red lines) and population average $\langle a \rangle$ (blue lines). Gray dots are raw data points and red lines are smoothed curves (10th order median filter; 10 sec time window). Green shading indicates when saturating stimuli (1 mM MeAsp and 100 μM serine) were applied, which were followed by the removal of the attractants (i.e., [MeAsp] = [serine] = 0 M) for 5 seconds before returning to the background of 100 μM MeAsp. **B)** Autocorrelation function of the kinase activity $a$ of the cells shown in (A). Gray error bars are the average and standard error of the autocorrelation function computed from six segments of the time series with identical length. The black lines are the autocorrelation function of the Ornstein-Uhlenbeck (OU) process $C(t) = \tau_n D_n \exp(-t/\tau_n)$ with the extracted parameters using a Bayesian-filtering method (see SI). In blue at the bottom is the autocorrelation function of the population-averaged time series, with error bars. **C)** (Top) Values of extracted parameters sorted by the mean (for $a_0$) or MAP estimates (for $\tau_n$, $D_n$, and $\sigma_n^2 = \tau_n D_n$). The error bars are the standard deviation (for $a_0$) or 25 and 75 percentiles of the posterior distributions (for $\tau_n$, $D_n$, and $\sigma_n^2 = \tau_n D_n$), and are shown for one in every five cells for visualization purposes. (Bottom) Marginal histograms of each parameter. 95% bootstrap confidence intervals are shown by the error bars.
10

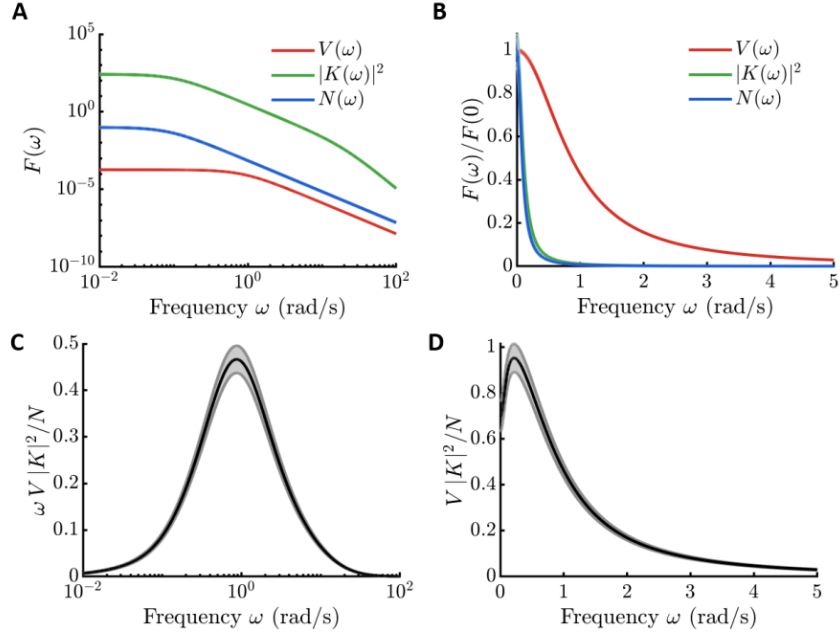

**Fig. S7. Frequency domain representations of signal, response, and noise**
**A)** Inferred models for up-gradient velocity power spectrum $V(\omega)$ (red; units $(mm/s)^2$), frequency response squared $|K(\omega)|^2$ (green), and noise power spectrum $N(\omega)$ (blue) are shown (SI). Shading in all panels indicates ± one standard error (SI). **B)** Normalized frequency-space quantities plotted with linear axis scales. **C-D)** The integrand for computing $\beta$, the proportionality constant between squared gradient steepness $g^2$ and information rate $\dot{I}_{s\to a}$, shown in main text Fig. 2A (SI Eqn. (102)). The integrand is plotted on (C) linear-log scale (bits/s) / $(mm^{-2})$ and (D) linear-linear scale (units (bits/s) / $(mm^{-2}\ rad/s)$). In (C), the integrand is multiplied by $\omega$ so that the area under the curve is equal to the integral.

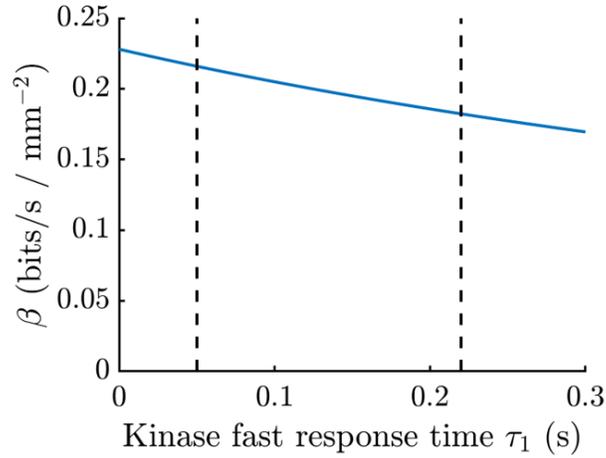

**Fig. S8. Dependence of the information rate on $\tau_1$**

The information rate from signal to kinase activity $\dot{I}_{s \to a} = \beta\, g^2$ depends on the kinase fast response time $\tau_1$ through $\beta$ (see main text and SI). Plotted here is $\beta$ for varying values of $\tau_1$, showing weak, monotonically decreasing dependence. The dashed line at $\tau_1 = 0.22$ s is the kinase response time we measured, which contains the kinetics of CheY-CheZ binding and catalysis. The dashed line near $\tau_1 = 0.05$ s indicates the previously-measured kinase response time. We used the latter value of $\tau_1$ to calculate the information rate. This can only make cells less information efficient, achieving the same chemotactic performance with more information. Regardless, the effect on the information rate is small, and the effect on the efficiency is even smaller due to the square root dependence of the bound in Eqn. 1 the information rate.

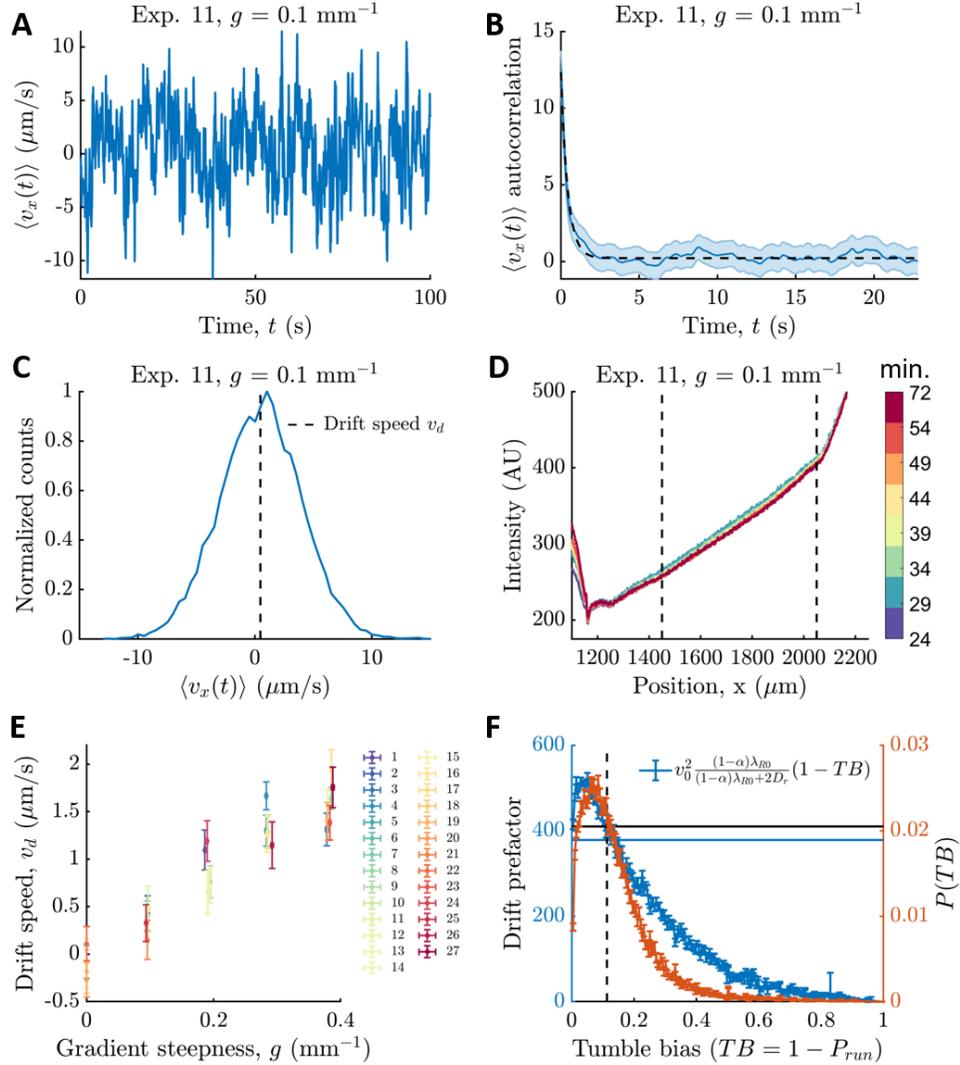

**Fig. S9. Computing average drift speeds**

**A)** Up-gradient velocity $\langle v_x(t) \rangle$ averaged over cells in the field of view in each frame of a particular movie of cells climbing a chemical gradient. Only the first 100 s of the movie are shown. **B)** Autocorrelation function of $\langle v_x(t) \rangle$ from (A) is shown in blue (shading is standard error), and an exponential fit is shown in black. **C)** Histogram of all data points in (B) (bin size 0.5 μm/s). Drift speed $v_d$ is the mean of the distribution. **D)** Example time course of fluorescein fluorescence intensity from a particular experiment. Color represents time since loading the cells into the device. In this experiment, a movie of swimming cells was recorded 54 minutes after loading them into the device, and a final fluorescein image was taken afterwards, 72 minutes after loading. Since this is raw fluorescence intensity, the profile deviates from a linear one outside of the region marked by black dashed lines because the depth of the device changes. Drift speed was computed from trajectories recorded inside the marked region. **E)** Drift speeds of individual, independent experiments, colored by experiment number. The same data is shown in gray in Fig. 3B of the main text. Uncertainties are standard errors. Colors in (D), (E) come from ref [4]. **F)** In theory, the population average drift speed could be different from the drift speed of the median phenotype (see SI). However, drift speed mostly depends on the combination of swimming parameters plotted here for different bins of tumble bias $TB = 1 - P_{run}$ (blue line; error bars from bootstrapping; see SI). Using this and the distribution of $P_{run}$ (shown in orange; same as in Fig. S1A), we compared the average value of this function to its value when evaluated at the median $P_{run}$ (black dashed line) (see SI). We find that these two are similar: the population average gives $\left\langle v_0^2 \frac{(1-\alpha)\lambda_{R0}}{(1-\alpha)\lambda_{R0}+2\,D_r} P_{run} \right\rangle \sim 375 \pm 1 \left(\frac{\mu m}{s}\right)^2$ (solid blue horizontal line), whereas the value of the blue curve at the median bin $P_{run}$ gives $v_0^2 \frac{(1-\alpha)\lambda_{R0}}{(1-\alpha)\lambda_{R0}+2\,D_r} P_{run} \sim 410 \pm 3 \left(\frac{\mu m}{s}\right)^2$ (solid black horizontal line). This justifies our comparison of population-average drift speeds to bounds quantified using a median cell's parameters.

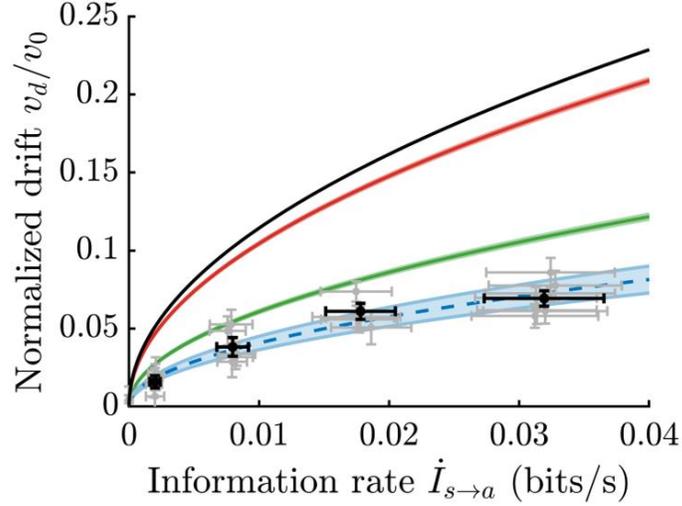

**Fig. S10. Information bounds on performance**

The upper bound on chemotactic drift speed set by information acquisition depends on the cell's behavioral response, $\lambda_R(\{s\})$, but also on its behavioral parameters, $\boldsymbol{\theta} = \{\lambda_{R0}, \alpha, P_{run}\}$ (alternative parameterization from the main text). The green line and blue lines/data points are reproduced from main text Fig. 3C. The green line is the maximum drift speed possible for a cell that has the same behavioral parameters as those we measure in RP437 *E. coli*. This bound was found by optimizing over responses $\lambda_R(\{s\})$. In principle, a cell with different behavioral parameters has a different bound on its drift speed. Optimizing the mean tumble rate $\lambda_{R0}$, but keeping the remaining behavioral parameters held fixed at their measured values, gives the red curve. Finally, if the remaining parameters are optimized, this gives the black curve. No cell's drift speed can exceed this bound. See SI for derivations and expressions for each bound. Shading and error bars indicate ± one standard error.

| Parameter | Value | Meaning | Source |
|---|---|---|---|
| $P_{run}$ | $0.89 \pm 0.01$ | Median run bias | This study |
| $\lambda_{R0}$ | $0.893 \pm 0.006$ s$^{-1}$ | Baseline tumble rate | This study |
| $\alpha$ | $0.06 \pm 0.01$ | Directional persistence of tumbles | This study |
| $v_0$ | $22.61 \pm 0.07$ μm/s | Speed during runs | This study |
| $D_r$ | $0.0441 \pm 10^{-4}$ rad$^2$/s | Rotational diffusion coefficient | This study |
| $a_v$ | $157.1 \pm 0.5 \left(\frac{\mu m}{s}\right)^2$ | Fit prefactor of the x-velocity autocorrelation function $V(t)$ | This study |
| $\lambda_{tot}$ | $0.862 \pm 0.005$ s$^{-1}$ | Fit decay rate of the x-velocity autocorrelation function $V(t)$ | This study |
| $G$ | $1.73 \pm 0.03$ | Gain of kinase output | This study |
| $\tau_2$ | $9.9 \pm 0.3$ s | Adaptation time | This study |
| $\tau_1$ | $0.22 \pm 0.01$ s | Time for kinase response and CheY/CheZ binding | This study |
| $\tau_1$ | $0.05$ s | Kinase response time (used to compute the information rate $\dot{I}_{s \to a}$) | Ref [5] |
| $D_n$ | $(7.2 \pm 0.3) \times 10^{-4}$ s$^{-1}$ | Diffusivity of kinase activity fluctuations | This study |
| $\tau_n$ | $11.75 \pm 0.04$ s | Correlation time of kinase activity fluctuations | This study |
| $a_0$ | $0.29 \pm 0.07$; $0.30 \pm 0.08$ | Baseline kinase activity | This study |
| $\beta$ | $0.22 \pm 0.03$ bits/s / mm$^{-2}$ | *E. coli*'s information rate from signal to kinase activity per squared gradient steepness, $\dot{I}_{s \to a} = \beta\, g^2$ | This study |
| $\chi$ | $4300 \pm 150$ μm$^2$/s | Drift speed per unit gradient steepness (chemotactic coefficient) | This study |

**Table S1.**

Implicit in the units of parameters $G$ and $D_n$ are the dimensionless units of kinase activity. Uncertainties for $a_0$ are its standard deviation over an isogenic population. The first value comes from the linear response experiments, and the second one from the noise measurement experiments.

**Supplementary Text**

Relationship between mutual information and transfer entropies

Consider two time-dependent random variables $X(t)$ and $Y(t)$, and let time be discrete with $X(k) = X(t_k) = X(k\,\Delta t)$. The set of values of $X$ from time $t = 0$ to time $t = t_{k-1}$ (inclusive) will be denoted $X_0^{k-1}$. Then, the mutual information between $X(t)$ and $Y(t)$ from $t = 0$ to $t = t_N$ is:

$$I(X_0^N; Y_0^N) = \int dX_0^N\, dY_0^N\, P(X_0^N, Y_0^N) \log\left(\frac{P(X_0^N, Y_0^N)}{P(X_0^N)\, P(Y_0^N)}\right) \tag{1}$$

where the probability distributions and integral are over trajectories. Each of the probability distributions can be decomposed into a product of conditional distributions, chosen such that the probability of each $X$ and $Y$ is conditioned only on those at earlier times:

$$P(X_0^N, Y_0^N) = \prod_{k=0}^{N} P(X(k), Y(k) | X_0^{k-1}, Y_0^{k-1}) \tag{2}$$

where the $k = 0$ term is just $P(X(0), Y(0))$. Doing this for each of the distributions in Eqn. (1) gives:

$$I(X_0^N; Y_0^N) = \int dX_0^N\, dY_0^N\, P(X_0^N, Y_0^N) \log\left(\frac{\prod_{k=0}^{N} P(X(k), Y(k) | X_0^{k-1}, Y_0^{k-1})}{\prod_{k'=0}^{N} P(X(k') | X_0^{k'-1})\, \prod_{k''=0}^{N} P(Y(k'') | Y_0^{k''-1})}\right). \tag{3}$$

Writing the log of the products as a sum of the logs and grouping probabilities of $X$ and $Y$ at the same time step:

$$I(X_0^N; Y_0^N) = \int dX_0^N\, dY_0^N\, P(X_0^N, Y_0^N) \sum_{k=0}^{N} \log\left(\frac{P(X(k), Y(k) | X_0^{k-1}, Y_0^{k-1})}{P(X(k) | X_0^{k-1})\, P(Y(k) | Y_0^{k-1})}\right) \tag{4}$$

Moving the sum outside of the integral, the $k$'th term in the sum only depends on $X$ and $Y$ at earlier times; therefore $P(X_0^N, Y_0^N)$ can be marginalized for $X$ and $Y$ at all times larger than $t_k$:

$$I(X_0^N; Y_0^N) = \sum_{k=0}^{N} \int dX_0^k\, dY_0^k\, P(X_0^k, Y_0^k) \log\left(\frac{P(X(k), Y(k) | X_0^{k-1}, Y_0^{k-1})}{P(X(k) | X_0^{k-1})\, P(Y(k) | Y_0^{k-1})}\right) \tag{5}$$

Multiplying the numerator and denominator inside the log by $P(X(k) | X_0^{k-1}, Y_0^{k-1})$ and $P(Y(k) | X_0^{k-1}, Y_0^{k-1})$ and then separating the log of the product into a sum of logs gives:

$$I(X_0^N; Y_0^N)$$
$$= \sum_{k=0}^{N} \int dX_0^k \, dY_0^k \, P(X_0^k, Y_0^k) \log\left(\frac{P(X(k)|X_0^{k-1}, Y_0^{k-1})}{P(X(k)|X_0^{k-1})}\right)$$
$$+ \sum_{k=0}^{N} \int dX_0^k \, dY_0^k \, P(X_0^k, Y_0^k) \log\left(\frac{P(Y(k)|X_0^{k-1}, Y_0^{k-1})}{P(Y(k)|Y_0^{k-1})}\right) \quad (6)$$
$$+ \sum_{k=0}^{N} \int dX_0^k \, dY_0^k \, P(X_0^k, Y_0^k) \log\left(\frac{P(X(k), Y(k)|X_0^{k-1}, Y_0^{k-1})}{P(X(k)|X_0^{k-1}, Y_0^{k-1}) P(Y(k)|X_0^{k-1} Y_0^{k-1})}\right)$$
$$= I_{Y \to X} + I_{X \to Y} + \sum_{k=0}^{k=N} I(X(k); Y(k)|X_0^{k-1}, Y_0^{k-1}) \quad (7)$$
$$= I_{Y \to X} + I_{X \to Y} \quad (8)$$

The first term is the transfer entropy from $Y$ to $X$, and the second is the transfer entropy from $X$ to $Y$. The last term measures the conditional dependence of one variable at the current time on another at the current time, given their histories, and is zero for a stationary, causal, finite-order Markov system.

We use "transfer entropy" to refer to the cumulative amount of statistical influence of one variable onto another over some period of time. This is in contrast to Schreiber [6], who used transfer entropy to refer to a rate—Schreiber's "transfer entropy" is our "transfer entropy rate".

The steady state mutual information rate and transfer entropy rates are obtained by dividing by the total time $t_N = N \Delta t$ and taking $N$ to infinity.

$$\dot{I}_{X \to Y} = \lim_{N \to \infty} \frac{1}{N \Delta t} \sum_{k=0}^{N} \int dX_0^k \, dY_0^k \, P(X_0^k, Y_0^k) \log\left(\frac{P(Y(k)|X_0^{k-1}, Y_0^{k-1})}{P(Y(k)|Y_0^{k-1})}\right) \quad (9)$$

For a stationary process, after sufficient time has passed (sufficiently large $N$), each term in the sum is equivalent, and we can reassign the indices so that $k = 0$ is the current time, $k = 1$ is the next future time step, and negative indices indicate time in the past:

$$\dot{I}_{X \to Y} = \lim_{N \to \infty} \frac{1}{\Delta t} \int dX_{-N}^1 \, dY_{-N}^1 \, P(X_{-N}^1, Y_{-N}^1) \log\left(\frac{P(Y(k=1)|X_{-N}^0, Y_{-N}^0)}{P(Y(k=1)|Y_{-N}^0)}\right) \quad (10)$$

Taking $\Delta t \to dt$, the continuous time limit, $k = 0$ is replaced with the current time $t$, $k = 1$ becomes time $t + dt$, and we get

$$\dot{I}_{X \to Y} = \frac{1}{dt} \int d\{X(t+dt)\} \, d\{Y(t+dt)\} \, P(\{X(t+dt)\}, \{Y(t+dt)\}) \log\left(\frac{P(Y(t+dt)|\{X(t)\}, \{Y(t)\})}{P(Y(t+dt)|\{Y(t)\})}\right) \quad (11)$$

$$\dot{I}_{X \to Y} = \frac{1}{dt} \langle D_{KL}\big(P(Y(t+dt)|\{X(t)\},\{Y(t)\}) || P(Y(t+dt)|\{Y(t)\})\big) \rangle_{\{X(t)\},\ \{Y(t)\}} \quad (12)$$

where $\{X(t)\}$ is the full history of $X(t)$ up to time $t$ (inclusive), and $D_{KL}$ is the Kullback-Leibler divergence.

Drift speed in the regime of small chemotactic bias

The first step in connecting information transfer to chemotactic performance is to construct a model of chemotaxis and derive expressions for the two quantities. Here we draw on the derivation by Locsei [7] to derive the drift speed of a chemotactic *E. coli* cell in shallow, static gradients. In this section, rather than re-derive that result, we will cast it in a form that will be convenient later on, when we derive the upper bound that information transfer places on the maximum drift speed a cell can achieve.

We model chemotaxis, like others have done before [7–11], as follows. During runs, the cell swims with speed $v_0$ and is subject to rotational diffusion with coefficient $D_r$. Many experiments to date [12–16] are consistent with a Monod-Wyman-Changeux (MWC) model [17] of cells' transmembrane receptor activity. In this description, cells sense concentration through its effects on the free energy difference between the inactive and active states of the receptors, $f_c = \log\left(\frac{1+c(t)/K_i}{1+c(t)/K_a}\right)$. Here, $K_i$ and $K_a$ are the receptor dissociation constants for the ligand when the receptors are in the inactive or active state [18,19], and cells are log-sensing over a wide range of concentrations $K_i \ll c(t) \ll K_a$ [14,20]. In a static, shallow, exponential gradient, we have $c(x) = c_0 e^{g x}$, where $c_0$ is the background concentration and $g$ is the gradient steepness. Cells respond to time changes of concentration through time changes of receptor state $\frac{d}{dt} f_c = \frac{d}{dt} \log\left(\frac{1+c(t)/K_i}{1+c(t)/K_a}\right)$, which in the log-sensing regime is $\sim \frac{d}{dt} \log c(t) = v_x(t) \frac{d}{dx} \log\big(c(x(t))\big) = g\, v_0 \cos(\theta(t))$. Here, $\theta(t)$ is the angle between the cell's heading at time $t$ and the direction of the concentration gradient (the $x$-axis). Therefore, we define the signal to be $s(t) = \frac{d}{dt} \log c(t)$. In shallow gradients (small $g$), we could just as well define the signal as $s(t) = \frac{dc}{dt}$, lumping $c_0$ into the cell's response to the signal. Since signal and response always come as a pair, this distinction does not affect our results.

The cell's swimming state $m(t)$ can be either run ($R$) or tumble ($T$). We assume that the cell has zero speed during tumbles [21]. Previous studies have shown that slow fluctuations in signaling activity drive long tails in the counterclockwise rotation durations of single motors [22–25]. Here and in our experiments, we only consider shallow gradients, where these slow fluctuations are predicted to increase drift speed by at most 10% [26,27]. For simplicity, here we model run-tumble transitions as Poisson processes with rates that do not fluctuate in the absence of signal. As a result, the tumble rate should be understood as an effective tumble rate that lumps together the effects of ligand arrival noise, noisy internal state, and multiple motors. This model fits well our measurements of cells' velocity autocorrelation functions, which on the time scale we are able to measure exhibits an exponential decay (Fig. S2A; see also refs [1,21]). The run-to-tumble transition occurs with rate

$$\lambda_R(t) = \lambda_{R0}\left(1 - \epsilon(\{s(t)\})\right), \tag{13}$$

depending on the history of signal seen, $\{s(t)\}$. Curly brackets { }, such as $\{s(t)\}$, indicate a trajectory of a quantity up to and including the indicated time (in this case, $t$). We define $\lambda_{R0}$ as the average tumble rate in the absence of a gradient. Tumble to run transitions occur with constant rate $\lambda_T$, which is taken to be independent of $s(t)$, valid in shallow gradients [7,21]. As a result of the Poisson assumption, run and tumble transitions have no memory of how long the cell has been in its current state.

Tumbles can partially reorient the cell [21,28], quantified by $\alpha = \langle \cos(\gamma) \rangle$, where $\gamma$ is the angle between the cell's swimming direction before and after the tumble. When $\alpha = 0$, tumbles fully reorient the cell. We also assume that the cell is navigating a shallow gradient, therefore the tumble rate modulation is small, $\epsilon(\{s(t)\}) \ll 1$.

In general, $\epsilon(\{s(t)\})$ might be a complicated function of the past signals, but in shallow gradients, only a linear response approximation is needed to calculate the cell's drift speed. Therefore, we consider responses of the form

$$\epsilon(\{s(t)\}) = \epsilon_0 \int_{-\infty}^{t} K_b(t - t')\, s(t')\, dt', \tag{14}$$

with kernel $K_b(T)$ mapping signal to behavior. Without loss of generality, we normalize the response kernel, $\int_0^\infty K_b(T)\, dT = 1$, so that the response gain is contained in $\epsilon_0$. The parameter $\epsilon_0$ determines how strongly the tumble rate responds to the signal, which depends on the properties of the cell's signal transduction pathway. Ligand arrival noise and other factors likely constrain which combinations of $\epsilon_0$ and $K_b(T)$ a cell can physically implement, but since we are seeking an upper bound on cells' performance (at a given information rate), we can relax these constraints.

By integrating by parts, one can show that responses to $s(t)$ with kernel $K_b(T)$ are equivalent responses to (log) concentration with a different kernel:

$$\epsilon(\{c(t)\}) = \epsilon_0 \int_{-\infty}^{t} R_b(t - t') \log(c(t'))\, dt'. \tag{15}$$

These are equivalent for $s(t) = \frac{d}{dt}\log(c(t))$ and $R_b(t - t') = K_b(0)\,\delta(t - t') - \frac{dK_b(t-t')}{dt'}$. We also assume that cells adapt perfectly to the background concentration of attractant [29,30], which here means that $\int_0^\infty R_b(T)\, dT = 0$ or equivalently $K_b(\infty) = 0$ (eventually, to adapt perfectly, the cell has to stop responding to a signal that happened far in the past).

Locsei [7] showed that the drift speed is

$$v_d = \frac{v_0}{3} \frac{(1-\alpha)\lambda_{R0}}{(1-\alpha)\lambda_{R0} + 2 D_r} P_{run} \frac{\epsilon_0 v_0 g}{(1-\alpha)\lambda_{R0} + 2 D_r} \int_0^\infty R_b(T) \, e^{-((1-\alpha)\lambda_{R0} + 2 D_r)T} \, dT. \quad (16)$$

$P_{run} = P(m(t) = R)$ is the fraction of time the cell spends in the run state, or $P_{run} = 1 - TB$, with tumble bias $TB = \frac{\lambda_{R0}}{\lambda_{R0} + \lambda_T}$. Plugging in the relationship between $R_b(T)$ and $K_b(T)$ above, and integrating by parts, one finds:

$$v_d = \frac{v_0}{3} \frac{(1-\alpha)\lambda_{R0}}{(1-\alpha)\lambda_{R0} + 2 D_r} P_{run} \, \epsilon_0 \, v_0 \, g \int_0^\infty K_b(T) \, e^{-((1-\alpha)\lambda_{R0} + 2 D_r)T} \, dT. \quad (17)$$

Below, we will show that in this expression:

$$\langle \epsilon(\{s(t)\}) \cos(\theta(t)) \rangle = \frac{1}{3} \epsilon_0 \, v_0 \, g \int_0^\infty K_b(T) \, e^{-((1-\alpha)\lambda_{R0} + 2 D_r)T} \, dT, \quad (18)$$

where the angled brackets are an average over trajectories, conditioned on the cell being in the run state at the current time $t$, $m(t) = R$. Writing the drift speed in this form,

$$v_d = v_0 \frac{(1-\alpha)\lambda_{R0}}{(1-\alpha)\lambda_{R0} + 2 D_r} P_{run} \langle \epsilon(\{s(t)\}) \cos(\theta(t)) \rangle, \quad (19)$$

will simplify the derivations later when we upper bound the cell's drift speed at fixed information rate from signal to behavior.

To show this, we start from the definition of $\epsilon(\{s(t)\})$ in Eqn. (14) and split the average over trajectories into two expectations. The first is an average over past trajectories given that they end with signal $s(t)$, and the second is an average over current signals $s(t)$. The first expectation is:

$$\langle \epsilon(\{s(t)\}) \cos(\theta(t)) \, | \, s(t), m(t) = R \rangle$$
$$= \epsilon_0 \left\langle \cos(\theta(t)) \int_{-\infty}^t K_b(t - t') \, s(t') \, dt' \, \Big| \, s(t), m(t) = R \right\rangle \quad (20)$$

Given $s(t) = v_0 \, g \cos(\theta(t))$, only $\theta(t')$ is a random variable in this expression. Similarly, conditioning on $s(t)$ is the same as conditioning on $\cos(\theta(t))$, so we can simplify to:

$$= \epsilon_0 \, g \, v_0 \cos(\theta(t)) \int_{-\infty}^t K_b(t - t') \langle \cos(\theta(t')) \, | \, \cos(\theta(t)), m(t) = R \rangle \, dt'. \quad (21)$$

The heading at time $t'$ in the past is $\cos(\theta(t')) = \cos(\theta(t') - \theta(t) + \theta(t)) = \cos(\theta(t') - \theta(t)) \cos(\theta(t)) - \sin(\theta(t') - \theta(t)) \sin(\theta(t))$, using the angle sum formula. The product of sine terms will equal zero in the end, so we drop them here. Plugging this in:

$$\langle\epsilon(\{s(t)\})\cos(\theta(t))\,|s(t),m(t)=R\rangle$$
$$=\epsilon_0\,g\,v_0\cos^2(\theta(t))\int_{-\infty}^{t}K_b(t-t')\,\left\langle\cos\left(\theta(t')-\theta(t)\right)\Big|\cos(\theta(t)),m(t)=R\right\rangle\,dt'. \tag{22}$$

Since we assume the gradient is shallow, we will only keep terms up to first order in $g$. Therefore, $\langle\cos(\theta(t')-\theta(t))\rangle$ is an average over swimming trajectories that experience rotational diffusion and tumbles but that don't respond to the gradient. Rotational diffusion is symmetric around the initial heading, which is why we excluded the term $\langle\sin(\theta(t')-\theta(t))\rangle=0$ above. Since $\langle\cos(\theta(t')-\theta(t))\rangle=e^{-((1-\alpha)\lambda_{R0}+2\,D_r)\,(t-t')}$ for $t'<t$[28], we now have:

$$=\epsilon_0\,g\,v_0\,\cos^2(\theta(t))\int_{-\infty}^{t}K_b(t-t')\,e^{-((1-\alpha)\lambda_{R0}+2\,D_r)\,(t-t')}\,dt'. \tag{23}$$

Since the system is stationary and we conditioned on the state at time $t$, we can change variables to integrate over time into the past $T$, back from the current time $t$: $T=t-t'$ and $dT=-dt'$. This gives:

$$\langle\epsilon(\{s(t)\})\cos(\theta(t))\,|s(t),m(t)=R\rangle$$
$$=\epsilon_0\,g\,v_0\,\cos^2(\theta(t))\int_{0}^{\infty}K_b(T)\,e^{-((1-\alpha)\lambda_{R0}+2\,D_r)\,T}\,dT. \tag{24}$$

Then, taking the second expectation, over current signals $s(t)$, or current headings $\cos(\theta(t))$ (with the conditioning on $m(t)=R$ implied):

$$\langle\epsilon(\{s(t)\})\cos(\theta(t))\rangle=\epsilon_0\,g\,v_0\,\langle\cos^2(\theta(t))\rangle\int_{0}^{\infty}K_b(T)\,e^{-((1-\alpha)\lambda_{R0}+2\,D_r)\,T}\,dT. \tag{25}$$

Again, since we assume the gradient is shallow and only keep terms to first order in $g$, the expectation $\langle\cos^2(\theta(t))\rangle$ is with respect to a uniform distribution of headings in three-dimensional space, giving:

$$\langle\epsilon(\{s(t)\})\cos(\theta(t))\rangle$$
$$=\epsilon_0\,g\,v_0\,\frac{1}{4\pi}\int_{0}^{2\pi}\int_{0}^{\pi}\cos^2(\theta)\sin(\theta)\,d\theta\,d\phi\int_{0}^{\infty}K_b(T)\,e^{-((1-\alpha)\lambda_{R0}+2\,D_r)\,T}\,dT \tag{26}$$

or:

$$\langle\epsilon(\{s(t)\})\cos(\theta(t))\rangle=\frac{1}{3}\epsilon_0\,g\,v_0\int_{0}^{\infty}K_b(T)\,e^{-((1-\alpha)\lambda_{R0}+2\,D_r)\,T}\,dT. \tag{27}$$

Information rate from signal to behavior

To derive the transfer entropy rate from signal to behavior, or information rate for short, we will use the differential expression for the steady state transfer entropy rate,

$$\dot{I}_{s \to m} = \frac{1}{dt} \left\langle D_{KL}\left(P(m(t+dt)|\{m(t)\},\{s(t)\}) || P(m(t+dt)|\{m(t)\})\right)\right\rangle_{\{m(t)\},\{s(t)\}} \tag{28}$$

$$= \frac{1}{dt} \left\langle \log\left(\frac{P(m(t+dt)|\{m(t)\},\{s(t)\})}{P(m(t+dt)|\{m(t)\})}\right)\right\rangle_{\{m(t)\},\{s(t)\},m(t+dt)}. \tag{29}$$

Again, the cell's swimming state $m(t)$ can be either run ($R$) or tumble ($T$), $s(t)$ is the signal the cell perceives, and curly brackets { }, such as $\{s(t)\}$ indicate a trajectory of a quantity up to and including time $t$.

Deriving the information rate for a chemotactic *E. coli* requires deriving the transition probabilities above. Each of the transition probabilities can be written with the trajectories $\{m(t)\}$ and $\{s(t)\}$ separated by the current time and the past: $\{m(t)\} = (m(t), \{m(t-dt)\})$ and $\{s(t)\} = (s(t), \{s(t-dt)\})$. Furthermore, since we assumed in the previous section that transitions between runs and tumble are inhomogeneous Poisson processes given the trajectory $\{s(t)\}$, the transition probabilities are independent of how long the cell has been in its current state. Together, these give: $P(m(t+dt)|\{m(t)\},\{s(t)\}) = P(m(t+dt)|m(t), s(t), \{s(t-dt)\})$. In particular, the probability that a running cell transitions to tumbling in the next $dt$ is $P(m(t+dt) = T|m(t) = R, \{s(t)\}) = \lambda_R(\{s(t)\}) dt$ and the probability that a tumbling cell transitions to running in $dt$ is $P(m(t+dt) = R|m(t) = T, \{s(t)\}) = \lambda_T dt$.

We also need the conditional probability $P(m(t+dt)|\{m(t)\})$. When the cell is in the tumble state at time $t$, $P(m(t+dt) = R|m(t) = T, \{m(t-dt)\})$, reduces to $P(m(t+dt) = R|m(t) = T) = \lambda_T dt$, since tumble-to-run transitions are independent of the past and of the signal. When the cell is the run state, the transition probability $P(m(t+dt) = T|m(t) = R, \{m(t-dt)\})$ is more complicated. This quantity measures the inferred probability that the cell will tumble in $dt$, given the past of motor states $\{m(t)\}$. The past motor states are informative of whether the cell will tumble only because they allow inference of the past signals $\{s(t)\}$. Writing this out mathematically:

$$P(m(t+dt) = T|m(t) = R, \{m(t-dt)\}) \tag{30}$$

$$= \int P(m(t+dt) = T|\{s(t)\}, m(t) = R, \{m(t-dt)\}) \, P(\{s(t)\}|m(t) = R, \{m(t-dt)\}) \, d\{s(t)\} \tag{31}$$

$$= \int P(m(t+dt) = T|\{s(t)\}, m(t) = R) \, P(\{s(t)\}|m(t) = R, \{m(t-dt)\}) \, d\{s(t)\} \tag{32}$$

$$= \int \lambda_R(\{s(t)\}) \, dt \, P(\{s(t)\}|m(t) = R, \{m(t-dt)\}) \, d\{s(t)\}. \tag{33}$$

$P(\{s(t)\}|m(t) = R, \{m(t-dt)\})$ is the distribution of past signal trajectories that can be inferred from knowledge that the cell is currently running and from its past behavior. Since the behavioral response $\epsilon(\{s(t)\})$ is small and behavior only depends on the history of signal through it, $P(\{s(t)\}|m(t) = R, \{m(t-dt)\})$ can be written as an asymptotic series solution in $\epsilon(\{s(t)\})$:

$$P(\{s(t)\}|m(t) = R, \{m(t - dt)\}) \sim P_0(\{s(t)\}|m(t) = R)\big(1 + O(\epsilon(\{s(t)\})) + O(\epsilon^2(\{s(t)\})) + \cdots\big), \tag{34}$$

where $P_0(\{s(t)\}|m(t) = R)$ is the distribution of signal trajectories when the cell does not respond to the signal. This is a probability distribution, so it must integrate to one. Since the zeroth order term integrates to one, all higher order terms must either integrate to zero or equal zero. Since terms with $\epsilon(\{s(t)\})$ raised to an even power are nonnegative, their integrals can only equal zero if they are zero. Therefore, we have:

$$P(\{s(t)\}|m(t) = R, \{m(t - dt)\}) \sim P_0(\{s(t)\}|m(t) = R)\big(1 + O(\epsilon(\{s(t)\})) + \cdots\big). \tag{35}$$

We assumed that $\lambda_{R0}$ is the average tumble rate in the absence of a gradient, or:

$$\lambda_{R0} = \int P_0(\{s(t)\}|m(t) = R)\, \lambda_R(\{s(t)\})\, d\{s(t)\} \tag{36}$$

$$= \lambda_{R0} \int P_0(\{s(t)\}|m(t) = R)\, (1 - \epsilon(\{s(t)\}))\, d\{s(t)\} \tag{37}$$

$$= \lambda_{R0}\, (1 - \langle \epsilon(\{s(t)\}) \rangle) \tag{38}$$

$$\rightarrow \quad \langle \epsilon(\{s(t)\}) \rangle = 0 \tag{39}$$

Therefore, $\langle O(\epsilon(\{s(t)\})) \rangle = 0$. Below, angled brackets without a subscript will denote an average with respect to $P_0(\{s(t)\}|m(t) = R)$, which is the same usage as in the derivation of the drift speed in the section above.

Using the expression for $P(\{s(t)\}|m(t) = R, \{m(t - dt)\})$ above, we get:

$$P(m(t + dt) = T|m(t) = R, \{m(t - dt)\}) = \tag{40}$$
$$dt \int \lambda_{R0}(1 - \epsilon(\{s(t)\}))\, P_0(\{s(t)\}|m(t) = R)\big(1 + O(\epsilon(\{s(t)\})) + \cdots\big) d\{s(t)\}.$$

Keeping terms up to order $\epsilon^2(\{s(t)\})$ (and dropping $(t)$ for clarity):

$$= \lambda_{R0}\, dt \int P_0(\{s(t)\}|m(t) = R)\left(1 - \epsilon(\{s\}) + O(\epsilon(\{s\})) - O(\epsilon(\{s\}))^2\right) d\{s(t)\}. \tag{41}$$

Using $\langle \epsilon(\{s(t)\}) \rangle = 0$, this is:

$$= \lambda_{R0}\, dt\, \left(1 - \left\langle O(\epsilon(\{s\}))^2 \right\rangle\right) = \lambda_{R0}\, dt\, (1 - A), \tag{42}$$

where we define $A = \left\langle O\big(\epsilon(\{s\})\big)^2 \right\rangle > 0$ as the expectation of the first-order terms squared that come from $P(\{s(t)\}|m(t) = R, \{m(t - dt)\})$. This definition will be useful because $A$ will ultimately drop out in the final expression for $\dot{I}_{s \to m}$. To understand the expression for $P(\{s(t)\}|m(t) = R, \{m(t - dt)\})$ above, observing that the cell has been running for any finite amount of time increases the likelihood that it has been running up the gradient. The amount by which that likelihood increases is order $\epsilon(\{s(t)\})$. Then, since it is more likely going up-gradient than down-gradient, the likelihood that it will tumble in $dt$ should be strictly lower than average. Predicting whether the cell will tumble in $dt$ produces another factor of $\epsilon(\{s(t)\})$ in $A$ from the definition of $\lambda_R(t)$.

With these transition probabilities in hand, we now derive the transfer entropy rate. As a reminder, the information rate is:

$$\dot{I}_{s \to m} = \frac{1}{dt} \left\langle \log\left(\frac{P(m(t+dt)|\{m(t)\},\{s(t)\})}{P(m(t+dt)|\{m(t)\})}\right) \right\rangle_{\{m(t)\},\{s(t)\},m(t+dt)}. \tag{43}$$

Writing out the expectation with respect to $m(t)$:

$$= \frac{1}{dt}\left(P_{run}\left\langle \log\left(\frac{P(m(t+dt)|m(t)=R,\{m(t-dt)\},\{s(t)\})}{P(m(t+dt)|m(t)=R,\{m(t-dt)\})}\right)\right\rangle_{\{s(t)\},m(t+dt),\{m(t-dt)\}|m(t)=R} \right.$$
$$\left. + P_{tumble}\left\langle \log\left(\frac{P(m(t+dt)|m(t)=T,\{m(t-dt)\},\{s(t)\})}{P(m(t+dt)|m(t)=T,\{m(t-dt)\})}\right)\right\rangle_{\{s(t)\},m(t+dt),\{m(t-dt)\}|m(t)=T}\right), \tag{44}$$

where $P_{tumble} = 1 - P_{run}$. Since tumble durations are assumed to not depend on past behaviors or on the signal, the second term in the parentheses is zero. We also simplify the numerator in the remaining logarithm:

$$\dot{I}_{s \to m}$$
$$= \frac{1}{dt} P_{run} \left\langle \log\left(\frac{P(m(t+dt)|m(t)=R,\{s(t)\})}{P(m(t+dt)|m(t)=R,\{m(t-dt)\})}\right) \right\rangle_{\{s(t)\},m(t+dt),\{m(t-dt)\}|m(t)=R}. \tag{45}$$

Writing out the expectation with respect to $m(t+dt)$,

$$= \frac{1}{dt} P_{run} \left\langle P(m(t+dt)=T|m(t)=R,\{s(t)\}) \log\left(\frac{P(m(t+dt)=T|m(t)=R,\{s(t)\})}{P(m(t+dt)=T|m(t)=R,\{m(t-dt)\})}\right) \right.$$
$$\left. + P(m(t+dt)=R|m(t)=R,\{s(t)\}) \log\left(\frac{P(m(t+dt)=R|m(t)=R,\{s(t)\})}{P(m(t+dt)=R|m(t)=R,\{m(t-dt)\})}\right) \right\rangle_{\{s(t)\},\{m(t-dt)\}|m(t)=R}. \tag{46}$$

Plugging in the transition probabilities derived above:

$$= \frac{1}{dt} P_{run} \left\langle \lambda_{R0}(1-\epsilon(\{s(t)\}))dt \, \log\left(\frac{\lambda_{R0}(1-\epsilon(\{s(t)\}))\,dt}{\lambda_{R0}\,dt\,(1-A)}\right) \right.$$
$$\left. + (1-\lambda_{R0}(1-\epsilon(\{s(t)\}))dt) \log\left(\frac{(1-\lambda_{R0}(1-\epsilon(\{s(t)\}))\,dt)}{(1-\lambda_{R0}\,dt\,(1-A))}\right) \right\rangle_{\{s(t)\},\{m(t-dt)\}|m(t)=R}. \quad (47)$$

Next, expanding the second logarithm to first order in $dt$:

$$= \frac{1}{dt} P_{run} \left\langle \lambda_{R0}\,dt\,(1-\epsilon(\{s(t)\})) \log\left(\frac{1-\epsilon(\{s(t)\})}{1-A}\right) \right.$$
$$+ \left(1 - \lambda_{R0}\,dt\,(1-\epsilon(\{s(t)\}))\right)\left(-\lambda_{R0}\,dt\,(1-\epsilon(\{s(t)\}))\right. \quad (48)$$
$$\left.\left. + \lambda_{R0}\,dt\,(1-A)\right) \right\rangle_{\{s(t)\},\{m(t-dt)\}|m(t)=R}$$

$$= \frac{1}{dt} P_{run} \left\langle \lambda_{R0}\,dt\,(1-\epsilon(\{s(t)\})) \log\left(\frac{1-\epsilon(\{s(t)\})}{1-A}\right) \right. \quad (49)$$
$$\left. + \left(-\lambda_{R0}\,dt\,(1-\epsilon(\{s(t)\})) + \lambda_{R0}\,dt\,(1-A)\right) \right\rangle_{\{s(t)\},\{m(t-dt)\}|m(t)=R}$$

$$= \lambda_{R0}\, P_{run} \left\langle (1-\epsilon(\{s(t)\})) \log\left(\frac{1-\epsilon(\{s(t)\})}{1-A}\right) + \epsilon(\{s(t)\}) - A \right\rangle_{\{s(t)\},\{m(t-dt)\}|m(t)=R} \quad (50)$$

For shallow gradients and small $\epsilon(\{s(t)\})$, we can expand the remaining logarithm to second order in $\epsilon(\{s(t)\})$. Recalling that $A$ is already second order in $\epsilon(\{s(t)\})$, we get:

$$= \lambda_{R0}\, P_{run} \left\langle (1-\epsilon(\{s(t)\}))\left(-\epsilon(\{s(t)\}) - \frac{1}{2}\epsilon^2(\{s(t)\}) + A\right) + \epsilon(\{s(t)\}) \right.$$
$$\left. - A \right\rangle_{\{s(t)\},\{m(t-dt)\}|m(t)=R}. \quad (51)$$

Keeping to second order in $\epsilon(\{s(t)\})$:

$$= \lambda_{R0}\, P_{run} \left\langle -\epsilon(\{s(t)\}) - \frac{1}{2}\epsilon^2(\{s(t)\}) + A + \epsilon^2(\{s(t)\}) + \epsilon(\{s(t)\}) - A \right\rangle_{\{s(t)\},\{m(t-dt)\}|m(t)=R} \quad (52)$$

$$= \frac{1}{2}\lambda_{R0}\, P_{run} \langle \epsilon^2(\{s(t)\}) \rangle_{\{s(t)\},\{m(t-dt)\}|m(t)=R}. \quad (53)$$

Next we separate the remaining expectation into two parts:

$$= \frac{1}{2}\lambda_{R0}\, P_{run} \langle \langle \epsilon^2(\{s(t)\}) \rangle_{\{s(t)\}|\{m(t-dt)\},m(t)=R} \rangle_{\{m(t-dt)\}|m(t)=R}. \quad (54)$$

Since the term inside the angled brackets is already order $\epsilon^2(\{s(t)\})$, only the zeroth order term of $P(\{s(t)\}|\{m(t-dt)\}, m(t) = R)$ contributes to the inner expectation:

$$= \frac{1}{2} \lambda_{R0} \, P_{run} \, \langle\langle \epsilon^2(\{s(t)\})\rangle\rangle_{\{m(t-dt)\}|m(t)=R}. \tag{55}$$

The inner expectation is taken with respect to $P_0(\{s(t)\}|m(t) = R)$. What remains there does not depend on $\{m(t-dt)\}$, so we can integrate out $\{m(t-dt)\}$, leaving:

$$\dot{I}_{s\to m} = \frac{1}{2} \lambda_{R0} \, P_{run} \, \langle \epsilon^2(\{s(t)\})\rangle$$
$$= \frac{1}{2} \lambda_{R0} \, P_{run} \, \left\langle \left(\frac{\lambda_R(\{s(t)\}) - \lambda_{R0}}{\lambda_{R0}}\right)^2\right\rangle. \tag{56}$$

This equation indicates that in the linear regime, the information rate measures the signal-induced variation in the tumble rate $\lambda_R(\{s\})$. The factor of $\lambda_{R0} P_{run}$ indicates that $\dot{I}_{s\to m}$ measures the long-time, signal-induced variation in $\lambda_R(\{s\})$ times the frequency of runs. Larger variation of the tumble rate per run due to response to signals indicates larger information transfer from signal to behavior. When tumbles occur more frequently on average (larger $\lambda_{R0}$), the signal influences behavior more frequently, leading to a higher transfer entropy rate.

Information-performance bound with fixed behavioral parameters
Here we derive the bound in Eqn. 1 of the main text and in Figs. 1 and 3. As we've shown above, the drift speed $v_d$ and the information rate $\dot{I}_{s\to m}$ depend on the behavioral response $\lambda_R(\{s\})$, or $\epsilon(\{s\})$. However, while some responses correspond to high information rates, they don't necessarily generate high drift speed. We reiterate that we are looking for the behavioral response that achieves the highest drift speed with a given information rate—this is not necessarily the response that produces the highest information rate or the highest drift speed.

To find the behavioral response $\epsilon(\{s(t)\})$ that maximizes the drift speed at fixed information rate, we solve:

$$\max_{\epsilon(\{s(t)\})} \mathcal{L}(\epsilon(\{s(t)\}), \boldsymbol{\theta}) \tag{57}$$

where $\boldsymbol{\theta} = \{\alpha, \lambda_{R0}, P_{run}\}$ are the behavioral parameters characterizing the cell's swimming in absence of a gradient (this is an alternative parameterization from the one in the main text). The Lagrangian is:

$$\mathcal{L} = v_d - \beta_1 \dot{I}_{s\to m} - \beta_2 \langle \epsilon(\{s(t)\})\rangle. \tag{58}$$

$\beta_1$ and $\beta_2$ are Lagrange multipliers (and since $\dot{I}_{s\to m} \geq 0$, we must have that $\beta_1 \geq 0$). The second term constrains the information rate, while the third term enforces the constraint that $\langle \epsilon(\{s(t)\})\rangle = 0$ (see previous section). The full expressions for the quantities above are:

$$v_d = v_0 \frac{(1-\alpha)\lambda_{R0}}{(1-\alpha)\lambda_{R0} + 2D_r} P_{run} \langle \epsilon(\{s(t)\}) \cos(\theta(t)) \rangle, \tag{59}$$

$$\dot{I}_{s\to m} = \frac{1}{2} \lambda_{R0} P_{run} \langle \epsilon^2(\{s(t)\}) \rangle. \tag{60}$$

Our efforts in previous sections to write $v_d$ and $\dot{I}_{s\to m}$ in terms of $\epsilon(\{s(t)\})$ and to write expectations with respect to the same distributions make taking the functional derivative with respect to $\epsilon(\{s(t)\})$ easier here. Setting it equal to zero gives:

$$\frac{\delta \mathcal{L}}{\delta \epsilon(\{s(t)\})} = 0$$

$$\to v_0 \frac{(1-\alpha)\lambda_{R0}}{(1-\alpha)\lambda_{R0} + 2D_r} P_{run} \cos(\theta(t)) - \beta_1 \lambda_{R0} P_{run} \epsilon^*(\{s(t)\}) + \beta_2 = 0 \tag{61}$$

$$\epsilon^*(\{s(t)\}) = -\frac{\beta_2}{\beta_1 \lambda_{R0} P_{run}} + \frac{1}{\beta_1 \lambda_{R0}} v_0 \frac{(1-\alpha)\lambda_{R0}}{(1-\alpha)\lambda_{R0} + 2D_r} \cos(\theta(t)) \tag{62}$$

$$= -\frac{\beta_2}{\beta_1 \lambda_{R0} P_{run}} + \frac{1}{\beta_1} \frac{(1-\alpha)}{(1-\alpha)\lambda_{R0} + 2D_r} \frac{1}{g} s(t) \tag{63}$$

$$= A + B\, s(t), \tag{64}$$

with $B = \frac{1}{\beta_1} \frac{1}{g} \frac{(1-\alpha)}{(1-\alpha)\lambda_{R0} + 2D_r}$. Enforcing that $\langle \epsilon(\{s(t)\}) \rangle = 0$ gives that $A = 0$, $\beta_2 = 0$.

Comparing to Eqn. (14) of the drift speed derivation, $\epsilon(\{s(t)\}) = \epsilon_0 \int_{-\infty}^{t} K_b(t - t')\, s(t')\, dt'$, we see that the optimal strategy is achieved by a behavioral kernel that is a delta function at the current time:

$$K_b^*(T) = \delta(T), \tag{65}$$

and $B = \epsilon_0$. Putting this together, we have:

$$\epsilon^*(\{s(t)\}) = \epsilon_0\, s(t). \tag{66}$$

The strategy that achieves the highest drift at a given information rate, in the regime of shallow gradients, is to modulate the tumble rate proportionally to the current signal $s(t)$. This optimal strategy can't be achieved by a real cell, which measures the signal by comparing concentrations at different times [31,32]. Still, no cell or agent can outperform this optimal strategy.

At the optimum:

$$v_d^* = \frac{v_0}{3} \frac{(1-\alpha)\lambda_{R0}}{(1-\alpha)\lambda_{R0} + 2D_r} (v_0\, g\, \epsilon_0)\, P_{run}, \tag{67}$$

$$\dot{I}_{s\to m}^* = \frac{1}{2} \lambda_{R0} \frac{(v_0\, g\, \epsilon_0)^2}{3} P_{run}. \tag{68}$$

Expressing the drift speed in terms of the information rate, one finds that the drift speed cannot exceed:

$$\frac{v_d}{v_0} \leq \frac{(1-\alpha)\lambda_{R0}}{(1-\alpha)\lambda_{R0} + 2D_r} P_{run} \left(\frac{2}{3}\frac{\dot{I}_{s\to m}}{\lambda_{R0} P_{run}}\right)^{\frac{1}{2}}. \tag{69}$$

The term $\frac{\dot{I}^*_{s\to m}}{\lambda_{R0} P_{run}}$ inside the square root can be interpreted as the information the cell gets about the signal trajectory per run: higher information per run is necessary, but not sufficient, for higher drift. The factor outside of the square root determines how well the information is translated into drift in the presence of rotational diffusion. Therefore, in addition to information acquisition, navigation depends on the matching between the cell's behavioral parameters and the properties of the physical environment. Comparing to Eqn. 1 of the main text, we have

$$F(\boldsymbol{\theta}) = \frac{(1-\alpha)\lambda_{R0}}{(1-\alpha)\lambda_{R0} + 2D_r}\left(8\frac{D_r}{\lambda_{R0}}P_{run}\right)^{\frac{1}{2}}. \tag{70}$$

Relevant bits for bacterial chemotaxis

$\dot{I}_{s\to m}$ measures the rate of information transfer from the full history of signal to motor behavior. But the response strategy that maximizes drift speed given some information rate $\dot{I}_{s\to m}$ is one that only responds to the current signal. This suggests that the information rate can be partitioned into two parts: information about current signal, which drives chemotaxis, and information about the rest of the signal trajectory. We will now show that the drift speed is proportional to the square root of the former, *for any behavioral response*. Among strategies with *the same information rate* $\dot{I}_{s\to m}$, the optimal response achieves the highest drift speed by only transferring information about current signal $s(t)$. We will make this more precise below.

$\dot{I}_{s\to m}$ can be partitioned as follows:

$$\dot{I}_{s\to m} = \frac{1}{dt}\left\langle \log\left(\frac{P(m(t+dt)|\{m(t)\},\{s(t)\})}{P(m(t+dt)|\{m(t)\})}\right)\right\rangle_{\{m(t)\},\{s(t)\},m(t+dt)} \tag{71}$$

$$= \frac{1}{dt}\left\langle \log\left(\frac{P(m(t+dt)|\{m(t)\},s(t))}{P(m(t+dt)|\{m(t)\})}\right)\right\rangle_{\{m(t)\},\{s(t)\},m(t+dt)} \tag{72}$$

$$+ \frac{1}{dt}\left\langle \log\left(\frac{P(m(t+dt)|\{m(t)\},s(t),\{s(t-dt)\})}{P(m(t+dt)|\{m(t)\},s(t))}\right)\right\rangle_{\{m(t)\},\{s(t)\},m(t+dt)}$$

$$= \dot{I}_{s(t)\to m} + \dot{I}_{\{s(t-dt)\}\to m|s(t)}. \tag{73}$$

The first term measures the information transferred about current signal $s(t)$ to behavior. The second term measures how much information is transferred about past signals that aren't correlated with $s(t)$. In general, $\dot{I}_{s(t)\to m} \leq \dot{I}_{s\to m}$, because $\dot{I}_{s\to m}$ contains information about $s(t)$ as well as the full history of $s$. For the optimal response, $\dot{I}_{s\to m} = \dot{I}_{s(t)\to m}$ and $\dot{I}_{\{s(t)-dt\}\to m|s(t)} = 0$—no information is transferred about past signals that aren't correlated with the current signal. Note that there is a distinction between responding to a signal in the past (i.e. by making $\lambda(\{s\})$ depend on that past signal) and transferring information about a signal: responding to a signal at some time transfers information about all signals at other times that are correlated with it.

Next, we will show that the drift speed of any behavioral response is proportional to $\dot{I}_{s(t)\to m}$. The derivation of $\dot{I}_{s(t)\to m}$ is nearly identical to that of $\dot{I}_{s\to m}$, but with $\epsilon(\{s(t)\})$ replaced with $\epsilon(s(t))$, which is the deviation of the tumble rate in response to the *current* signal, after averaged over histories of signal that end with signal $s(t)$ at the current time $t$. Carrying out the same steps gives:

$$\dot{I}_{s(t)\to m} = \frac{1}{2}\lambda_{R0}\,P_{run}\,\langle \epsilon^2(s(t))\rangle. \tag{74}$$

The arguments in the drift speed derivation let us derive $\epsilon(s(t))$ in the linear regime as

$$\epsilon(s(t)) = \langle \epsilon(\{s(t)\})|s(t), m(t)=R\rangle$$
$$= \epsilon_0\, g\, v_0\, \cos(\theta(t))\int_0^\infty K_b(T)\, e^{-((1-\alpha)\lambda_{R0}+2\,D_r)\,T}\,dT. \tag{75}$$

Comparing to the drift speed section, the right-hand side of the equality above can be written as:

$$\langle \epsilon^2(s(t))\rangle = 3\,\langle \epsilon(\{s(t)\})\cos(\theta(t))\rangle^2. \tag{76}$$

Therefore, we have for the drift speed of *any* behavioral response in the linear regime:

$$v_d = v_0 \frac{(1-\alpha)\,\lambda_{R0}}{(1-\alpha)\,\lambda_{R0}+2\,D_r}\,P_{run}\left(\frac{1}{3}\langle \epsilon^2(s(t))\rangle\right)^{\frac{1}{2}} \tag{77}$$
$$= v_0 \frac{(1-\alpha)\,\lambda_{R0}}{(1-\alpha)\,\lambda_{R0}+2\,D_r}\,P_{run}\left(\frac{2}{3}\frac{\dot{I}_{s(t)\to m}}{\lambda_{R0}\,P_{run}}\right)^{1/2}. \tag{78}$$

The drift speed therefore depends on how much information about the current signal a cell communicates to its behavior. $\dot{I}_{s(t)\to m}$ is a kind of predictive information rate [33–35]: it measures the mutual information between the current signal and the cell's behavior state in the next $dt$, given the past of behavioral states, or $\dot{I}_{s(t)\to m} = \frac{1}{dt}MI(s(t); m(t+dt)|\{m(t)\})$. The optimal response strategy makes $\dot{I}_{s\to m} = \dot{I}_{s(t)\to m}$.

In static gradients, run-tumble transitions and rotational diffusion determine the signal statistics. We assumed that run-tumble transitions were Poisson processes, so to leading order in the gradient steepness $g$, the signal is Markovian. Therefore, signals farther in the past are less correlated with the current signal. However, if the signal were non-Markovian, for example if it were oscillatory, responding to past signals that are highly correlated with the current signal could be an effective way of transferring information about the current signal (for example, see Becker et al *PRL* 2015 [36], who solved a different but related optimization problem).

Information-performance bound with optimal behavioral parameters

The bound derived in the previous section is valid for any behavioral parameters $\alpha$, $\lambda_{R0}$, and $P_{run}$. In the main text, we measured how efficiently a typical RP437 *E. coli* uses information to climb gradients by comparing their performance to the maximum they could possibly achieve, which was set by measuring the behavioral parameters of the median phenotype $P_{run}$ and plugging them into the bound (right hand side of supplemental Eqn. (69), main text Eqn. 1). But individual cells in an isogenic population or different strains of *E. coli* can have different behavioral parameters. The performance of these cells is bounded by a different curve from the one in Fig. 3 of the main text, given by Eqn. (69) with those cells' behavioral parameter values. In principle, the Lagrangian $\mathcal{L}$ can be further optimized with respect to the behavioral parameters to find the bound that no individual cell can exceed.

First setting the derivative of the Lagrangian with respect to $\lambda_{R0}$ equal to zero, we get:

$$\frac{\partial \mathcal{L}}{\partial \lambda_{R0}} = \frac{v_0}{3} \frac{2 D_r (1-\alpha)}{\left((1-\alpha) \lambda_{R0}^* + 2 D_r\right)^2} (v_0\, g\, \epsilon_0)\, P_{run} - \frac{1}{6} \beta_1 (v_0\, g\, \epsilon_0)^2\, P_{run} = 0 \qquad (79)$$

Using $\beta_1 = \frac{1}{g\, \epsilon_0} \frac{(1-\alpha)}{(1-\alpha)\lambda_{R0}+2 D_r}$ from the expressions above, this reduces to:

$$(1-\alpha)\, \lambda_{R0}^* = 2 D_r. \qquad (80)$$

This indicates that at the optimum, the rate at which the ideal cell actively changes direction by tumbling equals the rate at which it passively changes direction by rotational diffusion. The optimal value of this behavioral parameter depends on the physical environment through $D_r$. Interestingly, others have arrived at the same optimal mean tumble rate, but by optimizing a different objective[8].

For the optimal value of $\lambda_{R0}$, the drift speed is:

$$\frac{v_d}{v_0} \leq \left(\frac{(1-\alpha)\, \dot{I}_{s\to m}}{12\, D_r}\, P_{run}\right)^{1/2}, \qquad (81)$$

and $F(\boldsymbol{\theta})$ reduces to $F(\boldsymbol{\theta}) = \left((1-\alpha)\, P_{run}\right)^{1/2}$.

Finally, the right-hand side of the inequality above is maximized when the cell's tumbles completely reorient its swimming direction, i.e. $\alpha = 0$, and tumbles are instantaneous, i.e. $P_{run} = 1$:

$$\frac{v_d}{v_0} \leq \left(\frac{\dot{I}_{s\to m}}{12\, D_r}\right)^{1.2}. \qquad (82)$$

This is the absolute maximum speed at which a cell could possibly climb a gradient given its information rate. With these optimal behavioral parameters, $F(\boldsymbol{\theta}) = 1$, its maximal value. But, in addition to responding instantaneously, achieving $\alpha = 0$ and $P_{run} = 1$ might not be physically realizable for *E. coli* cells. If reorienting completely during tumbles takes finite time, then having $\alpha = 0$ might require that $P_{run} < 1$.

With this, we have series of bounds:

$$\frac{v_d}{v_0} \leq \frac{(1-\alpha)\lambda_{R0}}{(1-\alpha)\lambda_{R0} + 2D_r} P_{run} \left(\frac{2}{3} \frac{\dot{I}_{s \to m}}{\lambda_{R0} P_{run}}\right)^{\frac{1}{2}} \leq \left(\frac{(1-\alpha)\dot{I}_{s \to m}}{12 D_r} P_{run}\right)^{\frac{1}{2}} \leq \left(\frac{\dot{I}_{s \to m}}{12 D_r}\right)^{\frac{1}{2}}. \tag{83}$$

The first inequality comes from optimizing the cell's behavioral response to signal; the second one comes from additionally optimizing the mean tumble rate $\lambda_{R0}$; and the third one comes from further optimizing $\alpha$ and $P_{run}$. These bounds are plotted in Fig. S10.

Although we derived the optimal parameters assuming the optimal (instantaneous) behavioral response, these parameters should still be optimal for any given kernel. We focus on $\lambda_{R0}$ in particular because, for the parameter values we measured in *E. coli*, it has the biggest effect on the bound. First, we found that the highest drift speed at fixed information rate is achieve by responding to the current signal only. However, if the cell responds to signals that are highly correlated with the current one, it can approach the bound. Longer runs make the signal correlation time longer, making the efficiency of any given kernel get closer to that of the instantaneous one. Second, runs longer than $1/(2 D_r)$ waste information because they lose direction before the signal elicits a change in run duration. As a result, the optimal $\lambda_{R0}$ should still be about $2 D_r$ for any fixed kernel.

Information rate from signal to kinase activity: approach
Measuring information transfer between two continuous, time-dependent variables is highly nontrivial because of the need to infer high-dimensional probability distributions. However, a simple analytical expression has been derived in the special case of linear, Gaussian systems. If $S(\omega)$ is the power spectrum of the input signal, $K(\omega)$ is the linear response function mapping input to output, and $N(\omega)$ is the power spectrum of spontaneous noise in the output, then the mutual information rate $\dot{I}(X;Y)$ between input $X$ and output $Y$ is [37,38]:

$$\dot{I}(X;Y) = \frac{1}{4\pi} \int_{-\infty}^{\infty} \log\left(1 + \frac{S(\omega)|K(\omega)|^2}{N(\omega)}\right) d\omega. \tag{84}$$

In our case, the input signal is $s(t) = \frac{d}{dt}\log(c(t))$ and the output is $a(t)$, the activity of CheA kinases.

The mutual information rate between signal and kinase activity $\dot{I}(s;a)$ is not generally equal to the transfer entropy rate from signal to kinase output $\dot{I}_{s \to a}$, because kinase activity affects the cell's behavior, which feeds back onto the signal. If this feedback were cut off, by decoupling motor behavior from kinase activity, then the two would be equal. One could imagine immobilizing a cell and feeding it a time course of signals it could experience if it were swimming. Since the immobilized cell's kinase responses do not affect this input signal, the mutual information rate and transfer entropy rate are the same, $\dot{I}_{s \to a} = \dot{I}(s;a)$. Finally, in shallow gradients and to leading order in gradient steepness $g$, the transfer entropy from signal to kinase activity $\dot{I}_{s \to a}$ for the immobilized cell is equal to that for a freely swimming cell.

Eqn. (84) is valid for Gaussian inputs and outputs related by a linear mapping. The response of $a$ to $s$ is approximately linear for the range of signals we used (see Fig. S5 and previous work [13,39]), and the fluctuations in $a$ are well-described by a Gaussian process (see noise autocorrelation functions in Fig. 2H, Fig. S6B, and previous work [24,25]), consistent with the assumptions of Eqn. (84) above. But the statistics of the signal $s$ are not Gaussian. Still, the expression above for the mutual information rate is a good approximation in shallow gradients. In static, one-directional, shallow exponential gradients, the signal is directly proportional to the component of the cell's velocity projected onto the gradient direction, $v_x(t)$. To see this, $s(t) = \frac{d}{dt}\log\big(c(x(t))\big) = \frac{dx}{dt}\frac{d}{dx}\log\big(c(x)\big) = v_x(t)\frac{d}{dx}\log\big(c(x)\big)$. With our assumptions, $\frac{d}{dx}\log\big(c(x)\big) = g$, making $s(t) = g\,v_x(t) = g\,v_0\cos\big(\theta(t)\big)$, where $\theta$ is the angle between the cell's swimming direction and gradient direction. In the long term and to leading order in $g$, the distribution of $\cos\big(\theta(t)\big)$ (in 3D space) is uniform between -1 and 1, therefore $s$ is uniform between $[-v_0\,g, v_0\,g]$. Eqn. (84) only keeps the second moment of the signal statistics; higher moments are zero for a Gaussian process. However, higher moments of $s$ are multiplied by higher powers of $g$, and therefore they only add small corrections to the Gaussian information rate in shallow gradients. This makes Eqn. (84) a good approximation of the information rate, even though the signal $s$ is not Gaussian.

We can simplify the expression in Eqn. (84). As noted in the paragraph above, in static, exponential concentration profiles that vary in one direction the signal is $s(t) = g\,v_x(t)$. Therefore, the power spectrum of the signal is just $S(\omega) = g^2\,V(\omega)$, where $V(\omega)$ is the power spectrum of $v_x(t)$. Plugging this in, we have:

$$\dot{I}_{s \to a} \sim \dot{I}(s;a) = \frac{1}{4\pi}\int_{-\infty}^{\infty}\log\left(1 + g^2\frac{V(\omega)\,|K(\omega)|^2}{N(\omega)}\right)d\omega. \tag{85}$$

In shallow gradients, the term inside the log multiplied by $g^2$ is small for all frequencies $\omega$. Therefore, we can further simplify this by linearizing the log:

$$\dot{I}_{s \to a} \sim g^2\,\frac{1}{4\pi}\int_{-\infty}^{\infty}\frac{V(\omega)\,|K(\omega)|^2}{N(\omega)}d\omega, \tag{86}$$

from which we see that the information rate is proportional to $g^2$ in shallow gradients. Calculating the information rate, as described in the next section, using Eqn. (86) overestimates the exact integral in Eqn. (85) by about 1% in the steepest gradients we used experimentally.

These expressions are analogous to the mutual information between two univariate Gaussian random variables related by a linear mapping. Say we have $X$ and $Y = k\,X + \xi$, where the signal $X \sim N(0, \sigma_x^2)$ and the noise $\xi \sim N(0, \sigma_n^2)$. Since noise is assumed to be independent of $X$, the variance of $Y$ is $\sigma_y^2 = k^2\sigma_x^2 + \sigma_n^2$. Then the mutual information between $X$ and $Y$ can be written in terms of the Pearson correlation $r$ between them:

$$I(X;Y) = -\frac{1}{2}\log(1 - r^2). \tag{87}$$

By definition,

$$r^2 = \frac{\sigma_{xy}^2}{\sigma_x^2 \sigma_y^2} = \frac{\langle x(kx+\xi)\rangle^2}{\sigma_x^2\left(k^2\sigma_x^2 + \sigma_n^2\right)} = \frac{\left(k\sigma_x^2\right)^2}{\sigma_x^2\left(k^2\sigma_x^2 + \sigma_n^2\right)} = \frac{k^2\sigma_x^2}{k^2\sigma_x^2 + \sigma_n^2}. \tag{88}$$

That is, $r^2$ measures the fraction of total variance in $Y$ that comes from $X$. Plugging this into the mutual information and rearranging,

$$I(X;Y) = \frac{1}{2}\log\left(\frac{1}{1-r^2}\right) = \frac{1}{2}\log\left(\frac{1}{1 - \frac{k^2\sigma_x^2}{k^2\sigma_x^2 + \sigma_n^2}}\right) = \frac{1}{2}\log\left(\frac{1}{\frac{\sigma_n^2}{k^2\sigma_x^2 + \sigma_n^2}}\right) = \frac{1}{2}\log\left(\frac{k^2\sigma_x^2 + \sigma_n^2}{\sigma_n^2}\right) \tag{89}$$

$$= \frac{1}{2}\log\left(1 + \frac{k^2\sigma_x^2}{\sigma_n^2}\right) \tag{90}$$

For small signal variance $\sigma_x^2$, this can be simplified by linearizing the log:

$$I(X;Y) \sim \frac{1}{2}\frac{k^2\sigma_x^2}{\sigma_n^2}. \tag{91}$$

Comparing this univariate mutual information to the mutual information rate in Eqns. (84)-(86), we see that the variances of the signal and the noise become their power spectra, $\sigma_x^2 \to S(\omega)$ and $\sigma_n^2 \to N(\omega)$, and the gain becomes the frequency response function, $k \to K(\omega)$. The univariate mutual information quantifies the ratio of signal-induced variance in $Y$ to noise variance, $\frac{k^2\sigma_x^2}{\sigma_n^2}$. Analogously, the mutual information rate (Eqn. (86)) quantifies the ratio of signal-induced power in kinase activity $a(t)$ to noise power, $\frac{S(\omega)|K(\omega)|^2}{N(\omega)}$, integrated over frequency components. Since the chemotactic signal power $S(\omega)$ is proportional to the gradient steepness squared, $S(\omega) = g^2 V(\omega)$, the signal-induced power in kinase activity and the information rate are both also proportional to $g^2$ when the gradient is shallow.

Returning to Eqn. (86), even with the dramatic simplification that this expression provides for estimating the mutual information rate, inferring the spectra $V(\omega)$, $K(\omega)$, and $N(\omega)$ nonparametrically from data still requires long trajectories. Instead, we used parameterized models for the signal and noise power spectra and for the signaling kernel that we constrained with data. These phenomenological models are described below.

Models of signal statistics and kinase activity noise and response
In this section we explain the models used for the signal statistics, the noise statistics, and the linear response function. We start with the swimming statistics. As described above, in static gradients, the signals a cell experiences are proportional to its up-gradient velocity $v_x(t)$. The statistics of $v_x(t)$ are characterized by their power spectrum $V(\omega)$, which can be computed from the Fourier transform of the autocorrelation function $V(t)$. Importantly, in shallow gradients and to leading order in gradient steepness $g$, the statistics of $v_x(t)$ are identical to those of a cell swimming in the absence of a gradient. Therefore:

$$V(t) = \langle v_x(t) v_x(0) \rangle = \frac{v_0^2}{3} P_{run} \, e^{-((1-\alpha)\lambda_{R0} + 2 D_r)|t|} = a_v \, e^{-\lambda_{tot} |t|}, \tag{92}$$

where the parameters have the same meaning as in earlier sections. $\lambda_{tot}$ is the total rate at which the cell loses its swimming direction. The factor of 1/3 results from measuring variations of $v_x(t)$ in 3D space. We define the Fourier transform as

$$\mathcal{F}[f(t)] = \int_{-\infty}^{\infty} f(t) \, e^{-i\omega t} \, dt. \tag{93}$$

Therefore, the power spectrum of $v_x(t)$ is:

$$V(\omega) = \mathcal{F}[V(t)] = 2 \frac{v_0^2}{3} P_{run} \frac{(1-\alpha)\lambda_{R0} + 2 D_r}{\left((1-\alpha)\lambda_{R0} + 2 D_r\right)^2 + \omega^2} \tag{94}$$

$$= a_v \frac{2 \lambda_{tot}}{\lambda_{tot}^2 + \omega^2}. \tag{95}$$

We explain how we compute the velocity autocorrelation function of a typical cell, as well as how we measure the behavioral parameters of the model, in the section **Estimating behavioral parameters, run speed, and rotational diffusion**.

Next, we consider the linear response function, which can be inferred directly from the cells' kinase responses to an impulse (delta function) of signal. The cells' impulse responses have a stereotypical shape consisting of a fast change in kinase output, followed by slow relaxation to baseline due to adaptation. A simple phenomenological model that captures these features of the response to a unit impulse of stimulus, $\delta(t)$, is:

$$K(t) = G \left(1 - e^{-\frac{t}{\tau_1}}\right) e^{-\frac{t}{\tau_2}} H(t), \tag{96}$$

where $G$ is the gain, $\tau_2$ is the adaptation time, and $\tau_1$ is the fast response time, and $H(t)$ is the Heaviside step function.

We infer CheA kinase output from FRET between fluorescently-labeled and overexpressed CheY and CheZ (see below), as has been done before [13,16,24,25,39]. Therefore, the FRET signal we measure has a relaxation time arising from CheY-CheZ binding and CheY-p dephosphorylation, and the value of $\tau_1$ we infer includes these relaxation dynamics. The CheA autophosphorylation time has been measured before to be about 0.05 s [5]; we measure $\tau_1$ with CheY/CheZ relaxation to be about 0.22 s (Table S1; **Estimating linear response function parameters**). The information rate has a weak dependence on $\tau_1$, and decreasing the value of $\tau_1$ monotonically increases the information rate. We use the value of $\tau_1$ from the literature to compute $\dot{I}_{s \to a}$, which results in a lower information efficiency. Using $\tau_1 = 0.22$ s decreases the information rate by 16% relative to the literature value; using $\tau_1 = 0$ increases it by 6%. The square root dependence of the bound in Eqn. (69) (Eqn. 1 of the main text) makes the effects of these differences on the efficiency even smaller. In Fig. S8, we plot the dependence of the information rate on $\tau_1$. Our fit to $\tau_2$, the adaptation time, should not be affected by CheYp-CheZ binding because it is much longer than the measured $\tau_1$. The gain $G$ that we inferred should also not be affected by CheYp-CheZ binding.

The frequency response $K(\omega)$ is the Fourier transform of $K(t)$:

$$K(\omega) = \frac{G}{\tau_1} \frac{1}{\left(\frac{1}{\tau_2} + i\omega\right)\left(\frac{1}{\tau_1} + \frac{1}{\tau_2} + i\omega\right)}, \quad (97)$$

where $i^2 = -1$. $K(\omega)$ appears in Eqn. (86) for the information rate as $|K(\omega)|^2 = K(\omega)K^*(\omega)$, with $K^*(\omega)$ the complex conjugate of $K(\omega)$. This is:

$$|K(\omega)|^2 = \frac{G^2}{\tau_1^2} \frac{1}{\left(\left(\frac{1}{\tau_2}\right)^2 + \omega^2\right)\left(\left(\frac{1}{\tau_1} + \frac{1}{\tau_2}\right)^2 + \omega^2\right)}. \quad (98)$$

The noise in kinase output $a(t)$ is well-described by an Ornstein-Uhlenbeck process [24]. This process is characterized by two parameters, a diffusivity $D_n$ and a relaxation time scale $\tau_n$. The long-term distribution of the process is Gaussian with variance $\sigma_n^2 = D_n \tau_n$. The autocorrelation of this process, in the absence of signal, has the form:

$$\langle a(t)a(0)\rangle = \sigma_n^2 \, e^{-\frac{|t|}{\tau_n}} \quad (99)$$

and the power spectrum, $N(\omega)$ is:

$$N(\omega) = \frac{2 D_n}{\left(\frac{1}{\tau_n}\right)^2 + \omega^2}. \quad (100)$$

Using data from single-cell FRET experiments described in the following section, we constrained the parameters of the response function $K(\omega)$ and the noise spectrum $N(\omega)$: $G$, $\tau_1$, $\tau_2$, $\tau_n$, and $D_n$. The parameter inference procedure is described in the sections **Estimating linear response function parameters** and **Estimating noise statistics parameters**.

Error bars for the parameterized models above that are plotted in Fig. 2 of the main text come from propagating errors in parameter estimation. To compute how errors in parameter estimation propagate to errors in a function $f(x, \boldsymbol{\theta})$ with argument $x$ (such as time $t$ or radial frequency $\omega$) and parameters $\boldsymbol{\theta}$, we used:

$$\sigma_f^2(x) = \sum_i \left|\frac{\partial f(x)}{\partial \boldsymbol{\theta}_i}\right|^2 \sigma_{\boldsymbol{\theta}_i}^2, \tag{101}$$

where $\sigma_f^2(x)$ is the variance in the function $f$ at argument value $x$, $\boldsymbol{\theta}_i$ is the $i$th parameter, and $\sigma_{\boldsymbol{\theta}_i}^2$ is the variance of the $i$th parameter (standard error squared).

At this point, we have models for all of the expressions needed to calculate the transfer entropy rate from signal to kinase output. For the models above, the integral in Eqn. (86) can be solved analytically, giving the following expression for the information rate:

$$\dot{I}_{s \to a} \sim \frac{1}{\log(2)} \frac{1}{4} \left(\frac{g}{\tau_1} G\right)^2 \frac{a_v}{D_n} \frac{\frac{\tau_2}{\tau_n^2}\left(\frac{1}{\tau_1} + \frac{2}{\tau_2} + \lambda_{tot}\right) + \lambda_{tot}\left(\frac{1}{\tau_1} + \frac{1}{\tau_2}\right)}{\left(\frac{1}{\tau_1} + \frac{1}{\tau_2}\right)\left(\frac{1}{\tau_1} + \frac{2}{\tau_2}\right)\left(\lambda_{tot} + \frac{1}{\tau_2}\right)\left(\frac{1}{\tau_1} + \frac{1}{\tau_2} + \lambda_{tot}\right)} = \beta\, g^2, \tag{102}$$

where $\lambda_{tot} = (1 - \alpha)\lambda_{R0} + 2\,D_r$ and $a_v = \frac{v_0^2}{3} P_{run}$. The factor of $1/\log(2)$ makes the units bits/s. We used the above expression to compute the information rate in the main text. By inferring model parameters from data, we inferred the prefactor, $\beta$.

Error bars for the information rate in Fig. 3 come from propagating uncertainties in the parameters to uncertainty in the information rate using Eqn. (102) above, with $f(x)$ being $\dot{I}_{s \to a}(g)$ in this case.

We emphasize that this is the total information transferred from the trajectory of signal to kinase output. Not all of this information is relevant to gradient climbing—only the part that is informative of the current signal contributes to gradient climbing.

Overview of E-FRET analysis method
The goal of our FRET analysis is to quantify the degree of interaction between the phosphorylated response regulator CheY-mRFP and its phosphatase CheZ-mYFP to infer the output of the histidine kinase CheA that phosphorylates CheY [39,40] (see below). For this purpose, we used a 3-filter cube FRET imaging method, E-FRET [41], rather than the FRET method previously used in bacterial chemotaxis studies [16,24,39,40]. The two methods aim to infer the same quantity as a measure of the degree of molecular interaction (see below), but with a different set of observables and assumptions. The key differences are: (i) E-FRET provides a principled photobleaching correction method, enabling us to estimate the degree of molecular interaction quantitatively even in presence of photobleaching, while the previous FRET method gives a biased estimate of the degree of molecular interaction if photobleaching is nonnegligible. (ii) E-FRET is a 3-cube FRET imaging method (i.e., observes three different fluorescent signals), while the previous FRET is a 2-cube method. This makes E-FRET require fewer assumptions. Crucially, unlike the previous method, E-FRET does not require measuring fluorescent signals in the absence of FRET interactions, which is generally hard to measure precisely.

To introduce some notation used below, we consider a general bimolecular FRET system with the donor fluorophore D and acceptor fluorophore A fused to two target molecules X and Y respectively. In the presence of photobleaching, the system contains eight chemical species: $D^*, D, A^*, A, D^*A^*, D^*A, DA^*$, and $DA$, where fluorescent and non-fluorescent (i.e., photobleached) molecules are represented with and without the star respectively, and free and complexed molecules are also distinguished. We denote background-subtracted fluorescence signals from single cells by $I_{DD}$, $I_{DA}$, and $I_{AA}$, which are respectively the signals obtained through the donor channel (donor excitation and donor emission), the FRET channel (donor excitation and acceptor emission), and the acceptor channel (acceptor excitation and acceptor emission). As in a typical 3-cube imaging setup, we selected filter sets (see Methods) such that the donor is not excited by the acceptor excitation wavelengths and the acceptor emission is not transmitted through the donor emission filter. Under these conditions, the three observables are linked to the concentrations of the chemical species as follows [41]:

$$I_{DD} = C_{DD}\left([D^*] + [D^*A] + (1 - E_{max})[D^*A^*]\right) + \xi_{DD},$$

$$I_{AA} = C_{AA}\left([A^*] + [D^*A^*] + [DA^*]\right) + \xi_{AA}, \quad (103)$$

$$I_{DA} = d_E \overline{I_{DD}} + a_E \overline{I_{AA}} + C_{DD} G_E E_{max} [D^*A^*] + \xi_{DA},$$

where $\xi_{DD}$, $\xi_{AA}$, and $\xi_{DA}$ are zero-mean shot noise in each channel, and $\overline{I_{DD}}$ and $\overline{I_{AA}}$ are, respectively, the noise-free $I_{DD}$ and $I_{AA}$ signals. $E_{max}$ is the maximum FRET efficiency, where FRET efficiency is defined as the probability of energy transfer from the donor to the acceptor per donor excitation event [41,42], and the maximum FRET efficiency is achieved when only the chemical species $D^*A^*$ is present. The first and second term on the right-hand side of the last equation, respectively, represent the bleedthrough of the donor emission into the acceptor emission filter and the cross-excitation of the acceptor by the donor excitation wavelengths. $C_{DD}$, $C_{AA}$, $a_E$, $d_E$, and $G_E$ are parameters dependent on the imaging system and the photophysical properties of the FRET pair, which are defined as:

$$C_{DD} \equiv \nu_D \epsilon_{DD} Q_D L_D S_D t_{DD},$$

$$C_{AA} \equiv \nu_A \epsilon_{AA} Q_A L_A S_A t_{AA},$$

$$a_E \equiv \frac{\nu_D \epsilon_{DA} t_{DA}}{\nu_A \epsilon_{AA} t_{AA}}, \quad (104)$$

$$d_E \equiv \frac{L_A S_A t_{DA}}{L_D S_D t_{DD}},$$

$$G_E \equiv \frac{Q_A L_A S_A t_{DA}}{Q_D L_D S_D t_{DD}},$$

where $\nu_D$ ($\nu_A$) is the intensity of illumination reaching the sample through the donor (acceptor) excitation filter, $\epsilon_{DD}$ is the absorption coefficient of the donor, $\epsilon_{DA}$ ($\epsilon_{AA}$) is the absorption coefficient of the acceptor at the donor-excitation (acceptor-excitation) wavelength, $Q_D$ ($Q_A$) is the quantum yield of donor (acceptor), $L_D$ ($L_A$) is the throughput of the donor (acceptor) emission light-path, $S_D$ ($S_A$) is the quantum sensitivity of the camera for donor (acceptor) emission, and $t_{DA}$, $t_{AA}$, and $t_{DD}$ are, respectively, the exposure time for the FRET, acceptor, and donor channels.

As a measure of the degree of molecular interaction, E-FRET, as well as the above-mentioned previous FRET method, aims to obtain the following quantity [41]:

$$\mathcal{E} = \frac{[XY]}{[X_{total}]} E_{max}, \quad (105)$$

where $[X_{total}]$ is the total concentration of the carrier molecule to which the donor is attached (i.e., $[X_{total}] = [X] + [XY]$). In the absence of photobleaching, $\mathcal{E}$ is the FRET efficiency. E-FRET maintains that, under some assumptions (see below), the quantity $\mathcal{E}$ can be estimated by computing the following FRET index from the observables:

$$E_{corr}(t|a_E, d_E, G_E) = \frac{F_c(t|a_E, d_E)}{F_c(t|a_E, d_E) + G_E\, I_{DD}(t)} \frac{I_{AA}(0)}{I_{AA}(t)}, \quad (106)$$

where the sensitized emission $F_c$ is

$$F_c(t|a_E, d_E) = I_{DA}(t) - d_E\, I_{DD}(t) - a_E I_{AA}(t). \quad (107)$$

In practice, the value of $I_{AA}(t)$, which is independent of FRET and thus changes slowly due to photobleaching, is interpolated from sparsely-sampled data points over time, while the values of $I_{DD}(t)$ and $I_{DA}(t)$ are directly measured more frequently in time [41]. In the limit that the system-dependent parameters $a_E$, $d_E$, and $G_E$ can be determined with infinite precision, and zero measurement noise of the observables, one can show that $E_{corr}$ converges to $\mathcal{E}$ (see below for proof):

$$E_{corr}(t|a_E, d_E, G_E) \to \mathcal{E} \equiv \frac{[XY]}{[X_{total}]} E_{max}. \quad (108)$$

Realistically though, $a_E$, $d_E$, and $G_E$ can only be estimated with finite uncertainty. In what follows, we will show that errors in estimating these parameters creates a bias in $E_{corr}$ that grows quasi-exponentially over the course of a time-lapse FRET experiment. We also discuss some predictable properties of the bias, and how one can exploit these properties to correct the bias.

Derivation of the E-FRET formulae
Based on the original paper [41], here we re-derive the E-FRET formula (Eq. (106)) to make the assumptions and their validation in our system clearer. We first assume that the total concentrations of the donor and acceptor molecules inside the cell are conserved during the experiment, which in our case lasts ≲ 40 minutes. The assumption is satisfied in our experiment because RP437 *E. coli* and their derivatives are auxotrophic for several amino acids, which we do not provide in the experimental media. Therefore, new proteins cannot be synthesized after we wash the cells (see Methods). Furthermore, the interval (~2 hrs) between cell washing and the beginning of the FRET experiment ensures that essentially all fluorescent proteins have matured [43]. Therefore, we have:

$$[D_{total}] = [D^*](t) + [D](t) + [D^*A^*](t) + [D^*A](t) + [DA^*](t) + [DA](t)$$
$$[A_{total}] = [A^*](t) + [A](t) + [D^*A^*](t) + [D^*A](t) + [DA^*](t) + [DA](t).$$
(109)

Again, $D$ is the donor molecule (CheZ-mYFP), $A$ is the acceptor molecule (CheY-mRFP), and species with a star are not yet photobleached. We also assume that all of the target molecules are labeled by the fluorescent proteins:
$$[D_{total}] = [X_{total}]$$
$$[A_{total}] = [Y_{total}].$$
(110)

Next, we assume that the photobleaching rate is a first-order decay process. Combined with the assumption that there's no synthesis of new fluorescent proteins, this leads to:

$$\frac{d([D^*] + [D^*A^*] + [D^*A])}{dt} = -\delta(t)([D^*] + [D^*A^*] + [D^*A])$$
$$\frac{d([A^*] + [D^*A^*] + [DA^*])}{dt} = -\alpha(t)([A^*] + [D^*A^*] + [DA^*]),$$
(111)

where $\delta(t)$ and $\alpha(t)$ are the bleaching rates of the donor and acceptor at time $t$, which can depend on the degree of FRET and hence can depend on $t$. Solving these equations, we get:

$$[D^*] + [D^*A^*] + [D^*A] = [X_{total}] e^{-\int_0^t \delta(t')dt'}$$
$$[A^*] + [D^*A^*] + [DA^*] = [Y_{total}] e^{-\int_0^t \alpha(t')dt'}.$$
(112)

Lastly, we assume that the system is in a quasi-steady state at each time point, i.e. the kinetics of the binding and unbinding of the target molecules X and Y and the diffusion timescales of the donor and acceptor over the enclosed compartment are sufficiently short compared to the time scale of photobleaching. In our system, the binding-unbinding kinetics between CheY-p and CheZ ($< 0.3$ s [44]) and the time it takes CheY to diffuse throughout the cytoplasm (~0.05 s [45]) are much shorter than the time scale of photobleaching ($\gtrsim 10^2$ s). With this assumption, the fraction of *each free or complexed species* that is not photobleached decays exponentially, analogous to Eqns. (112). From this, we get:

$$\frac{[D^*]}{[D^*] + [D]} = e^{-\int_0^t \delta(t')dt'},$$
$$\frac{[A^*]}{[A^*] + [A]} = e^{-\int_0^t \alpha(t')dt'},$$
(113)

Furthermore, defining $\gamma$ as the binding affinity constant between X and Y:
$$[D^*A^*] = \gamma [D^*][A^*] \quad (114)$$

$$= \gamma \left(([D^*] + [D])e^{-\int_0^t \delta(t')dt'}\right)\left(([A^*] + [A])e^{-\int_0^t \alpha(t')dt'}\right) \tag{115}$$

$$= \gamma \left([X]e^{-\int_0^t \delta(t')dt'}\right)\left([Y]e^{-\int_0^t \alpha(t')dt'}\right) \tag{116}$$

$$= [XY]\, e^{-\int_0^t \alpha(t')+\delta(t')dt'}. \tag{117}$$

Under these assumptions, Eqns. (103) become

$$\begin{aligned}
I_{DD} &\simeq C_{DD}\left([X_{total}]\, e^{-\int_0^t \delta(t')dt'} - E_{max}[XY]\, e^{-\int_0^t \alpha(t')+\delta(t')dt'}\right) \\
I_{AA} &\simeq C_{AA}[Y_{total}]\, e^{-\int_0^t \alpha(t')dt'} \\
I_{DA} &\simeq d_E\, I_{DD} + a_E\, I_{AA} + C_{DD}\, G_E\, E_{max}[XY]\, e^{-\int_0^t \alpha(t')+\delta(t')dt'},
\end{aligned} \tag{118}$$

and Eqn. (107) becomes

$$F_c(t) = C_{DD}\, G_E\, E_{max}[XY]\, e^{-\int_0^t \alpha(t')+\delta(t')dt'}. \tag{119}$$

By plugging these expressions in the formula for $E_{corr}$ in Eqn. (106), one gets

$$E_{corr}(t) = \frac{F_c(t)}{F_c(t) + G_E I_{DD}(t)} \frac{I_{AA}(0)}{I_{AA}(t)} \tag{120}$$

$$\simeq \frac{C_{DD}G_E E_{max}[XY]e^{-\int_0^t \alpha(t')+\delta(t')dt'}}{C_{DD}G_E E_{max}[XY]e^{-\int_0^t \alpha(t')+\delta(t')dt'} + C_{DD}G_E\left([X_{total}]\, e^{-\int_0^t \delta(t')dt'} - E_{max}[XY]e^{-\int_0^t \alpha(t')+\delta(t')dt'}\right)} \frac{1}{e^{-\int_0^t \alpha(t')dt'}} \tag{121}$$

$$= \frac{E_{max}[XY]}{[X_{total}]} = \mathcal{E}, \tag{122}$$

which is the degree of molecular interaction defined above.

Measurements of imaging system parameters

The imaging-system parameters $a_E$, $d_E$ and $G_E$ were determined in the following way. The cross-talk coefficients $a_E$ and $d_E$ can be estimated by observing the fluorescent signals from strains that express only the acceptor or the donor because [41]

$$\begin{aligned}
a_E &\equiv \frac{\nu_D \epsilon_{DA} t_{DA}}{\nu_A \epsilon_{AA} t_{AA}} \simeq \frac{I_{DA(A)}}{I_{AA(A)}}, \\
d_E &\equiv \frac{L_A S_A t_{DA}}{L_D S_D t_{DD}} \simeq \frac{I_{DA(D)}}{I_{DD(D)}},
\end{aligned} \tag{123}$$

where $A$ ($D$) in the parentheses in the lower index indicates that the corresponding fluorescent signals are obtained from the strain that only expresses the acceptor (the donor). The approximations above are equalities in the limit of zero measurement noise. The equations can be shown by noting that

$$\begin{aligned}
I_{DA(A)} &= [A^*]\nu_D \epsilon_{DA} Q_A L_A S_A t_{DA} + \xi_{DA(A)} \\
I_{AA(A)} &= [A^*]\nu_A \epsilon_{AA} Q_A L_A S_A t_{AA} + \xi_{AA(A)} \\
I_{DA(D)} &= [D^*]\nu_D \epsilon_{DD} Q_D L_A S_A t_{DA} + \xi_{DA(D)} \\
I_{DD(D)} &= [D^*]\nu_D \epsilon_{DD} Q_D L_D S_D t_{DD} + \xi_{DD(D)},
\end{aligned} \quad (124)$$

where $\xi_{DA(A)}$, $\xi_{AA(A)}$, $\xi_{DA(D)}$ and $\xi_{DD(D)}$ represent shot noise. We obtained estimates for the parameters $a_{E,est}$ and $d_{E,est}$ by linear least-squares fitting the background-subtracted fluorescence signals from hundreds of cells (Fig. S4AB):

$$\begin{aligned}
a_{E,est} &= \arg\min_{a_E} \sum_i \left(a_E I_{AA(A),i} - I_{DA(A),i}\right)^2, \\
d_{E,est} &= \arg\min_{d_E} \sum_i \left(d_E I_{DD(D),i} - I_{DA(D),i}\right)^2,
\end{aligned} \quad (125)$$

where subscript $i$ indicates different cells. The values we obtained were $a_{E,est} = 0.3369$ ($\pm 0.0006$), and $d_{E,est} = 0.0891$ ($\pm 0.0001$).

The parameter $G_E$ quantifies the change in sensitized emission $F_c$ per change in $I_{DD}$ due to FRET, $G_E = \left|\frac{dF_c}{dI_{DD}}\right|$, which in principle can be measured by using a FRET strain expressing both donor and acceptor as [41]

$$G_E \equiv \frac{Q_A L_A S_A t_{DA}}{Q_D L_D S_D t_{DD}} \simeq \frac{F_c'}{I_{DD}^{post} - I_{DD}'}, \quad (126)$$

where $I_{DD}^{post}$ is the intensity of donor fluorescence in the condition where the acceptor is completely photobleached while the donor remains intact, and $F_c'$ and $I_{DD}'$ correspond to $F_c$ and $I_{DD}$, respectively, in the absence of photobleaching. Again, the equality is exact in the limit of zero measurement noise. The relation can be shown by noting

$$\begin{aligned}
F_c' &= [XY]\nu_D \epsilon_{DD} E_{max} Q_A L_A S_A t_{DA} + \xi_{F_c} \\
I_{DD}^{post} - I_{DD}' &= [XY]\nu_D \epsilon_{DD} E_{max} Q_D L_D S_D t_{DD} + \xi_{DD}',
\end{aligned} \quad (127)$$

where $\xi_{F_c}$ and $\xi_{DD}'$ represent the effects of shot noise. To avoid issues associated with acceptor photobleaching [42], we can take advantage of the fact that, in our system and microfluidic device, FRET changes can be induced rapidly by applying a step change of chemoattractant stimulus. Doing so lets us measure changes in $F_c$ and $I_{DD}$ before substantial photobleaching occurs, and we can estimate $G$ from:

$$G_E \simeq \frac{|\Delta F_c|}{|\Delta I_{DD}|}. \quad (128)$$

We estimated the value of $G_E$ by least-squares fitting the fluorescence signals from multiple cells (Fig. S4C), i.e.,

$$G_{E,est} = \arg\min_{G_E} \sum_i \left(G_E |\Delta I_{DD,i}| - |\Delta F_{c,i}|\right)^2, \tag{129}$$

where $|\Delta I_{DD,i}|$ and $|\Delta F_{c,i}|$ for the $i$-th cell were obtained by applying a saturating stimulus (1 mM MeAsp and 1 µM serine) on top of a background stimulus 100 µM MeAsp, and then removing the stimulus and the background (i.e., 0 M chemoattractants) in the microfluidic chamber because it induces the maximal FRET change [24,39]. The obtained value was $G_{E,est} = 0.3497\ (\pm 0.0018)$.

Effects of parameter-estimation error on the FRET signal

The FRET index $E_{corr}$ computed from observables (Eqn. (106); see Fig. S4D) provides an unbiased estimator of $\mathcal{E}$ (Eqn. (105)) in the presence of photobleaching, given the true values of the system-dependent parameters $a_E, d_E$ and $G_E$. However, the parameters are always estimated with finite precision. Furthermore, although it is often assumed that the values are invariant given a system, they may not be constant over the course of measurements. It has been reported that some fluorescent proteins change their fluorescent properties upon photobleaching [42]. Zal and Gascoigne only explored how errors in the parameter estimates bias $E_{corr}$ in the absence of photobleaching [41]. Here we study the effect of the parameter-estimation error in the presence of photobleaching and propose a method to correct for the effects.

The estimated values of the parameters can be written as
$$\begin{aligned} a_{E,est} &= a_E + \Delta a_E, \\ d_{E,est} &= d_E + \Delta d_E, \\ G_{E,est} &= G_E + \Delta G_E, \end{aligned} \tag{130}$$
where true values of the parameters are denoted by $a_E$, $d_E$, and $G_E$ and the deviations from them by $\Delta a_E, \Delta d_E$, and $\Delta G_E$. First, we note that $E_{corr}$ can be approximated as

$$E_{corr} = \frac{\frac{F_c}{I_{DD}}}{\frac{F_c}{I_{DD}} + G_E} \frac{I_{AA}(0)}{I_{AA}} \tag{131}$$

$$\approx \frac{I_{AA}(0)}{G_E} \frac{F_c}{I_{DD} I_{AA}} \tag{132}$$

$$= \frac{I_{AA}(0)}{G_E} \frac{I_{DA} - d_E I_{DD} - a_E I_{AA}}{I_{DD} I_{AA}}. \tag{133}$$

In the second line, we used $\frac{F_c}{I_{DD}} \ll G_E$ to simplify the following calculation, but it is not essential. This assumption is valid in a typical low FRET-efficiency experiment where the value of $F_c$ is sufficiently lower than $I_{DD}$ $\left(\frac{F_c}{I_{DD}} \ll 1\right)$ and yet, to be able to detect FRET signals, the parameter $G_E$ needs to be $\sim \mathcal{O}(1)$ [41]. In our setup, $\frac{F_c}{I_{DD}} \lesssim 0.05$ and $G_E \simeq 0.35$.

The error in $E_{corr}$ due to the error in the estimated parameters $\Delta a_E$, $\Delta d_E$, and $\Delta G_E$ can be written as

$$\Delta E_{corr} = E_{corr}(a_E + \Delta a_E, d_E + \Delta d_E, G_E + \Delta G_E) - E_{corr}(a_E, d_E, G_E) \quad (134)$$

$$\simeq \frac{\partial E_{corr}(a_E, d_E, G_E)}{\partial a_E}\Delta a_E + \frac{\partial E_{corr}(a_E, d_E, G_E)}{\partial d_E}\Delta d_E + \frac{\partial E_{corr}(a_E, d_E, G_E)}{\partial G_E}\Delta G_E \quad (135)$$

$$\simeq -\frac{I_{AA}(0)}{G_E I_{DD}}\Delta a_E - \frac{I_{AA}(0)}{G_E I_{AA}}\Delta d_E - \frac{E_{corr}}{G_E}\Delta G_E. \quad (136)$$

Thus, the fraction of error in $E_{corr}$ can be written as

$$\frac{\Delta E_{corr}}{E_{corr}} = -\frac{I_{AA}\Delta a_E}{F_c} - \frac{I_{DD}\Delta d_E}{F_c} - \frac{\Delta G_E}{G_E}. \quad (137)$$

From the assumptions about the molecular interactions and photobleaching that underly the E-FRET method, the observables $I_{DD}$ and $I_{AA}$ and the sensitized emission $F_c$ can be written as

$$\frac{\Delta E_{corr}}{E_{corr}} = -\frac{I_{AA}\Delta a_E}{F_c} - \frac{I_{DD}\Delta d_E}{F_c} - \frac{\Delta G_E}{G_E}. \quad (138)$$

$$I_{DD} \simeq C_{DD}\left([X_{total}]\, e^{-\int_0^t \delta(t')dt'} \right. \quad (139)$$

$$\left. - E_{max}[XY]\, e^{-\int_0^t \alpha(t')+\delta(t')dt'}\right) \sim C_{DD}[X_{total}]\, e^{-\int_0^t \delta(t')dt'}$$

$$I_{AA} \simeq C_{AA}[Y_{total}]\, e^{-\int_0^t \alpha(t')dt'} \quad (140)$$

$$F_c(t) \simeq C_{DD}\, G_E\, E_{max}[XY]\, e^{-\int_0^t \alpha(t')+\delta(t')dt'}, \quad (141)$$

where $\delta(t) > 0$ and $\alpha(t) > 0$ are, respectively, the (time-dependent) rates of photobleaching of the donor and acceptor, and the final approximation for $I_{DD}$ is valid under the assumption $\frac{F_c}{I_{DD}} \ll G_E$.

Using these expressions, we get

$$\frac{\Delta E_{corr}}{E_{corr}} \sim A\, e^{\int_0^t \delta(t')dt'}\Delta a_E + D\, e^{\int_0^t \alpha(t')dt'}\Delta d_E - \frac{\Delta G_E}{G_E}, \quad (142)$$

where $A = \frac{C_{AA}[Y_{total}]}{C_{DD}\, G_E\, E_{max}[XY]} > 0$ and $D = \frac{[X_{total}]}{G_E E_{max}[XY]} > 0$. The first and the second terms grow quasi-exponentially as the fluorescent proteins photobleach; thus, the measured value of $E_{corr}$, at baseline levels of molecular interaction, changes over time. Note that the time scale of this change is governed by the time scale of photobleaching. We discuss how we corrected the baseline of $E_{corr}$ in the next section below.

The remaining question is how uncertainty in the parameters $a_E$, $d_E$, and $G_E$, in the presence of photobleaching, affects the mapping between *changes* in molecular interactions and the corresponding *change* in $E_{corr}$. If this mapping is sensitive to photobleaching, it severely limits the reliability of the $E_{corr}$ measurements because the same changes in molecular interactions would lead to different changes in $E_{corr}$ at different times in the experiment. To address this, we analyze the sensitivity of $E_{corr}$ to the change in the degree of molecular interaction and its dependence on photobleaching.

The degree of molecular interaction is dictated by the time-dependent binding affinity $\gamma(t)$ between the two target molecules X and Y. Therefore, the sensitivity of $E_{corr}$ to changes in $\gamma$ at a given time can be quantified by $\frac{\partial E_{corr}(\gamma|a_E,d_E,G_E)}{\partial \gamma}$. With errors in the parameters, this quantity can be written as

$$\frac{\partial E_{corr}(\gamma|a_E + \Delta a_E, d_E + \Delta d_E, G_E + \Delta G_E)}{\partial \gamma} = \frac{\partial E_{corr}(\gamma|a_E, d_E, G_E)}{\partial \gamma} + \frac{\partial \Delta E_{corr}(\gamma)}{\partial \gamma} \quad (143)$$

$$= \frac{\partial E_{corr}(\gamma|a_E, d_E, G_E)}{\partial \gamma}\left(1 + \frac{\frac{\partial \Delta E_{corr}(\gamma)}{\partial \gamma}}{\frac{\partial E_{corr}(\gamma|a_E, d_E, G_E)}{\partial \gamma}}\right) \quad (144)$$

$$\equiv \frac{\partial E_{corr}(\gamma|a_E, d_E, G_E)}{\partial \gamma}(1 + \Delta).$$

Thus, $\Delta \equiv \frac{\frac{\partial \Delta E_{corr}}{\partial \gamma}}{\frac{\partial E_{corr}(\gamma|a_E, d_E, G_E)}{\partial \gamma}}$ characterizes the bias error, and the question is how this quantity behaves with photobleaching. To compute this, we note

$$\Delta = \frac{\frac{\partial}{\partial \gamma}\left(-\frac{I_{AA}(0)}{G_E I_{DD}(\gamma)}\Delta a_E - \frac{I_{AA}(0)}{G_E I_{AA}}\Delta d_E - \frac{E_{corr}(\gamma)}{G_E}\Delta G_E\right)}{\frac{\partial E_{corr}(\gamma)}{\partial \gamma}} \quad (145)$$

$$= \frac{-\frac{I_{AA}(0)\Delta a_E}{G_E}\frac{\partial}{\partial \gamma}\left(\frac{1}{I_{DD}(\gamma)}\right) - \frac{\Delta G_E}{G_E}\frac{\partial E_{corr}}{\partial \gamma}}{\frac{\partial E_{corr}}{\partial \gamma}} \quad (146)$$

$$= \frac{-\frac{I_{AA}(0)\Delta a_E}{G_E}\frac{\partial}{\partial \gamma}\left(\frac{1}{I_{DD}(\gamma)}\right)}{\frac{\partial E_{corr}}{\partial \gamma}} - \frac{\Delta G_E}{G_E}, \quad (147)$$

where we used the fact that $I_{AA}$ is independent of $\gamma$, i.e., $\frac{\partial I_{AA}}{\partial \gamma} = 0$. We note

$$\frac{\partial E_{corr}}{\partial \gamma} \simeq \frac{\partial}{\partial \gamma}\left(\frac{1}{G_E}\frac{F_c(\gamma)}{I_{DD}(\gamma)}\frac{I_{AA}(0)}{I_{AA}}\right) \tag{148}$$

$$= \frac{I_{AA}(0)}{G_E I_{AA}}\frac{\partial}{\partial \gamma}\left(\frac{F_c}{I_{DD}}\right) \tag{149}$$

$$= \frac{I_{AA}(0)}{G_E I_{AA}}\left(\frac{1}{I_{DD}}\frac{\partial F_c}{\partial \gamma} - \frac{F_c}{I_{DD}^2}\frac{\partial I_{DD}}{\partial \gamma}\right) \tag{150}$$

$$= -\frac{I_{AA}(0)}{G_E I_{AA}}\frac{1}{I_{DD}}\frac{\partial I_{DD}}{\partial \gamma}\left(-\frac{\frac{\partial F_c}{\partial \gamma}}{\frac{\partial I_{DD}}{\partial \gamma}} + \frac{F_c}{I_{DD}}\right) \tag{151}$$

$$= -\frac{I_{AA}(0)}{G_E I_{AA}}\frac{1}{I_{DD}}\frac{\partial I_{DD}}{\partial \gamma}\left(G_E + \frac{F_c}{I_{DD}}\right), \tag{152}$$

where at the final step we used $-\frac{\partial F_c}{\partial \gamma}/\frac{\partial I_{DD}}{\partial \gamma} = |\Delta F_c|/|\Delta I_{DD}| = G_E$. By plugging this to the expression for $\Delta$, we get

$$\Delta = \frac{\frac{I_{AA}(0)\Delta a_E}{G_E}\frac{1}{I_{DD}^2}\frac{\partial I_{DD}}{\partial \gamma}}{-\frac{I_{AA}(0)}{G_E I_{AA}}\frac{1}{I_{DD}}\frac{\partial I_{DD}}{\partial \gamma}\left(G_E + \frac{F_c}{I_{DD}}\right)} - \frac{\Delta G_E}{G_E} \tag{153}$$

$$= \frac{-I_{AA}\Delta a_E}{I_{DD}\left(G_E + \frac{F_c}{I_{DD}}\right)} - \frac{\Delta G_E}{G_E} \tag{154}$$

$$\simeq -\frac{I_{AA}}{I_{DD}}\frac{\Delta a_E}{G_E} - \frac{\Delta G_E}{G_E} \tag{155}$$

$$= H\, e^{\int_0^t \delta(t')-\alpha(t')dt'}\Delta a_E - \frac{\Delta G_E}{G_E}, \tag{156}$$

where $H = \frac{C_{AA}[Y_{total}]}{C_{DD}[X_{total}]G_E} > 0$ and in the third line we used $\frac{F_c}{I_{DD}} \ll G_E$. This expression tells us that the relative error in the mapping from molecular interaction to $E_{corr}$, $\Delta$, is small if $\Delta a_E$ and $\Delta G_E$ are small. Furthermore, this relative error grows slower than the relative error in the baseline of $E_{corr}$, $\frac{\Delta E_{corr}}{E_{corr}}$, because only the difference of the donor and acceptor photobleaching rates appears in the exponential. Additionally, the coefficient $H$ is typically smaller than the coefficients in $\frac{\Delta E_{corr}}{E_{corr}}$, $A$ and $D$. In fact, assuming $C_{AA} \approx C_{DD}$ and $[Y_{total}] \approx [X_{total}]$, one can show that both $H/A$ and $H/D$ are bounded by $\frac{E_{max}[XY]}{[X_{total}]} < 1$.

In summary, the estimator of the degree of molecular interaction $E_{corr}$ computed from the observables has the following properties due to uncertainties in estimating the system parameters: (i) the baseline level of $E_{corr}$ changes on the timescale of fluorescence photobleaching, even if there is no change in the level of molecular interactions; and (ii) biologically-induced changes in molecular interactions are relatively well-preserved in changes in $E_{corr}$. This means that although the baseline level of $E_{corr}$ changes over time, deviations from that baseline faithfully represent changes in molecular interactions.

We tested whether these properties are observed in actual FRET data by investigating both the absolute levels of and changes in $E_{corr}$ before and after photobleaching (Fig. S4D-G). Under the assumption that each cell retains approximately identical FRET-response properties over the course of an experiment, our analyses predict that the baseline level of $E_{corr}$ monotonically changes over time, while the changes in $E_{corr}$ upon stimuli remain approximately invariant over the course of a measurement. Consistent with these predictions, we observed: (i) the absolute level of $E_{corr}$ slowly increased over time (Fig. S4E) and the values after photobleaching were higher than those before photobleaching (Fig. S4F), although the degree of the change was moderate (Fig. S4F; roughly 15% increase after >1200 frames of image acquisition); (ii) the changes in $E_{corr}$ induced by fast-switching, identical stimuli showed essentially no bias after photobleaching (Fig. S4G). Thus, the behavior of our FRET data due to photobleaching can be consistently explained by the properties of $E_{corr}$.

Correcting baseline FRET signal

The analyses above have established that the baseline level of $E_{corr}$ monotonically changes on the timescale of photobleaching because of parameter uncertainties, but changes in $E_{corr}$ due to fast biological FRET interactions are more reliable. This suggests that the slow artifact in the baseline level $E_{corr}$ can be corrected for by estimating and subtracting the slowly-varying trend. We obtained a corrected signal $\overline{E_{corr}}$ in the following way. First, since the drift in baseline $E_{corr}$ is slow compared to the durations of our experiments (Fig. S4EF), we assumed that the error in $E_{corr}$ changes approximately linearly with frame number. The slope of the linear function was estimated by measuring the minimum values of $E_{corr}$ at the beginning (frame number $i_1$) and the end (frame number $i_2$) of each measurement by applying a saturating stimulus, and by fitting a linear function to $\{(i_1, E_{corr}(i_1)), (i_2, E_{corr}(i_2))\}$. That is, the corrected $E_{corr}$ is written as

$$\overline{E_{corr}}(i) = E_{corr}(i) - \frac{E_{corr}(i_2) - E_{corr}(i_1)}{i_2 - i_1} i, \qquad (157)$$

where $E_{corr}(i_1)$ and $E_{corr}(i_2)$ were estimated as medians of 18 consecutive frames to mitigate the effect of measurement noise. With this correction, we observed population-average FRET signals that show essentially no trends, only stimulus-induced changes (see, e.g., Fig. S5A and S6A), validating the assumption of the linearity of the error-induced trend.

Converting FRET signal to kinase activity

Using the corrected FRET signal $\overline{E_{corr}}$, we defined the kinase activity as

$$a(t) = \frac{\overline{E_{corr}}(t) - \overline{E_{corr}^{min}}}{\overline{E_{corr}^{max}} - \overline{E_{corr}^{min}}}, \qquad (158)$$

where $\overline{E_{corr}^{min}}$ and $\overline{E_{corr}^{max}}$ are respectively the minimum and maximum values of $\overline{E_{corr}}$ obtained from responses to a saturating stimulus and following removal of background at the beginning of each measurement (see Methods). Note that the mutual information rate between chemoattractant signal $s(t)$ and the kinase output of the chemotaxis signaling pathway, which is the ultimate goal of our FRET analysis, is invariant to this linear conversion—we could have computed it from, e.g., $\overline{E_{corr}}(t)$. However, we computed $a(t)$ to facilitate comparison with preceding works where the same normalized measure was used [13,24]. The distributions of the steady-state values of $a$, $a_0$, across cells in isogenic populations were evaluated under two different stimulus and illumination conditions, giving: $a_0 = 0.29 \pm 0.07$ (mean and standard deviation, estimated from the experiments for response-function extraction; Fig. S5C), and $a_0 = 0.30 \pm 0.08$ (mean and standard deviation, estimated from the experiments for signaling-noise characterization; Fig. S6C). These values are comparable to previously reported values: $a_0 \approx 1/3$ from population-averaged FRET measurements [13] and $a_0 = 0.30 \pm 0.07$ (mean and standard deviation) from single-cell FRET measurements [24].

Biochemical interpretations of the kinase activity $a(t)$ are the following. First, from Eqn. (105), we write [39,40]

$$\overline{E_{corr}}(t) \approx \frac{E_{max}\,[\text{CheYp}\cdot\text{CheZ}]}{[\text{CheZ}_{total}]}, \tag{159}$$

where the equality holds in the limit of zero measurement noise. Using this expression, $a$ can be written as

$$a(t) \approx \frac{[\text{CheYp}\cdot\text{CheZ}](t) - [\text{CheYp}\cdot\text{CheZ}]^{min}}{[\text{CheYp}\cdot\text{CheZ}]^{max} - [\text{CheYp}\cdot\text{CheZ}]^{min}}, \tag{160}$$

and therefore $a(t)$ is primarily the relative degree of CheYp-CheZ formation. On time scales longer than the time scale of CheY-p hydrolysis by CheZ (~0.3 s [44]), phosphorylation and dephosphorylation of CheY equilibrate [39,40], and therefore

$$[\text{CheYp}\cdot\text{CheZ}](t) = A(t)\frac{k_A}{k_Z}[\text{CheA}] \approx A(t)\frac{k_A}{k_Z}[\text{CheA}_{total}], \tag{161}$$

where $k_A$ and $k_Z$ are, respectively, the rate constants for autophosphorylation of CheA and for hydrolysis of CheY-p by CheZ, $A(t)$ ($0 < A(t) < 1$) is the fraction of active CheA, and $[\text{CheA}_{total}]$ is the total concentration of CheA. Given the conservation equation $[\text{CheA}_{total}] = [\text{CheA}] + [\text{CheAp}]$, the last step of the above equations holds when $[\text{CheAp}] \ll [\text{CheA}]$. This is achieved when the total concentrations of CheY-mRFP and CheZ-mYFP in the cell are large, which we achieve by overexpressing them in our experimental conditions (see Methods), as done before [24]. Assuming this, $a$ can be written as

$$a(t) \approx \frac{A(t) - A^{min}}{A^{max} - A^{min}}, \tag{162}$$

and therefore $a(t)$ can also be interpreted as the relative autophosphorylation activity of the kinase CheA, which we have referred to as kinase activity.

Estimating linear response function parameters

A kinase activity time series from a single cell (labeled by $k$) consisted of 10 step-up responses $a_{k,+,i}(t)$ ($i = 1,2 \ldots 10$), where the concentration of MeAsp was changed from $c_0 = 100$ μM to $c_+ = 110$ μM, and 10 step-down responses $a_{k,-,i}(t)$, where the concentration was changed from $c_0$ to $c_- = 90$ μM (Fig. 2E, and Fig. S5A). Each response $a_{k,s,i}(t)$ ($s = \{+,-\}$ for shorthand) consisted of a pre-stimulus measurement (7.5 s; 10 time points) and post-stimulus measurement (30 s; 40 time points) of kinase output, followed by a 60 second interval before the next step-response measurement (Fig. 2E, and Fig. S5A; Methods). The kinase response induced by an impulse of stimulus (step change in concentration) was thus defined as

$$\Delta a_{k,s,i}(t) = a_{k,s,i}(t) - \langle a_{k,s,i}(t) \rangle_{prestim}, \tag{163}$$

where $\langle a_{k,s,i}(t) \rangle_{prestep}$ is the time-averaged kinase activity in the moments before the step change of concentration was delivered. The steady-state kinase activity of the $k$-th cell $a_{0,k}$ was estimated by the average of the pre-stimulus kinase activities $\langle a_{k,s,i}(t) \rangle_{prestep}$, i.e.,

$$a_{0,k} = \frac{1}{2 N_{stim}} \sum_{s=\{+,-\}} \sum_{i=1}^{N_{stim}} \langle a_{k,s,i}(t) \rangle_{prestim}, \tag{164}$$

where $N_{stim} = 10$ is the number of stimuli of each sign delivered to the cells. The average and the standard error of the kinase activity responses in each cell were computed as

$$\langle \Delta a_{k,s} \rangle(t) = \frac{1}{N_{stim}} \sum_{i=1}^{N_{stim}} \Delta a_{k,s,i}(t), \quad SE_{k,s}(t)$$

$$= \frac{1}{\sqrt{N_{stim}}} \sqrt{\frac{1}{N_{stim} - 1} \sum_{i=1}^{N_{stim}} \left( \Delta a_{k,s,i}(t) - \langle \Delta a_{k,s} \rangle(t) \right)^2}. \tag{165}$$

To each single cell average response, we fitted the following function with 3 parameters, $G$, $\tau_1$, and $\tau_2$:

$$f_s(t|G_k, \tau_{1,k}, \tau_{2,k}) = -\log\left(\frac{c_s}{c_0}\right) K(t - t_0), \tag{166}$$

where $K(t)$ is the parameterized impulse-response function discussed earlier and $t_0$ denotes the timing at which the stimulus levels were changed. $f_s(t|G_k, \tau_{1,k}, \tau_{2,k})$ is the result of convolving the response function $K(t)$ with a delta function of signal $s(t)$ with amplitude $\log\left(\frac{c_s}{c_0}\right)$ at time $t_0$, or $s(t) = \frac{d}{dt}\log(c) = \log\left(\frac{c_s}{c_0}\right)\delta(t_0)$. The minus sign is needed because positive signals lead to drops in kinase activity. The time of each stimulus $t_0$ was inferred from the data in the following way. For each signal, the first time point at which the population response was more than 3 standard deviations below baseline $a_{0,k}$ was found. Then, $t_0$ for that signal was defined as the time half-way between that time point and the previous one. This was repeated for each stimulus. The same values of $t_0$ were used for all cells.

We fit this function to both the step-up and step-down responses simultaneously, since in the linear-response regime the up and down responses are expected to be symmetric. The measurements of $\langle \Delta a \rangle(t)$ were not smoothed before fitting. The fitting was done using a Bayesian framework [46]. Log-uniform distributions were used as priors for each parameter:

$$P(\tau_{1,k}) = \mathcal{U}(\tau_{1,k}; 10^{-2}, 5)$$

$$P(\tau_{2,k}) = \mathcal{U}(\tau_{2,k}; 10^{-1}, 10^2)$$

$$P(G_k) = \mathcal{U}\left(G_k; 10^{-1}, \frac{a_{0,k}}{\log\left(\frac{c_+}{c_0}\right)}\right),$$

(167)

where $\mathcal{U}(x; a, b) = \frac{1}{x \log b/a}$ for $a \leq x \leq b$, and $\mathcal{U}(x; a, b) = 0$ otherwise. Note the upper bound for $P(G_k)$ comes from the positivity of the kinase activity $a_{k,s,i}(t) > 0$. The log posterior distribution is defined as

$$\begin{aligned}\log P(G_k, \tau_{1,k}, \tau_{2,k}|\mathcal{D}) \\ = -\frac{1}{2}\sum_{n,s}\frac{\left(\langle\Delta a_{k,s}\rangle(t_n) - f_s(t_n|G_k, \tau_{1,k}, \tau_{2,k})\right)^2}{SE^2_{k,s}(t_n)} + \log(P(G_k)) \\ + \log\left(P(\tau_{1,k})\right) + \log\left(P(\tau_{2,k})\right) + C,\end{aligned}$$

(168)

where the first term on the right-hand side is the log-likelihood function and $C$ comes from a normalization constant. A maximum a posteriori probability (MAP) estimate, defined as the mode of the posterior distribution was obtained using a MATLAB optimization function (*fminunc*) (Fig. S5C). As a measure for the parameter-estimation uncertainty, we computed 25 and 75 percentiles of the marginalized posterior distribution of each parameter ($P(G_k|\mathcal{D})$, $P(\tau_{1,k}|\mathcal{D})$, and $P(\tau_{2,k}|\mathcal{D})$) obtained by a Markov chain Monte Carlo algorithm, slice sampling [46].

The population-level representative value of each parameter was defined as the median of the MAP estimates, and the uncertainty of the value was evaluated using the median absolute deviations (MAD) from the population's median parameter values:

$$\text{SE} = \frac{1.4826 \times \text{median}(|X_i - \text{median}(X)|)}{\sqrt{N_{Data}}}, \quad (169)$$

where $X = \{X_1, X_2, ..., X_{N_{Data}}\}$ is the MAP estimates of each parameter. The MAD in the numerator equals the standard deviation of $X$ when it is Gaussian distributed, but is robust to outliers.

Estimating noise statistics parameters
One of the first direct measurements of the fluctuation of the chemotaxis signaling pathway, or signaling noise, showed that the dynamics approximately follow an Ornstein-Uhlenbeck (OU) process. The OU process is specified by two parameters, $\tau_n$ and $D_n$, and obeys the following Langevin equation

$$\frac{dx}{dt} = -\frac{1}{\tau_n}x + \sqrt{2 D_n}\,\xi(t) \quad (170)$$

where $\xi(t)$ is a Gaussian white noise with average zero and a delta correlation in time

$$\langle \xi(t) \rangle = 0, \quad \langle \xi(t)\xi(t') \rangle = \delta(t-t'), \quad (171)$$

and $\delta(t)$ is the Dirac delta function.

Following preceding work [24], we modeled the signaling noise measured by FRET (Fig. S6A) as an OU process, and estimated the process parameters (and their uncertainties) using a Bayesian filtering-based method [47]. To introduce some notation, the measurements were conducted at time points $t_1, t_2, ..., t_T$, and we obtained a series of observables (i.e., FRET signals) $y_{1:T} = \{y_1, y_2, ..., y_T\}$, which are noise-corrupted measurements of the true, hidden state of the system $x_{1:T} = \{x_1, x_2, ..., x_T\}$, which obeys Eqn. (170) with unknown parameters $\boldsymbol{\theta} = \{\tau_n, D_n\}$. Using the framework of a state-space model, the parameter-estimation problem can be written in the form

$$\begin{aligned} \boldsymbol{\theta} &\sim p(\boldsymbol{\theta}) \\ x_k &\sim p(x_k|x_{k-1}, \boldsymbol{\theta}) \\ y_k &\sim p(y_k|x_k) \end{aligned} \quad (172)$$

where $p(\boldsymbol{\theta})$ is the prior distribution of the parameters $\boldsymbol{\theta}$, $p(x_k|x_{k-1}, \boldsymbol{\theta})$ the transition probability distribution specified by the dynamical model, and $p(y_k|x_k)$ is the measurement model.

Our goal is to evaluate the posterior distribution of the parameters $\boldsymbol{\theta}$ given the data $y_{1:T}$, $p(\boldsymbol{\theta}|y_{1:T})$. Using Bayes' rule, this can be written as

$$p(\boldsymbol{\theta}|y_{1:T}) \propto p(y_{1:T}|\boldsymbol{\theta})\, p(\boldsymbol{\theta}), \tag{173}$$

where $p(y_{1:T}|\boldsymbol{\theta})$ is the likelihood function. The likelihood function can be written as

$$p(y_{1:T}|\boldsymbol{\theta}) = \prod_{k=1}^{T} p(y_k|y_{1:k-1}, \boldsymbol{\theta}) = \prod_{k=1}^{T} \int p(y_k|x_k) p(x_k|y_{1:k-1}, \boldsymbol{\theta}) dx_k, \tag{174}$$

where we define $p(y_1|y_{1:0}, \boldsymbol{\theta}) \equiv p(y_1|\boldsymbol{\theta})$. The predictive distribution of the state $p(x_k|y_{1:k-1}, \boldsymbol{\theta})$ can be written as

$$p(x_k|y_{1:k-1}, \boldsymbol{\theta}) = \int p(x_k|x_{k-1}, \boldsymbol{\theta}) p(x_{k-1}|y_{1:k-1}, \boldsymbol{\theta}) dx_{k-1}. \tag{175}$$

The posterior distribution of the hidden state in the integral $p(x_{k-1}|y_{1:k-1}, \boldsymbol{\theta})$ can be written, using Bayes' rule, as

$$p(x_k|y_{1:k}, \boldsymbol{\theta}) = \frac{p(y_k|x_k) p(x_k|y_{1:k-1}, \boldsymbol{\theta})}{p(y_k|y_{1:k-1}, \boldsymbol{\theta})}, \tag{176}$$

where we use $p(y_k|x_k, y_{1:k-1}) = p(y_k|x_k)$. Given the predictive distribution of the state at the initial time point $p(x_1|\boldsymbol{\theta})$, using the Eqns. (175) and (176) recursively, the predictive distributions at the following time points $\{t_k\}$ can be computed, which then gives the likelihood function using Eqn. (174).

Under the assumption that the sequential states $x_{1:T}$ follow an OU process with parameters $\boldsymbol{\theta}$, we can write the transition probability distribution as

$$p(x_k|x_{k-1}, \boldsymbol{\theta}) = \mathcal{N}(x_k; x_{k-1}v, V_x), \tag{177}$$

where $v \equiv e^{-\frac{t_k - t_{k-1}}{\tau_n}}$, $V_x \equiv D_n \tau_n (1 - v^2)$, and $\mathcal{N}(x; \mu, \sigma^2)$ is a Gaussian distribution with mean $\mu$ and variance $\sigma^2$. Also, assuming that the measurement noise follows the Gaussian with zero mean and variance $\sigma_m^2$, we can write the measurement model as

$$p(y_k|x_k) = \mathcal{N}(y_k; x_k, \sigma_m^2). \tag{178}$$

Under these assumptions, one can find closed-form expressions for Eqns. (174)-(176) [47]. Specifically, the likelihood function (Eqn. (174)) can be written as

$$\begin{aligned} p(y_{1:T}|\boldsymbol{\theta}) &= \prod_{k=1}^{T} \int p(y_k|x_k) p(x_k|y_{1:k-1}, \boldsymbol{\theta}) dx_k \\ &= \prod_{k=1}^{T} \mathcal{N}\left(y_k; \mu(x_k|\boldsymbol{\theta}), \sigma_m^2 + V(x_k|\boldsymbol{\theta})\right), \end{aligned} \tag{179}$$

where $\mu(x_k|\boldsymbol{\theta})$ and $V(x_k|\boldsymbol{\theta})$ are the mean and variance of the predictive distribution $p(x_k|y_{1:k-1}, \boldsymbol{\theta})$. Evaluating Eqn. (175), the predictive distribution is written as

$$p(x_k|y_{1:k-1}, \boldsymbol{\theta}) = \mathcal{N}(x_k; \mu(x_k|\boldsymbol{\theta}), V(x_k|\boldsymbol{\theta})) \tag{180}$$

$$= \int p(x_k|x_{k-1}, \boldsymbol{\theta}) p(x_{k-1}|y_{1:k-1}, \boldsymbol{\theta}) dx_{k-1} \tag{181}$$

$$= \mathcal{N}(x_k; v\mu'(x_{k-1}|\boldsymbol{\theta}), V_x + v^2 V'(x_{k-1}|\boldsymbol{\theta})), \tag{182}$$

where $\mu'(x_{k-1}|\boldsymbol{\theta})$ and $V'(x_{k-1}|\boldsymbol{\theta})$ are the mean and variance of the posterior distribution of the state $x_{k-1}$, $p(x_{k-1}|y_{1:k-1}, \boldsymbol{\theta})$. Evaluating Eqn. (176), the posterior distribution $p(x_k|y_{1:k}, \boldsymbol{\theta})$ is written as

$$p(x_k|y_{1:k}, \boldsymbol{\theta}) \equiv \mathcal{N}(x_k; \mu'(x_k|\boldsymbol{\theta}), V'(x_k|\boldsymbol{\theta})) \tag{183}$$

$$= \frac{p(y_k|x_k) p(x_k|y_{1:k-1}, \boldsymbol{\theta})}{p(y_k|y_{1:k-1}, \boldsymbol{\theta})} \tag{184}$$

$$= \mathcal{N}\left(x_k; v \frac{y_{k-1} V(x_k|\boldsymbol{\theta}) + \mu(x_k|\boldsymbol{\theta}) \sigma_m^2}{\sigma_m^2 + V(x_k|\boldsymbol{\theta})}, \frac{\sigma_m^2 V(x_k|\boldsymbol{\theta})}{\sigma_m^2 + V(x_k|\boldsymbol{\theta})}\right), \tag{185}$$

which is dependent on the mean and variance of the predictive distribution $p(x_k|y_{1:k-1}, \boldsymbol{\theta})$. As the predictive distribution at the initial time point, $p(x_1|\boldsymbol{\theta})$, we chose

$$p(x_1|\boldsymbol{\theta}) = \delta(x_1 - y_1). \tag{186}$$

In the above formulation, we assumed that the variance of the measurement noise $\sigma_m^2$ was known. In our case, since the measurement noise is dominated by the shot noise of fluorescent signals, even after the photobleaching correction, $\sigma_m^2$ can slowly increase as more fluorescent proteins photobleach over the course of measurement, $\sigma_m^2 = \sigma_m^2(t_k)$. To estimate $\sigma_m^2(t_k)$, we first segmented the FRET time series into short 12 segments of identical length (100-sec interval with 1 sec sampling interval) and estimated the measurement noise in each segment $\sigma_{m,i}^2$ assuming that the noise level is approximately constant within the segment. The measurement noise at each segment was estimated by

$$\sigma_{m,i}^2 = C(\tau = 0) - A e^{-\frac{\tau}{B}}|_{\tau=0}, \tag{187}$$

where $C(\tau)$ is the autocorrelation function computed from the (mean-subtracted) raw FRET time series and $A e^{-\frac{\tau}{B}}$ is a fit to $C(\tau)$ over the range of $0 < \tau \leq 10$ s (note we exclude the zero lag time point $\tau = 0$). This gives an estimation of the variance of shot noise because $C(\tau = 0)$ estimates the sum of the shot noise and biological noise at lag time zero, while $A e^{-\frac{\tau}{B}}|_{\tau=0}$ estimates the biological noise at lag time zero. In fitting $A e^{-\frac{\tau}{B}}$ to $C(\tau)$, we only used the first 10 seconds because the statistical uncertainty of $C(\tau)$ is relatively large for $\tau > 10$ s. Once the measurement noise of each segment is obtained, we then estimated the measurement noise level at each time point $\sigma_m^2(t_k)$ by fitting a linear function $\sigma_m^2(t_k) = a\, t_k + b$ to $(\{\bar{t}_i\}, \{\sigma_{m,i}^2\})$ (segments $i = 1, \ldots, 12$), where $\bar{t}_i$ is the center of the time interval of segment $i$.

The log posterior distribution $\log p(\boldsymbol{\theta}|y_{1:T}) = \log p(y_{1:T}|\boldsymbol{\theta}) + \log p(\boldsymbol{\theta}) + \text{Const.}$ was approximated by using a Markov chain Monte Carlo method, a slice sampling (*45*), and the mode (i.e., MAP estimate) and 25 and 75 percentiles of the posterior distribution of each parameter was estimated (Fig. S6C). The prior distributions used were

$$P(\tau_n) = \mathcal{U}_{lin}(\tau_n; 0, 10^2)$$
$$P(D_n) = \mathcal{U}_{lin}(D_n; 0, 10^{-2})$$
(188)

where $\mathcal{U}_{lin}(x; a, b) = \frac{1}{b-a}$ for $a \leq x \leq b$ and $\mathcal{U}_{lin}(x; a, b) = 0$ otherwise. The autocorrelation function of the Ornstein-Uhlenbeck process $C_{OU}(\tau) = D_n \tau_n\, e^{-\tau/\tau_n}$ with the MAP estimates of the parameters $\tau_n$ and $D_n$ closely match the autocorrelation function $C(\tau)$ directly computed from the FRET time series. The representative value of each parameter in the population and its uncertainty was evaluated in the same way as those for the signaling-response parameters.

Cell detection
Movies of swimming cells were recorded to measure their behavioral parameters, run speed, rotational diffusion coefficient, and chemotactic drift in varying gradients. All but the chemotactic drift measurements were recorded with a 4X objective, whereas the chemotaxis measurements were recorded with a 10X objective (Methods). All movies were recorded at 20 frames per second by phase-contrast imaging for 1000 seconds. All of the steps and parameters used below were applied to trajectories recorded with both the 10X and 4X objectives, unless noted otherwise. In the chemotaxis experiments, each movie's frames were rotated slightly to align the x-axis with the direction of increasing concentration.

Movies were analyzed using custom MATLAB code. To detect cells in each frame of the movie, first, 5-second blocks of frames were loaded at a time, and the pixel-wise median of those frames was computed as a background image. The background image was then subtracted from each frame in the block. Most of the pixels did not contain a cell; therefore, the histogram of pixel intensities was fit in the vicinity of its peak with a Gaussian distribution to extract the mean and standard deviation of the noise pixel intensities. Since cells appear as dark spots in phase contrast imaging, each frame was subtracted from the background image, making cells become bright spots in the background-adjusted images. Each background-adjusted frame was then smoothed with a weak Gaussian filter, the standard deviation of which depended on the objective used for imaging. The pixel size of the 10X objective is ~0.65 μm, and the filter's standard deviation was set to 1 pixel; The pixel size of the 4X objective is ~1.62 μm, and the filter's standard deviation was set to 0.4 pixels. All pixel intensities in the block that were less than 2 noise standard deviations above the noise mean were set to zero. This left patches of non-zero pixels in each frame, most of which contained cells. Patches consisting of one non-zero pixel were set to zero intensity, because cells typically occupy several pixels. Patches in which the highest-intensity pixel was less than 25 (or 15) noise standard deviations above the noise mean at 10X (at 4X) were also set to zero, only leaving pixel patches that contained cells in focus. In case more than one cell fell within a pixel patch, we used MATLAB's *imregionalmax* function to find local intensity peaks after smoothing each frame again with a Gaussian filter of standard deviation 0.5 pixels, and each local intensity maximum was considered a different cell. Finally, the cells' positions were refined to sub-pixel resolution using the method in [48].

Cell tracking

Linking cell detections into tracks was also done using custom MATLAB code. Before linking, detections were pruned for quality. If two detections occurred within $v_0 \Delta t$ of each other, both detections were removed. For tracking purposes, $v_0$ was set to 50 μm/s and $\Delta t = 50$ ms was the time between frames. In some movies, a cell became immobilized and spun in place, causing it to still be visible after background subtraction. Therefore, pixels in each movie with >15 times more detections than the median number of detections per pixel (among pixels that contained detections) were identified. Detections within 7 (or 3) pixels of these high-detection pixels at 10X (at 4X) were removed.

As a first pass, starting from the first frame, for each detection in the frame, the closest detection in the next frame within a distance of $r = v_0 \Delta t$ was found. Then, for each detection in frame $i$, the closest detection in frame $i + 1$ within distance $r$ was considered the same cell. This process was repeated for each cell in the frame, and then for all frames, in chronological order. After this, tracks that lasted only one frame were considered false positive detections and were removed.

This assignment procedure could leave gaps in the track of a given cell if it disappeared from view for more than one frame, for example by swimming vertically out of the depth of field. Linking tracks across these gaps increases the average time we observe a given cell, allowing us to better estimate its behavioral parameters or drift speed. To close these gaps, we looped over tracks from longest to shortest for a given gap size $k$, in frames. For each track, we searched for tracks that started $k$ frames after (before) the current one within a distance $r(k) = \min(v_0 \Delta t \, k, \sqrt{D \, \Delta t \, k})$ of the current track's end (start). For tracking, we set $D = 300$ μm²/s. If nearby tracks were found, the closest ones were linked to the current track. The loop then continues to the next track. Tracks for which links were made are then revisited. Once no more links were made at gap size $k$, the gap size was increased by 1 frame. This process was repeated up to a maximum of 10 frames (0.5 s). Following this procedure, tracks that lasted only two frames were considered false positives and were removed.

When two cells came close to each other, it was possible that we misassigned their identities after that event. To try to identify and "untangle" these crossings, we first found all events in which two trajectories came within 5 μm of each other. We then generated the hypothetical trajectories resulting from swapping the assignments of the trajectories for all time after the event. Splines were fit to the original two trajectories and the hypothetical trajectories within $\pm 0.2$ s of the event. Then, the average acceleration magnitude along the splines of the original trajectories and along those of the hypothetical trajectory assignments were computed. If the average acceleration of the hypothetical trajectory assignments was less than 90% of the original ones, we performed the swap—doing so preferred trajectory assignments with fewer sharp turns, which could indicate a misassignment, but had a slight preference against making swaps. This process identified sharp turns when two trajectories came close, which could indicate two trajectories crossing each other almost perpendicularly but that were misassigned, and then it attempted to correct those misassignments. This process was repeated for all events when two trajectories came close, in chronological order. The same cell's trajectory could be swapped with other trajectories at multiple points in the movie.

Trajectories were smoothed by convolving their x- and y-positions, separately, with a Gaussian in time of width $\Delta t/2$. Velocities in the x and y directions were computed using first-order central differences: $v_x(t_i) = \big(x(t_i + \Delta t) - x(t_i - \Delta t)\big)/(2\,\Delta t)$ (likewise for $v_y(t_i)$). The swimming speeds projected onto the x-y plane (the z-direction being the objective point of view) at each time point were computed from the magnitudes of the velocity vectors at that time. Projected angular velocities were computed by calculating the angle between the vector incoming and outgoing at each position—i.e. the angle between $\big(x(t_i) - x(t_{i-1}), y(t_i) - y(t_{i-1})\big)$ and $\big(x(t_{i+1}) - x(t_i), y(t_{i+1}) - y(t_i)\big)$—and divided by $\Delta t$. These were then corrected so that the branch cuts at $\pi$ and $-\pi$ did not lead to artificially large angular velocities.

Tumble detection
Tumbles were detected using a variant of the algorithm introduced by Masson et al. [9]. We did not use a speed criterion to detect tumbles, only a criterion based on the projected angular speed. This is because we only see a projection of the cell's trajectory; therefore, if rotational diffusion turns the cell's heading towards or away from the viewing direction, artificial speed variations could be observed. Angular speed is still useful for detecting tumbles because the likelihood of a tumble not causing a change of heading in the x- or y-direction is very small. Briefly, we found each peak in the angular speed of a cell's trajectory and its surrounding troughs in angular speed. The peak angular speed of a tumble had to be greater than 25 rad/s, and the troughs of angular speed surrounding a tumble had to be less than 15 rad/s. For any points within this segment of time to be considered a tumble, the cumulative absolute change in angle from trough to trough had to be larger than $\beta\sqrt{D_r\,\Delta t}$, where $\beta = 4$, $D_r = 0.1$ rad$^2$/s (only for tumble detection), and $\Delta t$ was the time between the troughs. The beginning and end of the tumble were chosen to be the times when the angular speed fell below the smaller of 1/2 the peak angular speed or 15 rad/s, and all points in between were set to the tumble state. Runs that lasted one frame were set to tumbles.

Estimating behavioral parameters, run speed, and rotational diffusion
Behavioral parameters $\theta = \{\lambda_{R0}, \alpha, P_{run}\}$ and run speed $v_0$ were extracted from trajectories of wild-type (RP437) E. coli swimming in the absence of a gradient and recorded with a 4X objective. Our goal was to estimate the parameters of a typical cell. Previous work has shown that behavioral parameters in E. coli are correlated with their tumble bias [23,49,50], $TB = 1 - P_{run}$, and therefore correlated with $P_{run}$. Therefore, we binned tracks by $P_{run}$ (bin size 0.02), and then computed average behavioral parameters within each bin. The average trajectory duration among all cells was 4.7 seconds, and the total trajectory time was $6.3 \times 10^5$ seconds. Cells in bins $P_{run} = 0$ and $P_{run} = 1$ were excluded; these were trajectories for which we could not infer $P_{run}$.

To estimate the median "run bias" $P_{run}$ in the population, we first computed the single-cell run biases from $P_{run} = \frac{t_{run}}{t_{run} + t_{tumble}}$. Here, $t_{run}$ and $t_{tumble}$ are the total time that the cell spent in the run and tumble states, respectively. Then, we computed the time-weighted distribution of $P_{run}$, or the fraction of total trajectory time corresponding to cells of phenotype $P_{run}$. This is necessary to get an unbiased phenotype distribution because cells with high diffusivity (large $P_{run}$ for example) enter *and* leave the field of view more frequently that cells with low diffusivity, whereas cells with low diffusivity spend more time in the field of view but enter less frequently. As a result, the raw counts of each phenotype over the whole movie are biased samples of the population's phenotype distribution, over-counting high-diffusivity phenotypes. However, the distribution of phenotypes in a given frame is unbiased. Therefore, one could average the distribution of phenotypes seen in each frame over all frames of the movie, but this is equivalent to weighting each phenotype by the total time it was observed.

The swimming speed during runs $v_0$ was computed by taking the average of the speed during runs among cells with $P_{run}$ near the population-median value, with each cell given weight proportional to its duration. This estimate alone would be biased by our projected viewpoint. Assuming that in shallow gradients the headings are uniformly distributed over all possible directions, $v_0$ is expected to be underestimated by a factor of $\pi/4$. Therefore, to correct for the projection bias, we multiplied the computed value of $v_0$ by $4/\pi$. This $v_0$ was used to compute $v_d/v_0$ in Fig. 3C of the main text. Mean speeds in the behavior experiments ($v_0 = 22.61 \pm 0.07$ µm/s) were comparable to those in the gradient experiments ($v_0 = 21.9 \pm 0.2$ µm/s).

The persistence of tumbles $\alpha$ was computed from the average of the cosine of the angle between cells' projected headings before and after tumbles. Since we observe projections of the true headings, this slightly overestimates the true value of $\alpha$. We consider a model to correct for this bias. In the absence of a gradient, cells don't have a preference for one direction over another, so the distribution of new headings after a tumble must be centered around the heading before the tumble. This is encoded in a model by making the distribution of post-tumble headings $\hat{u}$, given the pre-tumble heading $u$, depend only on $u \cdot \hat{u}$. Celani and Vergassola[8] used the tumble angle distribution $P(\hat{u}|u) = \frac{1}{4\pi}(1 + u \cdot \hat{u})$, which is the simplest approximation of a general tumble angle distribution that could have more complex dependence on $u \cdot \hat{u}$. In this model, the true directional persistence is $\alpha = \langle u \cdot \hat{u} \rangle = 1/3$, but if only projected trajectories are observed, the observed value is $\alpha = \pi^2/32$ (~0.31). Therefore, we multiplied our computed value of $\alpha$ by $1/3 * 32/\pi^2$ to correct for projection biases.

We estimated the mean tumble rate $\lambda_{R0}$ in two ways. First, assuming exponentially-distributed runs, the probability that a cell tumbles in time step $dt$ is $p_{tumble,dt} = \lambda_{R0} \Delta t$, where $\Delta t = 50$ ms is the frame interval of the movie. We calculated $p_{tumble}$ as the total number of tumbles observed divided by the total number of cell-frames in which runs were observed, again only among cells with $P_{run}$ near the population-median. Then we compute the mean tumble rate from $\lambda_{R0} = p_{tumble,dt}/\Delta t$, giving $\lambda_{R0} \sim 0.893 \pm 0.003$ s$^{-1}$. This was the value used to compute $F(\theta)$ in the efficiency and $\beta$ in the information rate $\dot{I}_{s \to a}$. Another method was to compute the mean tumble rate from the median $P_{run}$ and average tumble time $\tau_T$ over cells with $P_{run}$ near the median, using $P_{run} \sim 1/(1 + \lambda_{r0} \tau_T)$, which gave $\lambda_{R0} \sim 0.912 \pm 0.003$ s$^{-1}$.

Behavioral parameters also appear in the up-gradient velocity autocorrelation function $V(t)$, which is used to estimate the information rate $\dot{I}_{s \to a}$. Although the cells were not swimming in a gradient, their velocity statistics in the absence of a gradient are approximately the same as in shallow gradients. Therefore, we estimated $V(t)$ by computing the autocorrelation of velocity along one dimension, chosen arbitrarily to be the x-axis of our field of view, which is the gradient direction in the gradient experiments, and averaged over cells within bins of $P_{run}$. The resulting correlation functions decayed roughly exponentially (Fig. 2D; Fig. S2A), and we fit the correlation function of the median bin of $P_{run}$ with the functional form $V(t) = a_v e^{-\lambda_{tot} |t|}$ by nonlinear least squares using MATLAB's *fit* function. In the fit, each time delay $t$ was given weight proportional to the number of samples observed at that time delay. The first few time points of $V(t)$ were excluded from the fit because they include sharp drops in velocity correlation that result from tumbles inevitably having finite velocity that rapidly decorrelates. In the median bin of $P_{run}$, the average trajectory duration was 7.0 seconds, and the total trajectory time was $1.3 \times 10^4$ seconds. Comparing to the model for $V(t)$ earlier, in theory, $\lambda_{tot} = (1-\alpha)\lambda_{R0} + 2 D_r$ and $a_v = \frac{v_0^2}{3} P_{run}$. The fit values were comparable to these theoretical expressions using estimates of the individual parameters above and an estimate of $D_r$ (below): $\lambda_{tot} = 0.862 \pm 0.005$ s$^{-1}$, compared to $(1-\alpha)\lambda_{R0} + 2 D_r = 0.93 \pm 0.01$ s$^{-1}$; $a_v = 157.1 \pm 0.5 \left(\frac{\mu m}{s}\right)^2$, compared to $\frac{v_0^2}{3} P_{run} = 151 \pm 2 \left(\frac{\mu m}{s}\right)^2$. We directly used the fit values $a_v$ and $\lambda_{tot}$ in $V(\omega)$ when computing the information rate $\dot{I}_{s \to a}$.

Rotational diffusion was measured from trajectories of cells lacking the gene for *cheY* and therefore shouldn't be able to tumble. For these cells, $\lambda_{R0} = 0$, so we extracted the rotational diffusion coefficient again by fitting their average velocity autocorrelation function. We filtered out trajectories that were shorter than 5 seconds or that appeared to tumble. After filtering, the average trajectory duration was 9.5 seconds, and the total trajectory time was $9.7 \times 10^4$ seconds. The resulting velocity autocorrelation function was well-fit by a single decaying exponential (Fig. S2B), whose decay rate $\lambda_{tot, \Delta cheY} = 2 D_r$ was consistent with a previously reported[21] value for $D_r$ (here, $D_r = 0.0441 \pm 0.0001$ rad$^2$/s; previously $D_r = 0.062$ rad$^2$/s).

Uncertainties in $v_0$, $\alpha$, and $\lambda_{R0}$ and their dependence on $P_{run}$ were computed by bootstrapping tracks from within each bin of $P_{run}$, excluding cells with $P_{run} = 0$ or $P_{run} = 1$ (bin size 0.02). The bootstrapped sample for each bin was the same size as the number of tracks in the bin, and samples were drawn with replacement 100 times. From each bootstrapped sample, the average of each parameter was computed. The standard deviation of the average of each parameter among the bootstrapped samples was taken to be the uncertainty. Uncertainties of the median $P_{run}$ were smaller than the bin size and therefore were taken to be half the size of a bin. The uncertainties of $a_v$, $\lambda_{tot}$, and $D_r$ were determined from the uncertainty of the exponential fit: the 68% confidence interval of each parameter was determined from MATLAB's *fit* function; dividing by 2 gave uncertainties equivalent to one standard deviation, assuming normally-distributed parameter uncertainties.

Estimating population-average drift speeds

To compute the drift speeds, we first computed the average $x$-velocity among cells in each frame of a given movie, $\langle v_x(t) \rangle$. This time course of "ensemble average" $\langle v_x(t) \rangle$ resembled an OU process with correlation time of about 1.5 s, similar to the cells' typical reorientation time $1/\lambda_{tot}$ (Fig S9). The drift speed $v_d$ was computed as the average of the time course $\langle v_x(t) \rangle$. This is equivalent to computing the drift speed of each cell and taking a weighted average, with weights given by the duration of each cell's trajectory. Time-weighted averaging like this is necessary to get an unbiased estimate of the population average drift speed, even for a population without diversity. This is because, with a finite depth of field, there should be disproportionately more short trajectories swimming vertically, with small up-gradient displacement, than long up-gradient trajectories. However, the distribution of swimming directions in a given frame is unbiased, and time-averaging takes advantage of this.

Since the time course of $\langle v_x(t) \rangle$ has a finite correlation time, the values of $\langle v_x(t) \rangle$ at consecutive time points are not independent. To compute the uncertainty of $v_d$, we first computed the autocorrelation function of $\langle v_x(t) \rangle$ in each movie and fit the result with a decaying exponential $a + b\, e^{-t/\tau_c}$ to get the correlation time of fluctuations $\tau_c$. The effective number of independent frames was then $n_I = n \frac{\Delta t}{\Delta t + 2\,\tau_c}$ [3], where $n$ is the total number of frames in the movie (minus outliers). The uncertainty (standard error) of $v_d$ in a given experiment $i$ was then computed as $\sigma_i = \sigma_{v_x}/\sqrt{n_I}$, where again $\sigma_{v_x}$ was the variance of $\langle v_x(t) \rangle$ in that experiment.

Then, we computed the weighted average of drift speeds $v_{d,i}$ in the same gradient steepness to determine $v_d(g)$. Each $v_{d,i}$ was weighted inversely proportional to its squared standard error $1/\sigma_i^2$. To compute the uncertainty of this average, we first note that there were experiment-to-experiment variations in the mean drift speeds $v_{d,i}$ that exceeded the typical within-experiment uncertainty $\sigma_i$. This suggested that there was another random, experiment-to-experiment source of variation. This could result from observing a different sample of phenotypes from the population in each experiment. To account for these effects in the uncertainty of the average $v_d(g)$, we assume that a random effect is added to the drift speed in each experiment, with variance $\sigma_r^2$. This random effect is assumed to be uncorrelated with our measurement errors, and we estimated $\sigma_r^2$ using the variance of $v_{d,i}$ among experiments with the same gradient steepness. Together, the uncertainty of the average $v_d(g)$ is then given by $\sigma^2 = 1/\left(\sum_i \frac{1}{\sigma_i^2}\right) + \frac{1}{(N-1)}\sigma_r^2$, where $N$ is the number of experiments performed in that gradient condition. The first term results from weighting each $v_{d,i}$ in the mean by the inverse of its measurement variance. The second term results from averaging over $N$ realizations of the random effect, which all have the same statistics.

The average trajectory duration was 1.52 seconds, the average trajectory time per experiment was $2 \times 10^4$ seconds, and the total trajectory time was $5.5 \times 10^5$ seconds. Multiple experiments were performed for each gradient condition ($N \geq 5$ each).

The gradient steepness $g$ in each experiment was estimated from fluorescein fluorescence images as follows. The background fluorescence intensity was measured by taking an image of a device filled with water but no fluorescein. The average intensity of this image was similar to that of an image with no sample mounted on the microscope. The average intensity of this background image was subtracted from all of the fluorescein gradient images, making zero-intensity regions correspond to regions where the MeAsp attractant concentration was 100 µM. To correct the influence of relative depth variations across the width of the gradient region and for spatial variations in illumination, an image of a device filled with fluorescein solution was taken. We will refer to this as the "blank" image. The average background intensity was subtracted from this image, as well. Then, in each gradient image and the blank image, the average intensity in a horizontal strip was computed ($\pm 300$ pixels around row 1024; image size 2048 x 2048 pixels). This produced fluorescence intensity profiles across the gradient region, $I(x,t)$. The fluorescence profile from the blank image, $I_b(x,t)$, was normalized to a maximum value of 1. The fluorescence profiles at each time point of each experiment were aligned to the blank profile, and divided point by point by the normalized blank profile: $i(x,t) = I(x,t)/I_b(x,t)$. We separately quantified the effects of photobleaching by imaging a device full of fluorescein using the same protocol as a gradient experiment. Changes in intensity due to photobleaching were negligible (<0.2%).

At this point, we had fluorescence profiles that were corrected for background intensity, illumination variations, and relative variations of device depth. What remained was to determine the linear transformation from intensity, $i(x,t)$, to MeAsp concentration, $c(x,t)$, which could be experiment-dependent due to variations in absolute dimensions of different devices. In addition to knowing that $i(x,t) = 0$ corresponds to $c(x,t) = 100$ µM, we also know that the maximum value of $i(x,t)$, $i_{max}$, corresponds to $c(x,t) = c_1$, where $c_1$ is the concentration of MeAsp in the high-concentration reservoir. To estimate the maximum $i(x,t)$ in each experiment, we used the earliest fluorescein image from a given experiment, and used the value of $i(x,t)$ at the location in the high-concentration reservoir that was furthest from the gradient region. The reasoning for this was that diffusion or flow between the reservoirs could transport fluorescein (and MeAsp) from the high-concentration reservoir to the low one. The location furthest from the gradient region at the first time point is least affected by diffusion or flow, providing the best estimate of the maximum $i(x,t)$ before any mixing between reservoirs occur. With this, we could estimate concentrations from intensities using $c(x,t) = c_0 + \frac{(c_1-c_0)}{i_{max}} i(x,t)$, where $c_0 = 100$ µM.

Finally, we used this transformation to estimate the concentration of MeAsp in each reservoir in the image taken just before the tracks movie and the time point just after it. We estimated $g$ in each image using $g_{est} = \frac{\log(c_{1,est}) - \log(c_{0,est})}{\Delta x}$, where $c_{1,est}$ and $c_{0,est}$ are the estimated MeAsp concentrations in the two reservoirs, and $\Delta x = 1$ mm is the width of the gradient channel (see Methods). The final estimate of the gradient during the movie was the average of these two estimates. We took the uncertainty of the estimate to be the difference between the estimates at the two time points divided by two.

In all experiments, $g_{est} < g$; the estimated $g$ was less than the intended one, consistent with mixing between the reservoirs making the gradient shallower. The amount by which $g_{est}$ was smaller than $g$ scaled with $g$. This is consistent with the fact that diffusive and convective fluxes scale linearly with concentration, so in the same amount of time (the typical time it took for the gradient to form), steeper gradients should have proportionally more molecules of MeAsp transported from the high reservoir to the low reservoir. The average percent reduction in gradient steepness relative to the intended steepness was 4.4%.

The chemotactic coefficient was estimated by linearly fitting $v_d = \chi g$ to the scatter plot of $v_{d,i}$ versus $g_{est}$ from all gradient experiments, with each data point given weight inversely proportional to the squared standard error of $v_{d,i}$, $1/\sigma_i^2$, using MATLAB's *fit* function. Since the error bars on $g_{est}$ were small, performing the fit in a way that accounted for errors in both $v_{d,i}$ and $g_{est}$ had little effect on the result. Allowing the y-intercept of the fit to be non-zero, i.e. fitting $v_d = a + \chi g$, only increases $\chi$ by 3%, but increases its uncertainty by 70%.

Throughout, we were careful to estimate parameters of a median cell. But here, we are computing the population-average drift speed. But drift speed depends nonlinearly on behavioral parameters, and the average of a nonlinear parameter combination does not in general equal the nonlinear function evaluated with median parameters. Therefore, the properties of the phenotype that achieves the population-average drift speed could be different from the median phenotype. To address this, we note that from theory (i.e. Eqn. (17)), most of the dependence of the drift speed on behavioral parameters is captured by $v_0^2 \frac{(1-\alpha)\lambda_{R0}}{(1-\alpha)\lambda_{R0}+2 D_r} P_{run}$. From swimming trajectories, we computed the average of each parameter in bins of $P_{run}$, i.e. $v_0(P_{run})$, $\lambda_{R0}(P_{run})$, etc (see Fig. S1). Then, we computed this expression for each bin (Fig. S9), as well as the distribution of $P_{run}$. Finally, we compared the average of this expression with respect to the distribution of $P_{run}$ to the value one gets from plugging in the parameters corresponding to the median $P_{run}$. We find that these two are similar: the population average gives $\left\langle v_0^2(P_{run}) \frac{(1-\alpha)\lambda_{R0}}{(1-\alpha)\lambda_{R0}+2 D_r} P_{run} \right\rangle \sim 375 \pm 1 \left(\frac{\mu m}{s}\right)^2$, whereas plugging parameters in the median bin of $P_{run}$ gives $v_0^2 \frac{(1-\alpha)\lambda_{R0}}{(1-\alpha)\lambda_{R0}+2 D_r} P_{run} \sim 410 \pm 3 \left(\frac{\mu m}{s}\right)^2$. This justifies our comparison of population-average drift speeds to bounds quantified using a median cell's parameters. Uncertainties in the distribution of $P_{run}$ and in these quantities was determined by bootstrapping, as described in the **Estimating behavioral parameters, run speed, and rotational diffusion** section.